\documentclass[iop]{emulateapj}

\usepackage{longtable}
\usepackage{rotating}
\def\simlt{\lower.5ex\hbox{$\; \buildrel < \over \sim \;$}}

\begin{document}

\shorttitle{Weak lensing in CFHTLS-Wide}
\shortauthors{Shan et al.}
\title{Weak lensing measurement of galaxy clusters in the CFHTLS-Wide survey}

   \author{HuanYuan Shan\altaffilmark{$\dagger$1,2,3}, Jean-Paul Kneib\altaffilmark{2},
   Charling Tao\altaffilmark{1,3}, Zuhui Fan\altaffilmark{4}, Mathilde Jauzac\altaffilmark{2}, Marceau Limousin\altaffilmark{2, 5},
   Richard Massey\altaffilmark{6}, Jason Rhodes\altaffilmark{7,8}, Karun Thanjavur\altaffilmark{9,10,11} and Henry J. McCracken\altaffilmark{12}}
   \altaffiltext{1}{Department of Physics and Tsinghua Center for Astrophysics, Tsinghua University, Beijing, 100084, China}
   \altaffiltext{2}{Laboratoire d'Astrophysique de Marseille, CNRS-Universit\'e de Provence, 38 rue Fr\'ed\'eric Joliot-Curie, 13\,388 Marseille Cedex 13, France}
   \altaffiltext{3}{Centre de Physique des Particules de Marseille, CNRS/IN2P3-Luminy and Universit\'e de la M\'editerran\'ee, Case 907, F-13288 Marseille Cedex 9, France}
   \altaffiltext{4}{Department of Astronomy, Peking University, Beijing, 100871, China}
   \altaffiltext{5}{Dark Cosmology Centre, Niels Bohr Institute, University of Copenhagen, Juliane Maries Vej 30, DK-2100 Copenhagen, Denmark}
   \altaffiltext{6}{Institute for Astronomy, Royal Observatory, Blackford Hill, Edinburgh EH9 3HJ, UK}
   \altaffiltext{7}{California Institute of Technology, MC 350-17, 1200 East California Boulevard, Pasadena, CA 91125, USA}
   \altaffiltext{8}{Jet Propulsion Laboratory, California Institute of Technology, Pasadena, CA 91109, USA}
   \altaffiltext{9}{Canada France Hawaii Telescope, 65-1238 Mamalahoa Hwy, Kamuela, HI 96743, USA}
   \altaffiltext{10}{Department of Physics \& Astronomy, University of Victoria, Victoria, BC, V8P 1A1, Canada}
   \altaffiltext{11}{National Research Council of Canada, Herzberg Institute of Astrophysics, 5071 West Saanich Road, Victoria, BC, V9E 2E7, Canada}
   \altaffiltext{12}{Institude d'Astrophysique de Paris, UMR 7095, 98 bis Boulevard Arago, 75014 Paris, France}
   \altaffiltext{$\dagger$}{Email address: {\tt shanhuany@gmail.com}}

\begin{abstract}
We present the first weak gravitational lensing analysis of the completed
Canada-France-Hawaii Telescope Legacy Survey (CFHTLS).
We study the $64~{\rm deg}^2$ W1 field, the largest of the CFHTLS-Wide survey fields,
and present the largest contiguous weak lensing convergence ``mass map'' yet made.

2.66 million galaxy shapes are measured, using a Kaiser, Squires and Broadhurst (KSB) pipeline verified against
high-resolution {\em Hubble Space Telescope} imaging that covers part of the CFHTLS.
Our $i'$-band measurements are also consistent with an analysis of independent $r'$-band imaging.
The reconstructed lensing convergence map contains $301$ peaks with signal-to-noise ratio $\nu>3.5$,
consistent with predictions of a $\Lambda$CDM model.
Of these peaks, $126$ lie within $3\arcmin.0$ of a brightest central galaxy identified from multicolor optical imaging in an independent, red sequence survey.
We also identify seven counterparts for massive clusters previously seen in X-ray emission within $6~\rm deg^{2}$ XMM-LSS
survey.

With photometric redshift estimates for the source galaxies, we use a tomographic lensing method
to fit the redshift and mass of each convergence peak.
Matching these to the optical observations, we confirm $85$ groups/clusters with $\chi^2_\mathrm{reduced}<3.0$,
at a mean redshift $\langle z_c\rangle=0.36$ and velocity dispersion $\langle\sigma_c\rangle=658.8~\rm km~s^{-1}$.
Future surveys, such as DES, LSST, KDUST and EUCLID, will be able to apply these techniques to
map clusters in much larger volumes and thus tightly constrain cosmological models.

\end{abstract}


\section{Introduction}

Clusters of galaxies are the largest
gravitationally bound structures in the universe.
The number and mass of the biggest clusters are highly sensitive to cosmological parameters
including the mass density $\Omega_m$, the
normalization of the mass power spectrum $\sigma_8$ (e.g., \ Press \& Schechter
1974; Frenk et al.\ 1990; Eke et al.\ 1996; Sheth \& Tormen 1999), and the dynamics of dark energy
(e.g., \ Bartelmann et al.\ 2006; Francis et al.\ 2009; Grossi \& Springel 2009).
Understanding the properties of clusters is vital to test theories of structure formation
and to map the distribution of cosmic matter on scales of $\sim$1--10\,Mpc.

Theoretical predictions of structure formation deal directly with the total mass of clusters;
measurements are restricted to indirect proxies that can be observed.
Contaminating the translation between theory and observation are
large uncertainties in the interpretation of galaxy richness, X-ray luminosity/temperature
and the Sunyaev-Zeldovich decrement (e.g.\ Bode et al.\ 2007; Leauthaud et al.\ 2010).
Weak gravitational lensing, the coherent
distortion of galaxies behind a cluster, can potentially provide direct measurements of the
total mass regardless of its baryon content, dynamical state, and star formation history.

By measuring the shear (coherent elongation) of many background galaxies, we can reconstruct
the two-dimensional (2D) weak lensing convergence map, which is proportional to density projected along
each line of sight. Peaks in the convergence map with high signal-to-noise ratio $\nu$ generally
correspond to massive clusters
(Hamana et al.\ 2004; Haiman et al.\ 2004).
Since the three-dimensional (3D) shear signal should increase behind those clusters in
a predictable way that depends upon only the lens-source geometry, we can also use photometric redshift
estimates of the background galaxies (from multi-band imaging) to measure the redshift and mass
of each foreground cluster
(Wittman et al.\ 2001, 2003; Hennawi \& Spergel 2005; Gavazzi \& Soucail 2007).

Systematic weak lensing cluster searches have only recently become practicable.
Miyazaki et al.\ (2002) used Subaru/Suprime-Cam
in excellent seeing conditions to find an excess of $4.9 \pm
2.3$ convergence peaks with $\nu>5$ in an area of $2.1~\rm deg^2$.
Dahle et al. (2003) and Schirmer et al. (2003) each identified several
shear-selected clusters with redshifts $z\sim0.5$ determined from two-color photometry.
Hetterscheidt et al.\ (2005)
reported the detection of five cluster candidates over a set of
$50$ disconnected Very Large Telescope/FORS images covering an effective area of
$0.64~\rm deg^2$, while Wittman et al.\ (2006) found eight
detections in the first $8.6~\rm deg^2$ of the {\em Blanco} Deep Lens Survey.
Gavazzi \& Soucail
(2007) presented a weak lensing analysis of initial Canada-France-Hawaii Telescope Legacy Survey (CFHTLS) Deep data
covering $4~\rm deg^2$. They demonstrated that the image
quality at CFHT is easily sufficient for cluster finding.
Miyazaki et al.\ (2007) presented the first large sample of
weak lensing-selected clusters in the Subaru weak lensing
survey, with $100$ significant convergence peaks in a $16.7~\rm
deg^2$ effective survey area. Hamana et al.\ (2009) reported results
from a multi-object spectroscopic campaign to target $36$ of these cluster
candidates, of which $28$ were confirmed (and $6$ were projections
along a line of sight of multiple, small groups).
%Hamana, Takada \& Yoshida (2004) and Haiman et al.\ (2004) have made
%interesting predictions for weak lensing cluster surveys with future telescopes.
%(Moved to preceding paragraph since these now-old papers were predictions for the *current* generation of surveys!)

The main astrophysical systematic effect afflicting weak lensing cluster
surveys is the projection of large-scale structure along the line of sight.
Random noise is also added due to the finite density of resolved source galaxies and the scatter of their intrinsic shapes.
Numerical studies (White et al.\ 2002; Hamana et al.\ 2004)
show that these contaminants significantly reduce the purity of cluster detection.
To improve our analysis, we shall combine our weak lensing results with multi-wavelength imaging.
Simultaneous detection of a weak lensing signature plus an overdensity of galaxies
with a single red sequence provides an unambiguous cluster identification.
Furthermore, 3D lensing tomography using photometric redshifts from the multi-wavelength data
can remove the other potential hurdles of: lensing signal dilution by cluster member galaxies,
and identifying the redshift of weak lensing peaks when no corresponding galaxy overdensity is
apparent.

Here we present a weak gravitational lensing analysis of the
$64~\rm deg^2$ CFHTLS-Wide W1 field, which is sufficiently
large to contain several hundred galaxy clusters.
Compared to the analysis of the CFHTLS-Deep survey by Gavazzi \& Soucail (2007),
our shallower CFHTLS-Wide imaging (and lower source galaxy density) will favor the detection of
higher mass, nearby clusters. The huge increase in survey area over any previous survey is expected to yield many more
systems overall.
%; and when constraints are drawn on cosmological parameters via cluster number counts, \com{in general
%it is area rather than depth that most efficiently reduces Poisson shot noise (e.g.\ Amara \& Refregier 2007) ???
%does this comment make sense? The purity for the Deep survey was much better than for the W1 survey because of the
%"much lower galaxy density".}.
In this paper, we shall primarily study the properties of the detected clusters, rather than the cosmology in which they are embedded.
For this purpose, we adopt a default cosmological model with $\Omega_m=0.27$, $\Omega_{\Lambda}=0.73$,
$\sigma_8=0.809$, $H_0=100~h~\mathrm{km}~{\mathrm s}^{-1}~\mathrm{Mpc}^{-1}$, and $h=0.71$.

This paper is organized in the following way.
In Section~2, we describe the CFHT and {\em Hubble Space Telescope} ({\em HST}) data used.
In Section~3, we present the measurement of galaxy shapes in the CFHT imaging, and their calibration
against measurements of the same galaxies in the {\em HST} imaging.
In Section~4, we reconstruct the 2D lensing convergence ``mass map'' signal,
and extract a catalog of local maxima that represent cluster candidates.
In Section~5, we search for optical counterparts of these candidates, dramatically cleaning the catalog.
In Section~6, we investigate the full 3D lensing signal around each cluster,
further cleaning the catalog when the lensing signal behind spurious peaks does not increase as expected with redshift -
but obtaining an independent estimate of the cluster redshift when it does.
We finally explore global scaling relations between cluster mass observables, then conclude in Section~7.

\section{Data} \label{sec:data}

\subsection{CFHTLS-Wide T0006 imaging}

The CFHT Legacy Survey is a joint Canadian-French program to make efficient use
of the CFHT wide field imager MegaPrime, simultaneously addressing
several fundamental questions in astronomy. Each MegaPrime/MegaCam
image consists of an array of $9 \times 4$ e2v CCDs with a pixel scale of $0\arcsec.187$
and a total field of view of $\sim1~\mathrm{deg}^2$. The survey used most of the telescope
dark and gray time from 2003 to 2008. We analyze CFHTLS-Wide imaging from the Terapix T0006 processing
run, which is the first to include the complete survey and was publicly released on 2010 November 15 (Goranova et al. 2009).
These data cover $\sim 171~\rm deg^2$ in four
fields (W1, W2, W3, and W4) of which the $72$-pointing, $\sim64~\rm deg^2$ W1 field is the largest,
and in five passbands ($u',g',r',i'$,and $z'$) down to $i \sim 24.5$ and $r \sim 25.0$.

Fu et al.\ (2008) showed that the $i'$-band exposures, taken in sub-arcsecond seeing conditions,
provide the best image quality and resolve the galaxy population with highest median redshift. % for a fixed exposure time.
Resolving the shapes of more distant galaxies is vital for weak lensing analysis, since the strength of
the shear signal is proportional to the ratio of the Lens--Source and Observer--Source distances.
We therefore choose to analyze the $i'$-band images
(mean seeing $0".73$) in the contiguous W1 field.
We also analyze the independent $r'$-band imaging to check the calibration of our shear measurements.
We also use photometric redshift estimates for source galaxies obtained from the multicolor imaging
(Ilbert et al.\ 2006; Coupon et al.\ 2009; Arnouts et al.\ 2010).

Early releases of smaller regions of the CFHTLS have also been used to measure the weak lensing cosmic shear signal
(Semboloni et al.\ 2006; Hoekstra et al.\ 2006; Fu et al.\ 2008). As the survey size has increased, the
statistical errors have shrunk, and difficulty measuring shapes at a precision better than the statistical error
has so far prevented publication of a cosmic shear analysis of the complete survey. However, weak lensing
cluster searches are restricted by construction to regions of the survey where the signal is strongest, and
the circular symmetry of our analysis removes the negative impact of additive shear measurement errors
(cf.\ Mandelbaum et al.\ 2005).

\subsection{{\em HST} COSMOS Imaging}

The {\em HST} COSMOS survey (Scoville et al.\ 2007) is the largest contiguous optical
imaging survey ever conducted from space. High resolution ($0\arcsec.12$) imaging in the $I_{F814W}$ band
was obtained during 2003--2005 across an area of 1.64 $\rm deg^2$
that also corresponds to the CFHTLS D2 deep field.
Any galaxies resolved by CFHT are very easily resolved by {\em HST}, which therefore provides highly
accurate shape measurements almost without the need for point-spread function (PSF) correction.
We shall calibrate our CFHTLS shape measurements against those from COSMOS by Leauthaud et al.\ (2010).

Note that measurements of the shapes of individual galaxies from ground-based and space-based observations
need not necessarily match exactly, even without shape measurement errors, because
the different noise properties of the data sets may make them most sensitive to
different isophotes, which can be twisted relative to each other. The slightly different
passbands may also emphasize different regions of a galaxy's morphology. However, across a large population
of galaxies, these differences should average out, and a comparison of successful shear measurements between
the two data sets should agree.

\section{Galaxy Shape Measurement} \label{sec:smm}

\begin{table}
\centering
\caption{\rm SExtractor Configuration Parameters}
\begin{tabular}{lc}
\hline
\hline
\rm Parameter & Value \\
\hline
\rm DETECT\_MINAREA & 3 \\
\rm DETECT\_THRESH & 1.0 \\
\rm DEBLEND\_NTHRESH & 32 \\
\rm DEBLEND\_MINCONT & 0.002 \\
\rm CLEAN\_PARAM & 1.0 \\
\rm BACK\_SIZE & 512 \\
\rm BACK\_FILTERSIZE & 9 \\
\rm BACKPHOTO\_TYPE & \rm local \\
\rm BACKPHOTO\_THICK & 30 \\
\hline
\label{tab:tab1}
\end{tabular}
\end{table}

\subsection{Object Detection and Masking}

\begin{figure}
\begin{center}
\protect
\includegraphics[angle=0.0,width=1.00\columnwidth]{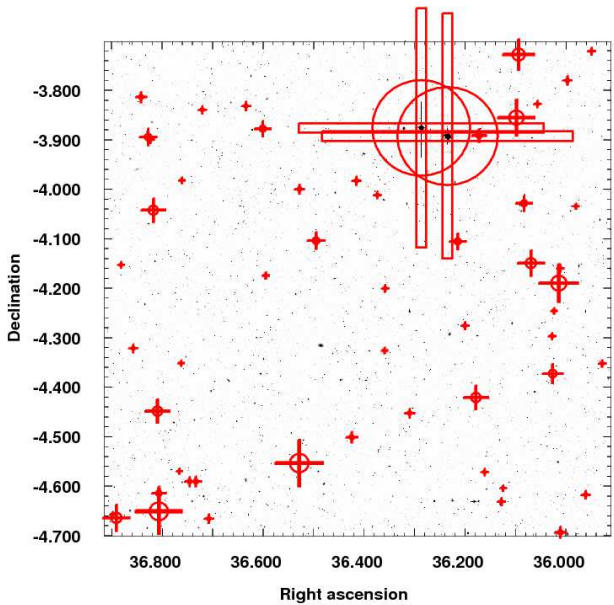}
\caption{``W1+2+3'' pointing from the CFHTLS-Wide W1 field in $i'$ band, showing masked regions.
This pointing is representative of those with fairly poor image quality:
the seeing of $0\arcsec.78$ is worse in only 24 of 72 (1 in 3) pointings.
In our automated algorithm for masking diffraction spikes around bright stars,
the basic shape of the star mask is predefined, and its size is scaled with the observed major axis of each star.
\label{fig:fig1}}
\end{center}
\end{figure}

We conduct shape measurement in both CFHTLS $i'$ and $r'$ bands.
We detect astronomical sources in the images using SExtractor (Bertin \& Arnouts 1996).
Our choice of the main SExtractor parameters is listed in Table~\ref{tab:tab1}, and
the data are filtered prior to detection by a $3$ pixel Gaussian kernel.

\begin{figure}
\centering
\begin{minipage}[t]{8.5cm}
\begin{center}
\includegraphics[trim=0 0 0 0,width=1.0\textwidth]{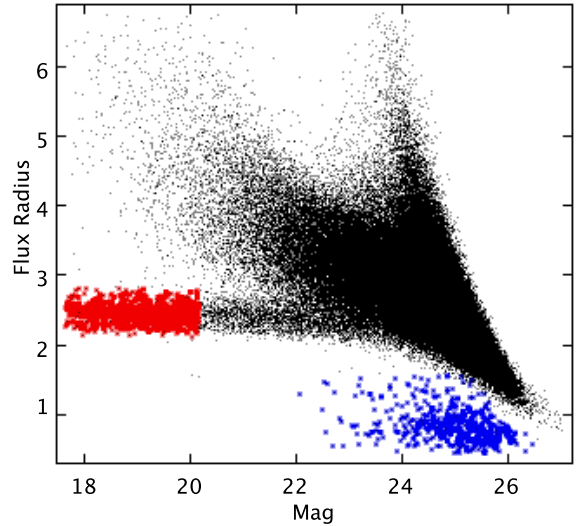}\\[0.1cm]
\includegraphics[trim=0 0 0 0,width=1.0\textwidth]{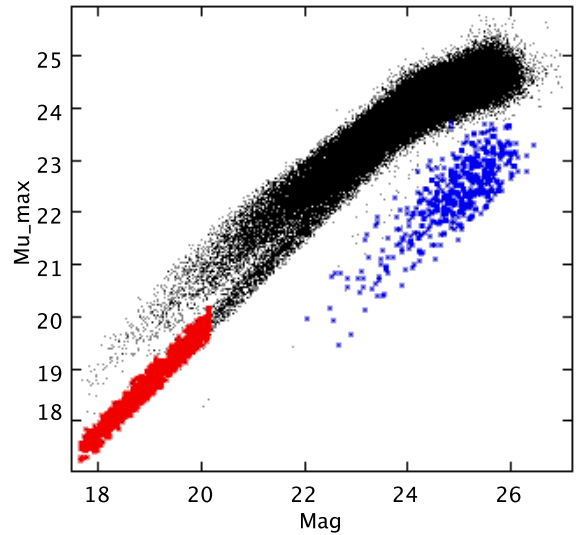}\\[0.1cm]
\end{center}
\end{minipage}
\caption{Star selection (red points) in the planes of magnitude vs. flux radius (top) and
magnitude vs. peak surface brightness (bottom). We find the latter more robust.
The red points denote objects selected as stars for PSF modeling. The blue
objects are spurious detection.
\label{fig:fig2}}
\end{figure}

Near saturated stars, many spurious objects are found due to detector effects and optical ghosting.
It would also be difficult to measure the shapes of real stars or galaxies in these regions, because of the steep background gradients.
We have developed an automatic pipeline to define polygonal-shaped
masks around saturated stars, and all objects inside the masks are removed from our catalog.
The masks in all images are then visually inspected; our automated pipeline fails in a few cases
(mainly very saturated stars for which the centroid of the star measured by {\tt SExtractor} was widely offset from the
diffraction spikes) and those stellar masks are corrected by hand.
An example of the masks for one CFHT pointing is shown in Figure~\ref{fig:fig1}.
This pointing has slightly worse than average image quality, so we shall use it
throughout this paper as a conservative representation of our analysis.
After applying all of our masks across the entire survey,
the final effective sky coverage drops
% from $72~\rm deg^2$ to $\sim57.9~\rm deg^2$ and $62.1~\rm deg^2$
from $64~\rm deg^2$ to $\sim51.3~\rm deg^2$ and $55.0~\rm deg^2$
for $i'$ and $r'$ bands, respectively.

We shall employ the popular KSB method for galaxy shear measurement (Kaiser
et al.\ 1995; Luppino \& Kaiser 1997; Hoekstra et al.\ 1998).
In this method, the observed galaxy shape is modeled as a convolution of the (sheared)
galaxy with the PSF, which is modeled as an isotropic,
circular profile convolved with a small anisotropy.

\subsection{PSF Modelling}

\begin{figure}
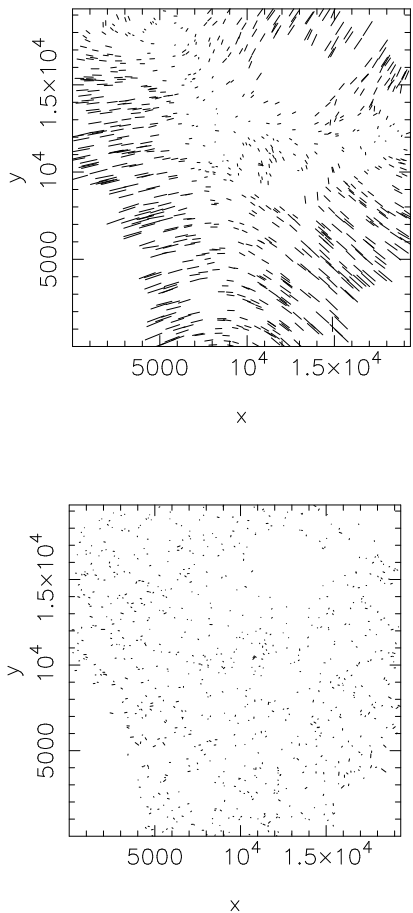

\centering
\begin{minipage}[t]{8.5cm}
\begin{center}
\includegraphics[angle=270.0,trim=0 0 0 0,width=1.0\textwidth]{fig3a.eps}
\includegraphics[angle=270.0,trim=0 0 0 0,width=1.0\textwidth]{fig3b.eps} \\[0.1cm]
\end{center}
\end{minipage}
\caption{Spatial variation of measured stellar ellipticities in the representative CFHTLS-Wide W1+2+3 field,
before (top) and after (bottom) PSF anisotropy correction.
%Black tick marks show raw measurements of the ellipticity and orientation of stars' major axis; red tick marks show our polynomial fit.
The longest tick marks represent ellipticities of $\sim11\%$.
The mean absolute ellipticity after correction is $0.62\%$.
\label{fig:fig3}}
\protect
\end{figure}

To measure the shapes of galaxies, it is first necessary to correct them for convolution with the
PSF imposed by the telescope optics and Earth's atmosphere.
The changing size and shape of the PSF across the field of view and between exposures can be
traced from stars, which are intrinsically point sources.
We identify stars from their constrained locus within the size-magnitude plane (Figure~\ref{fig:fig2}(a)).
Heymans et al.\ (2006) suggest using the full width at half maximum (FWHM).
However, we find that FWHM is not robustly measured by {\tt SExtractor}, so we
instead use the $\mu_{\rm max}$-magnitude plane (Bardeau et al.\ 2005, 2007; Leauthaud et al.\ 2007),
where $\mu_{\rm max}$ is the peak surface brightness (Figure~\ref{fig:fig2}(b)).
The red points in Figure~\ref{fig:fig2} indicate the selected stars; our chosen locus reflects a careful
balance between obtaining sufficient stars to model the small-scale variations that we
observe in the PSF pattern, and introducing spurious noise by including faint stars.
The blue points are spurious detections of noise, cosmic rays, etc.\ (cf.\ Leauthaud et al.\ 2007).

\begin{figure}
\begin{center}
\protect
\includegraphics[angle=270.0,width=0.85\columnwidth]{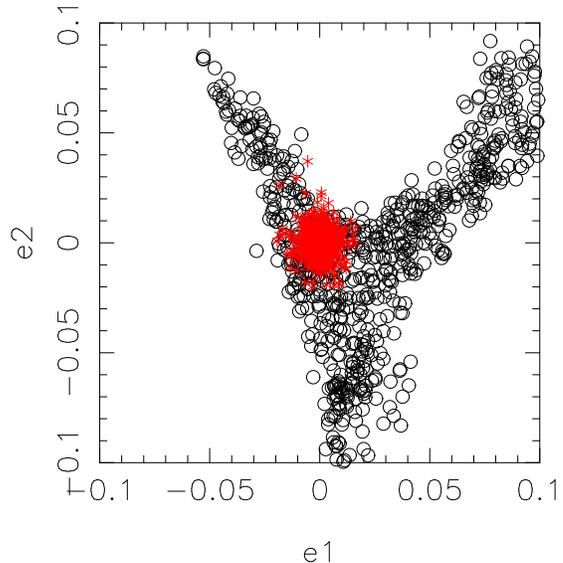}
\caption{Projection of the stellar ellipticities in the $(e_1,e_2)$
plane before (black) and after (red) PSF anisotropy correction.
The post-correction residuals are consistent with featureless white noise. \label{fig:fig4}}
\end{center}
\end{figure}

We then measure the Gaussian-weighted shape moments of the stars, and construct their ellipticity.
In addition to cuts in $\mu_\mathrm{max}$ and magnitude, we also exclude noisy outliers
with signal-to-noise $\nu<100$ or absolute ellipticity $e^*$ more than $2\sigma$ away from the mean local value,
and we iteratively remove objects very different from neighboring stars.
In 15 pointings with the worst image quality, including W1+2+3,
the PSF becomes larger than $r_g\sim0\arcsec.5$ in the corners of the
field of view, so we finally add these regions to the survey mask
(and exclude galaxies in them from our weak lensing analysis).

Having obtained our clean sample of stars, we
construct a spatially varying model of the PSF across the field of view.
In most pointings, we fit the $\sim30$ stars in each of the $36$ individual CCDs composing the MegaCam
focal plane, using a polynomial of second order in $x$ and $y$.
For stacked data with large dithers, we use a higher order polynomial.
Figures~\ref{fig:fig3} and \ref{fig:fig4} show
the stellar ellipticity before and after correction for the W1+2+3 pointing,
using a weight function of default size $r_g$ to measure the PSF shape moments.
The residual stellar ellipticity after correction is a consistent random scatter around zero,
of width $\sigma_{e_i} \sim 0.01$.

\begin{figure}
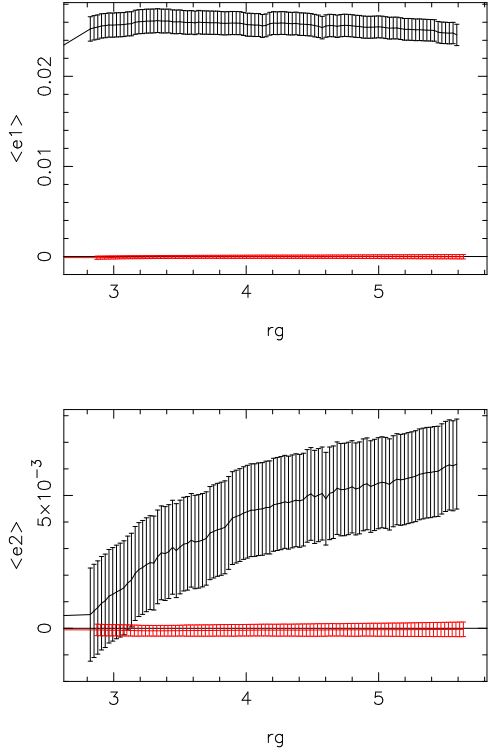

\centering
\begin{minipage}[t]{8.5cm}
\begin{center}
\includegraphics[angle=270.0,trim=0 0 0 0,width=0.75\textwidth]{fig5a.eps}\\[0.1cm]
\includegraphics[angle=270.0,trim=0 0 0 0,width=0.75\textwidth]{fig5b.eps}\\[0.1cm]
\end{center}
\end{minipage}
\caption{Ellipticity of the PSF changes from the core to the wings.
This shows the mean PSF ellipticity in the $i'$-band of the CFHTLS-Wide W1 pointing W1+2+3
as a function of the size of the Gaussian weight function with which it is measured,
before (black) and after (red) PSF anisotropy correction.
The error bars show the rms scatter throughout that pointing. \label{fig:fig5}}
\end{figure}

The ellipticity of the PSF changes from the core to the wings. We measure the PSF shape using differently-sized
weight functions and, when correcting each galaxy, use the same size weight function to measure both the PSF and galaxy shapes.
Figure~\ref{fig:fig5} shows the variation of mean stellar ellipticity as a function of the weight function size $r_g$,
before and after PSF anisotropy correction.

\subsection{Galaxy Shape Measurement}

\begin{figure}
\centering
\begin{minipage}[t]{8.5cm}
\begin{center}
\includegraphics[trim=0 0 0 0,width=1.0\textwidth]{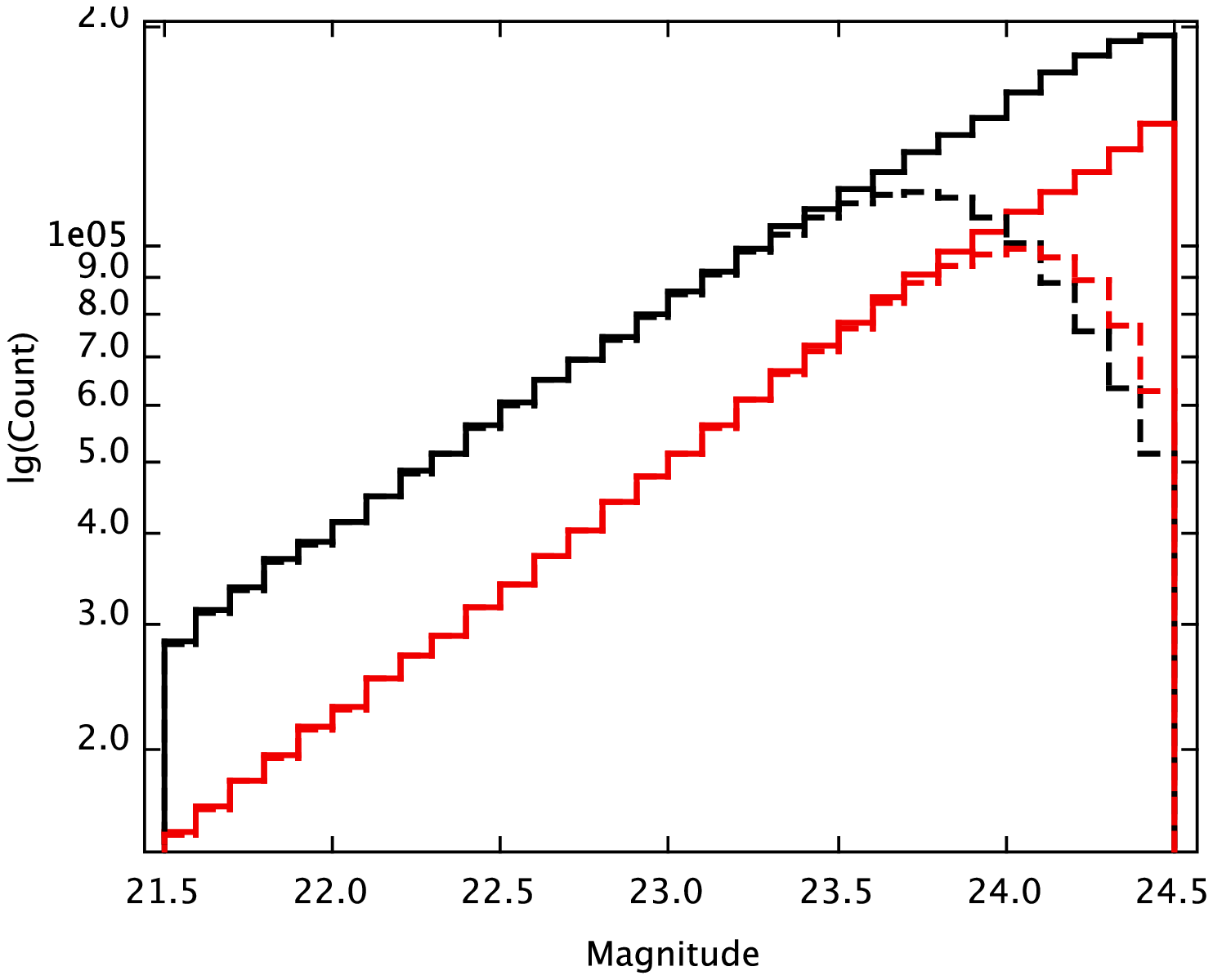}
\includegraphics[trim=0 0 0 0,width=1.0\textwidth]{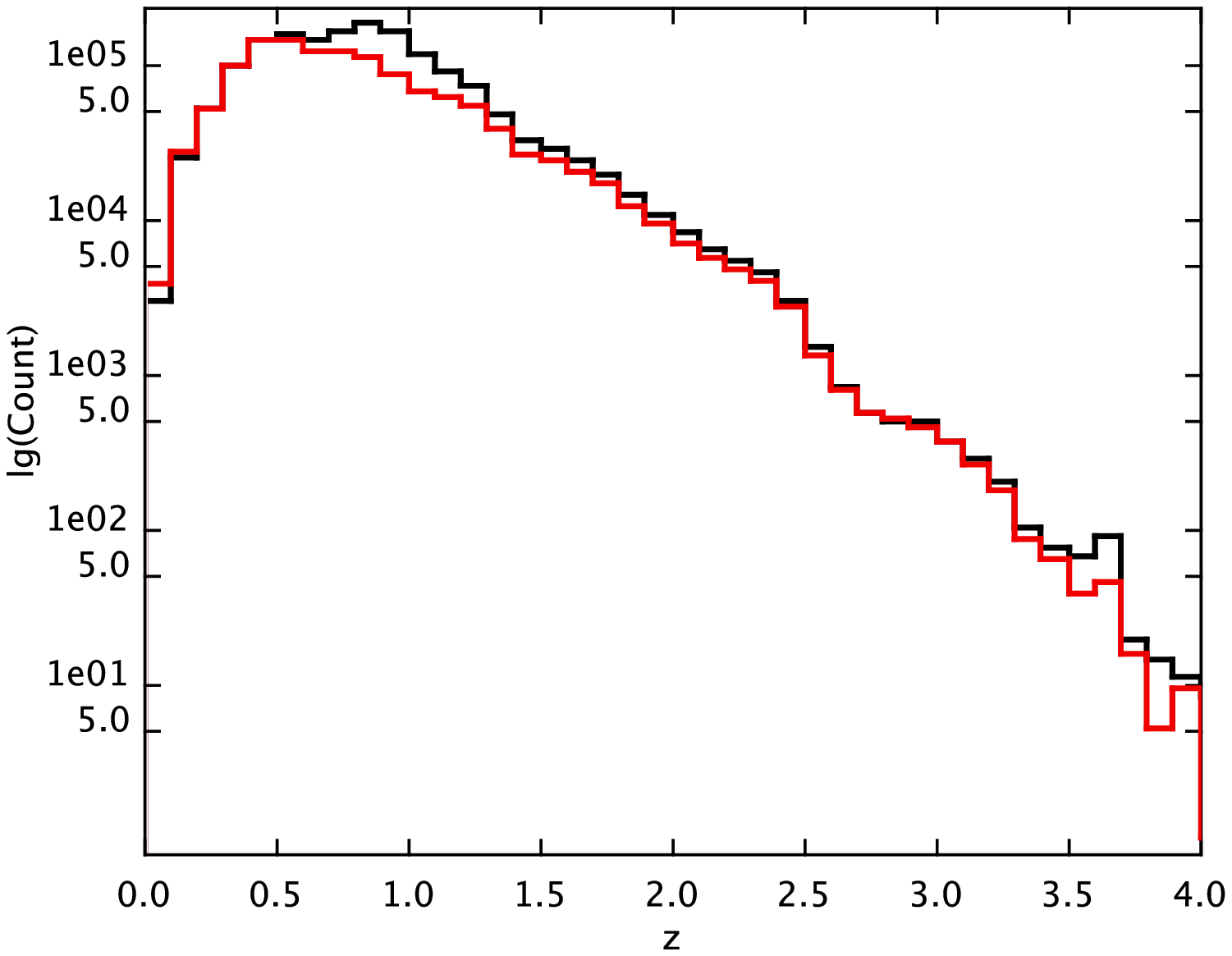}\\[0.1cm]
\end{center}
\end{minipage}
\caption{Galaxy magnitude and redshift distributions. Top: number counts of galaxies per
magnitude bin immediately after star-galaxy separation (solid histograms) and after all the lensing cuts
(dotted histograms) for imaging in $i'$ (black lines) and $r'$ (red lines) bands. Bottom:
the redshift distribution of galaxies used from the $i'$ (black lines) and $r'$ (red lines) bands.\label{fig:fig6}}
\protect
\end{figure}

Galaxies are selected as those objects
with half light radius $1.1r^{\mathrm{PSF}}_h<r_h<4$~pixels,
where $r^{\mathrm{PSF}}_h$ is the size of the largest star,
signal-to-noise $\nu>10$, magnitude $21.5<i'<24.5$ and {\tt SExtractor} flag $\rm {\tt FLAGS=0}$.
To also exclude blended or close pairs that could bias ellipticity measurements, we also
cut objects with corrected ellipticity $|e_{\rm cor}|>1$ and pairs of galaxies within $3\arcsec$.
After survey masking and catalog cuts, the galaxy number density is
%$n_g \sim 10.2~{\rm arcmin}^{-2}$ in an area of $A_{i'} \sim 57.9~\rm deg^2$ of $i'$-band imaging; and
%$n_g \sim 6.98~{\rm arcmin}^{-2}$ in $A_{r'} \sim 62.1~\rm deg^2$ of $r'$-band data.
$n_g \sim 11.5~{\rm arcmin}^{-2}$ in an area of $A_{i'} \sim 51.3~\rm deg^2$ of $i'$-band imaging; and
$n_g \sim 7.9~{\rm arcmin}^{-2}$ in $A_{r'} \sim 55.0~\rm deg^2$ of $r'$-band data.
Note that both are lower than the galaxy density $n_g \sim 38~{\rm arcmin}^{-2}$ obtained in
CFHTLS-Deep imaging by Gavazzi \& Soucail (2007).
This will restrict our detections to generally more massive clusters.
Figure~\ref{fig:fig6} shows the magnitude distribution of the galaxies,
and the redshift distribution of the $72~\%$ ($76~\%$) of
galaxies selected in the $i'$ ($r'$) bands that also have photometric redshift estimates by Arnouts et al.\ (2010).

We then measure the shapes of all the selected galaxies.
Our implementation of KSB is based on the KSBf90\footnote{\tt http://www.roe.ac.uk/~heymans/KSBf90/Home.html} pipeline (Heymans et al.\ 2006).
This has been generically tested on simulated images containing a known shear signal as part of the
Shear Testing Programme (STEP; Heymans et al.\ 2006; Massey et al.\ 2007) and the Gravitational
lensing Accuracy Testing (GREAT08; Bridle et al.\ 2010) challenge.
In all cases, the method was found to have small and repeatable systematic errors.

If the PSF anisotropy is small, the shear $\gamma$ can be recovered
to first-order from the observed ellipticity $e^{\rm obs}$ of each galaxies via
\begin{equation} \label{eqn:weight}
\gamma=P_{\gamma}^{-1}\left(e^{\rm obs}-\frac{P^{\rm sm}}{P^{\rm sm*}}e^{*}\right),
\end{equation}
where asterisks indicate quantities that should be measured from the PSF model interpolated to the position of the galaxy,
$P^{\rm sm}$ is the smear polarizability,
and $P_{\gamma}$ is the correction to the
shear polarizability that includes smearing
with the isotropic component of the PSF.
The ellipticities are constructed from a combination of each object's weighted quadrupole moments,
and the other quantities involve higher order shape moments. All definitions are taken from Luppino \& Kaiser (1997).
Note that we approximate the matrix $P_{\gamma}$ by a scalar equal to half its trace. %, $\mathrm{Tr}~P_{\gamma}/2$.
Since measurements of $\mathrm{Tr}~P_{\gamma}$ from individual galaxies are noisy,
we follow Fu et al.\ (2008) and fit it as a function of galaxy size and magnitude, which are
more robustly observable galaxy properties.

Following Hoekstra et al.\ (2000), we weight the shear contribution from each galaxy as
\begin{equation}
w=\frac{1}{\sigma_{e,i}^2}=\frac{P_{\gamma}^2}{\sigma_{0}^2 P_{\gamma}^2+\sigma_{e,i}^2},
\end{equation}
where $\sigma_{e,i}$ is the error in an individual ellipticity measurement
obtained via the formula in Appendix~A of Hoekstra et al.\ (2000),
and $\sigma_0\sim0.278$ is the dispersion in galaxies' intrinsic ellipticities.

\subsection{Calibration of Multiplicative Shear Measurement Biases}

\begin{figure}
\begin{center}
\protect\label{fig:fig7}
\includegraphics[width=1\columnwidth]{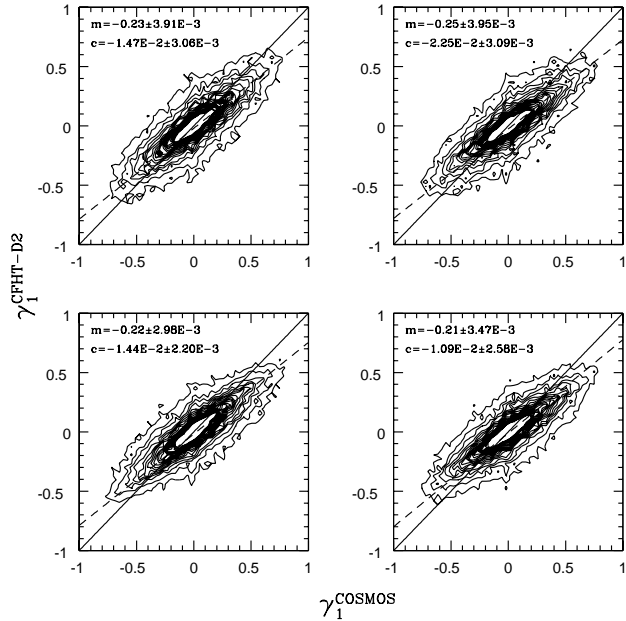}
\caption{Choices for the way shear polarizability $P_{\gamma}$ can be fitted to a galaxy population
in the CFHT KSBf90 pipeline, to reduce noise and bias in individual measurements.
Linearly spaced contours compare our shear measurements of galaxies in a subset of the
CFHTLS-Deep imaging, stacked to the depth of the CFHTLS-Wide survey, against
measurements from the {\em Hubble Space Telescope}.
Dashed lines show the best-fit relation $\gamma_1^\mathrm{CFHT}=(m-1)\gamma_1^\mathrm{HST}+c$.
The four panels illustrate various fitting schemes.
Top-left: raw (noisy) $P_{\gamma}$ measurements from each galaxy, without fitting.
Top-right: fitted as a polynomial in galaxy size $P_\gamma(r_h)$.
Bottom-left: fitting function $P_\gamma(r_h,\mathrm{mag})$ from Fu et al.\ (2008).
Bottom-right: best-fit rational function $P_\gamma(r_h,\mathrm{mag})$, as described in the text.
\label{fig:fig7}}
\end{center}
\end{figure}

\begin{figure}
\begin{center}
\protect\label{fig:fig8}
\includegraphics[width=1\columnwidth]{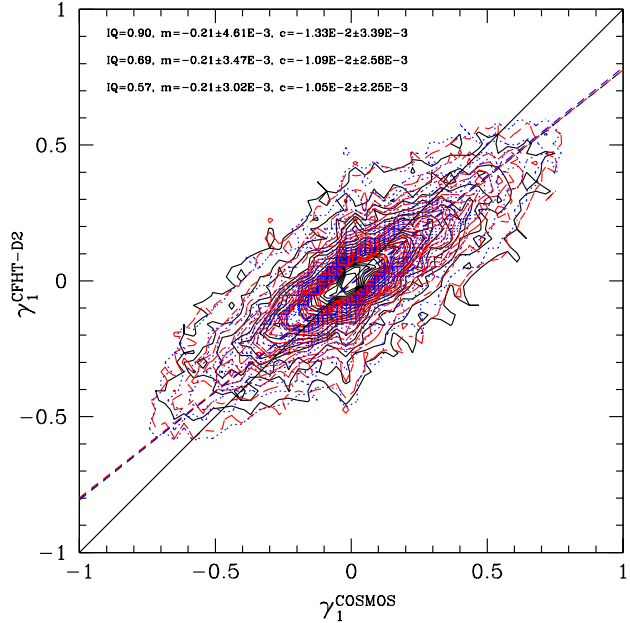}
\caption{Robustness of the calibration of our shear measurement as a function of image quality.
Linearly-spaced contours compare our shear measurements of galaxies in subsets of CFHTLS-Deep imaging
with varying mean seeing (black solid: $0\arcsec.90$, red dashed: $0\arcsec.69$, blue dotted: $0\arcsec.57$)
to measurements from the {\em Hubble Space Telescope}.
Shears are consistently underestimated by our pipeline, but the calibration is remarkably robust.
\label{fig:fig8}}
\end{center}
\end{figure}
%We also test the consistency of the shear measurement of matched galaxies
%in the independent $i'$ and $r'$-band imaging of the CFHTLS-Wide.

We exploit the opportunity that the CFHTLS-Deep D2 field includes the {\em HST} COSMOS survey field,
and verify the calibration of our shear measurement pipeline for ground-based data
against an independent analysis of the much higher resolution space-based data (Leauthaud et al.\ 2007, 2010).
We stack subsets of the CFHTLS-Deep D2 imaging to the same depth as the CFHTLS-Wide survey
and analyze it using the same pipeline applied to the CFHTLS-Wide W1 field.
Since any galaxy seen by CFHT is very well resolved by {\em HST}, and imaged
to very high signal-to-noise ratio by the COSMOS survey, the space-based shear measurements
require only negligible PSF correction and suffer from only negligible shot noise.
A consistent shear measurement between ground and space for this subset of galaxies
would therefore indicate a robust shear measurement across the CFHTLS.

Multiplicative shear measurement biases $m$ are the most problematic for circularly-symmetric cluster measurements.
Multiplicative biases cannot be internally diagnosed within a shear catalog, so our comparison against external data is
most useful for checking that $m$ is sufficiently small that it corresponds to a bias smaller than our statistical errors.
Within the KSB framework, difficulties in shear calibration mainly rest in measurement of the shear polarizability $P_{\gamma}$, so we first investigate different possibilities for fitting $P_{\gamma}$ across a galaxy population.
Figure~\ref{fig:fig7} compares shear measurements from a subset of the CFHTLS-Deep imaging with mean seeing $0\arcsec.69$ (similar to the mean seeing in our survey) stacked to the depth of the CFHTLS-Wide imaging against shear measurements obtained from {\em HST}.
The dashed lines show the best-fit linear relations $\gamma_1^\mathrm{CFHT}=(1+m)\gamma_1^\mathrm{HST}+c$,
which are obtained using a total least-squares fitting method (e.g.\ Kasliwal et al.\ 2008) that accounts for the noise present in both shear catalogs.
%Using COSMOS data, Leauthaud et al.\ (2007) found $\sigma_0 \sim 0.26$  independent of redshift.
The top-left panel shows the CFHT shear measurements with
$P_{\gamma}$ na\"ively obtained from each raw, noisy galaxy without any fitting.
This results in a large bias on shear measurements and a large amount of extra noise.
The top-right panel shows the shear measurements if $P_{\gamma}$ is fitted as a function of galaxy size, $P_{\gamma}(r_h)=a_0+a_1r_h+a_2r_h^2$.
The bottom-left panel shows shears if $P_{\gamma}(r_h,\mathrm{mag})=a_0+a_1 r_h+a_2 r_h^2 + a_3 m_{\nu}$ (Fu et al.\ 2008).
The bottom-right panel shows the matched shear with $P_{\gamma}(r_h,\mathrm{mag})$ the best-fit rational function
\begin{equation}
P_{\gamma}=\frac{a_0+a_1 m_{\nu}+a_2 m_{\nu}^2+a_3 r_h}{1+a_4 m_{\nu} +a_5 m_{\nu}^2+a_6 r_h +a_7 r_h^2}~.
\end{equation}
In this example, the coefficients are $a_0=25.07$, $a_1=-2.19$, $a_2=0.045$, $a_3=0.53$, $a_4=0.58$, $a_5=-0.022$, $a_6=-0.85$, and $a_7=0.14$.
The more sophisticated fits produce a shear catalog that is a marginally better match to the reliable {\em HST} measurements, and this is even more true if we redo the analysis using the full CFHTLS-Deep depth, in which galaxies are fainter and smaller.
We henceforth choose to adopt the rational function fit to $P_\gamma$ for all subsequent analyses, obtaining new best-fit coefficients for each pointing.

To quantify the performance of our shear measurement pipeline as a function of
image quality, we stack subsets of the CFHTLS-Deep D2 imaging with low-, medium-, and high-seeing
to the same depth as the CFHTLS-Wide survey, and analyze each separately (Figure~\ref{fig:fig8}).
We find that our CFHTLS pipeline consistently underestimates shear, but that the calibration is remarkably
robust to seeing conditions.
We can therefore simply recalibrate our pipeline for all images by multiplying all measured shears by $1/(1-0.21)$.

\subsection{Assessment of Residual Additive Shear Systematics}

\begin{figure}
\begin{center}
\includegraphics[angle=270.0,width=\columnwidth]{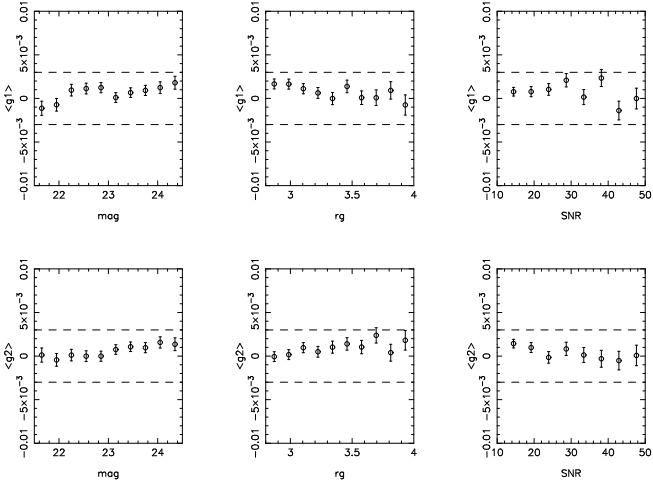}
\caption{Mean shear measurements from galaxies in $i'$-band observations of the entire CFHTLS-Wide W1 field.
In the absence of additive systematics, these should be consistent with zero.
In practice, they always remain within the dashed lines than indicate an order of magnitude lower than the
$1\%$--$10\%$ shear signal around clusters.
Upper and lower panels show components $\gamma_1$ and $\gamma_2$, respectively.
Left, middle, and right panels show trends as a function of galaxy magnitude, size, and detection signal-to-noise.
\label{fig:fig9}}
\end{center}
\end{figure}

Additive shear measurement systematics $c$ generally cancel out in circularly-symmetric cluster measurements (Mandelbaum et al.\ 2006).
However, to double-check for significant additive systematics, we first measure the mean shears
$\langle\gamma\rangle$ across all $72$ pointings of the CFHTLS-Wide W1 field.
Figure~\ref{fig:fig9} demonstrates that the mean shear is consistent with zero as expected,
for galaxies of all sizes, magnitudes, and signal-to-noise ratios.

\begin{figure}
\begin{center}
\includegraphics[angle=0.0,width=1.0\columnwidth]{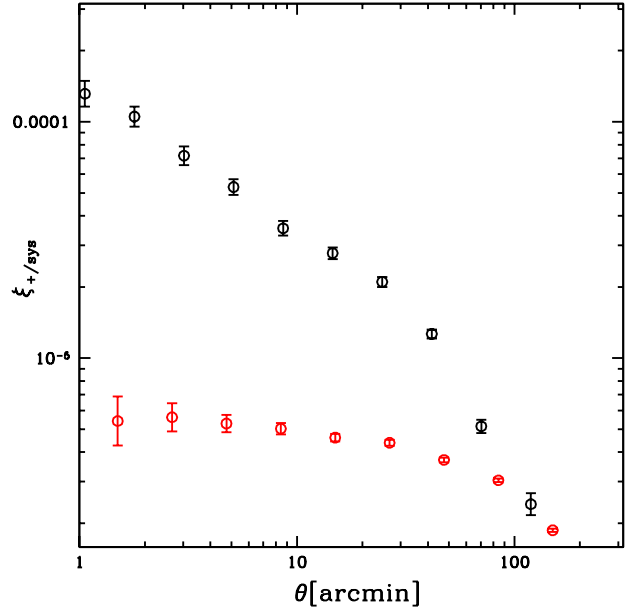}
\caption{The cross-correlation between shear measurements and stellar ellipticities, as a function of the separation
between galaxies and stars, averaged throughout the CFHTLS-Wide W1 field. If all residual influence of the
observational PSF has been successfully removed from the galaxy shape measurements, the red (lower) points should
be consistent with zero. \label{fig:fig10}}
\end{center}
\end{figure}

We also look for residual systematics left in the weak lensing cosmic shear signal due to imperfect PSF correction.
Figure~\ref{fig:fig10} shows the correlation $\xi_\mathrm{sys}$ between the corrected shapes of galaxies and the uncorrected shapes of stars.
Following Bacon et al.\ (2003) and Massey et al.\ (2005), we normalize the star-galaxy ellipticity correlation by the uncorrected star-star ellipticity correlation to assess its impact on shear measurements
\begin{equation}
\xi_{\rm sys}(\theta)=\frac{<e^*(\mathbf{x})\,\gamma(\mathbf{x}+\mathbf{\theta})>^2}{<e^*(\mathbf{x})\,e^*(\mathbf{x}+\mathbf{\theta})>},
\end{equation}
where $e^*$ is the ellipticity of the stars before PSF correction and $\gamma$ is the shear estimate from galaxies.
We find that our PSF correction is well within requirements for our analysis because on
cluster scales $1$--$5$~$\rm arcmin$, the amplitude of $\xi_{\rm sys}$
is at least one order of magnitude smaller than the cosmic shear signal
\begin{equation}
\xi_{\pm}=\xi_{\rm tt}(\theta)\pm\xi_{\rm xx}(\theta)=\frac{1}{2\pi}\int_0^\infty \ell\,P_{\kappa}(\ell)\,{\rm J}_{0,4} (\ell\theta)\,\mathrm{d}\ell~,
\end{equation}
where $\xi_{\rm tt}(\theta)$ ($\xi_{\rm xx}(\theta)$) are the correlation functions between components of shear rotated
tangentially (at $45^\circ$) to the line between pairs of galaxies separated by an angle $\theta$ and
$\rm J_{0}$, $\rm J_{4}$ are Bessel functions of the first kind.
%In particular, the amplitude of star-galaxy correlation function on cluster scales $1$--$5$~$\rm arcmin$ is completely negligible.

\begin{figure}
\begin{center}
\includegraphics[width=0.75\columnwidth]{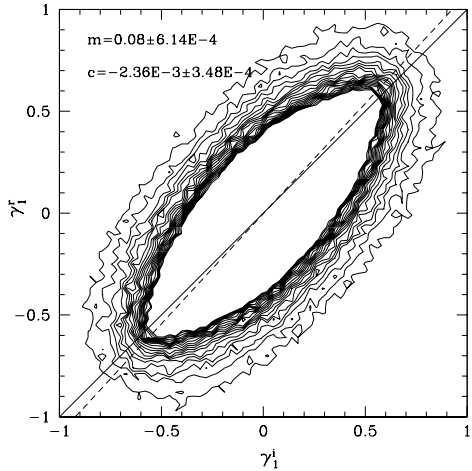}
\caption{Gravitational lensing is achromatic, so measurements of galaxy shapes from imaging in
different colors should on average be consistent. This shows a comparison of shear measurements
obtained from CFHTLS-Wide $r'$-band and $i'$-band imaging of the whole W1 field. The dashed line
shows the best-fit linear relation $\gamma_1^{i}=(m-1)\gamma_1^{r}+c$.
\label{fig:fig11}}
\end{center}
\end{figure}

\begin{figure*}
\centering
\begin{minipage}[t]{8.5cm}
\begin{center}
\includegraphics[angle=0.0,trim=0 0 0 0,width=1.0\textwidth]{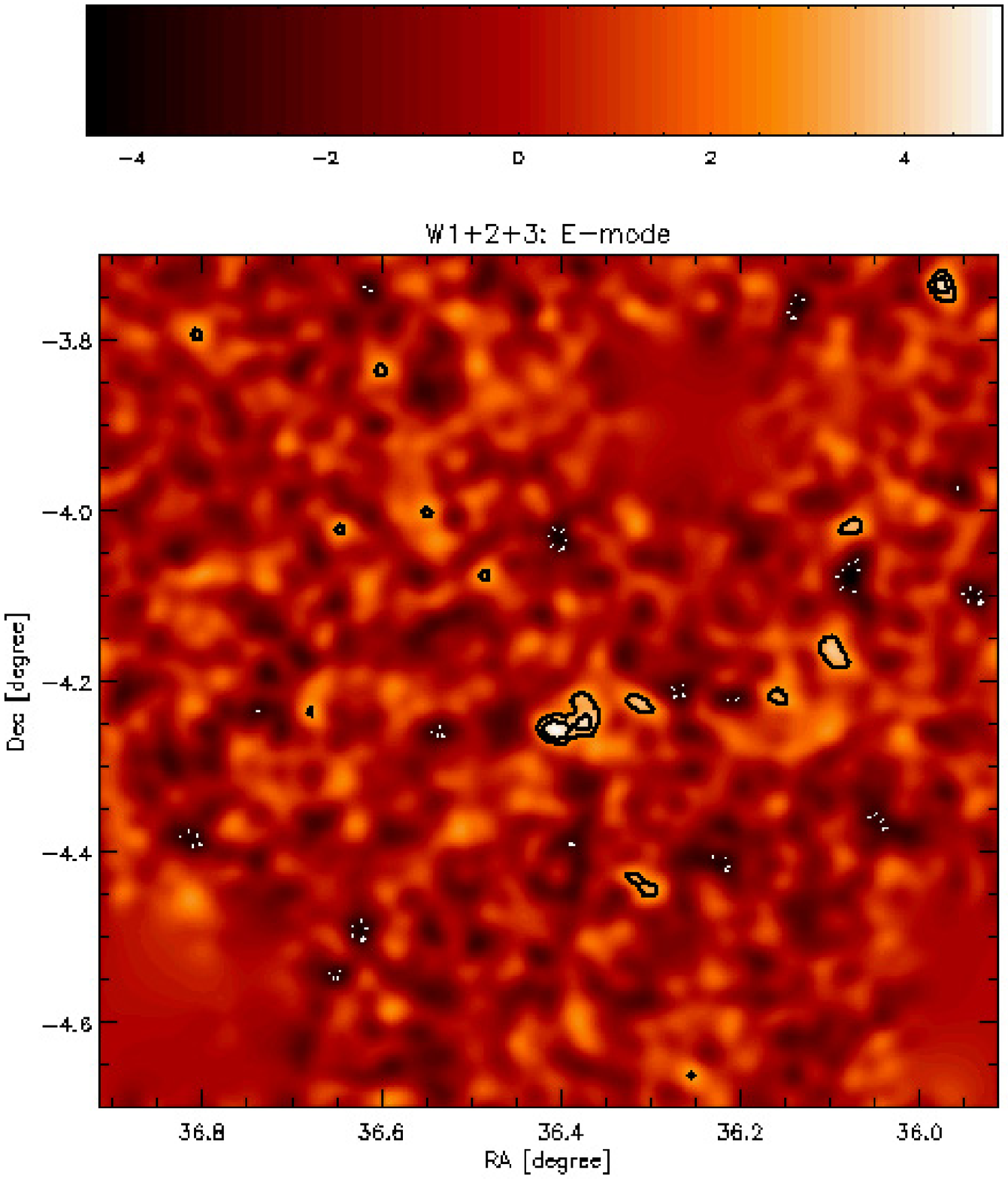}\\[0.1cm]
\end{center}
\end{minipage}
\qquad
\begin{minipage}[t]{8.5cm}
\begin{center}
\includegraphics[angle=0.0,trim=0 0 0 0,width=1.0\textwidth]{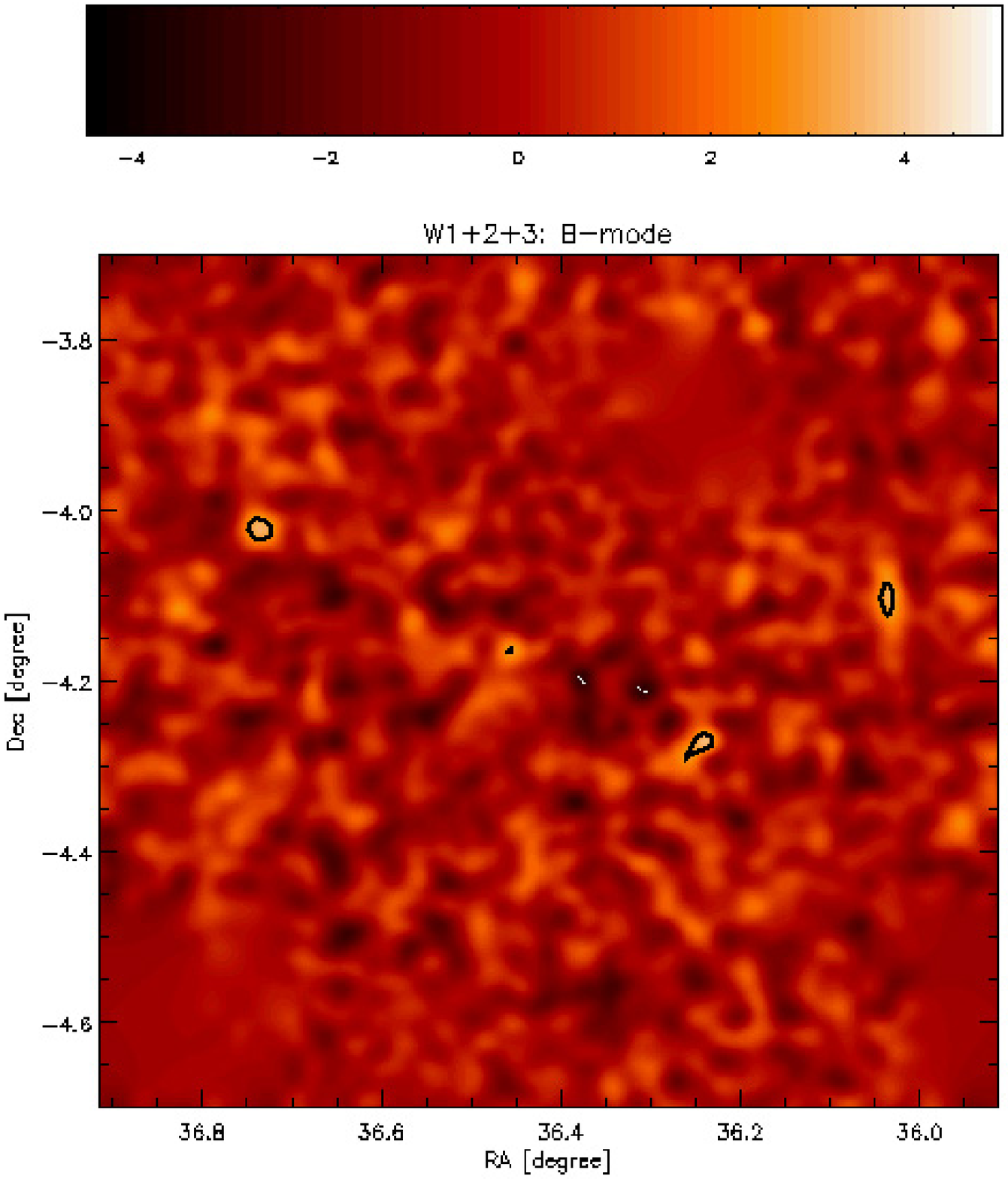}\\[0.1cm]
\end{center}
\end{minipage}
\caption{Distribution of foreground mass in the W1+2+3 pointing, reconstructed from shear measurements via the KS93 method.
Left: the physical $E$-mode convergence signal.
Right: the $B$-mode systematics signal, created by rotating the shears by $45^{\circ}$ then remaking the map.
Contours are drawn at detection significances of $3 \sigma$, $4 \sigma$, and $5 \sigma$, with dashed lines for negative values. \label{fig:fig12}}
\end{figure*}

\begin{figure*}
\begin{center}
\includegraphics[width=1.0\textwidth]{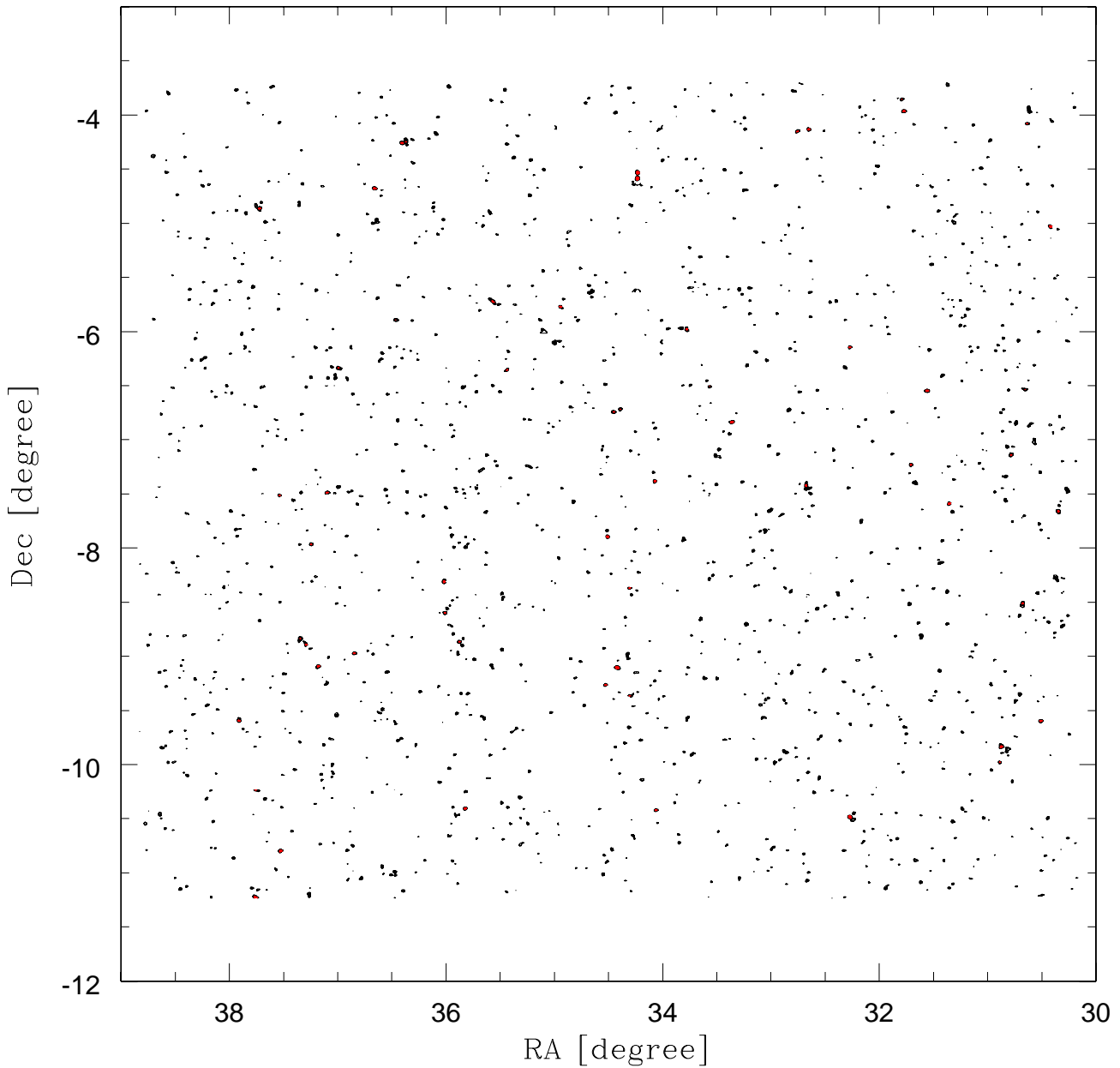}
\caption{Reconstructed ``dark matter mass'' convergence map for the entire $64$~$\rm deg^2$ CFHTLS-Wide W1 field.
This has been smoothed by a Gaussian filter of width $\theta_G=1\arcmin$.
Black contours are drawn at detection signal-to-noise ratios $\nu=3.0$, $3.5$, $4.0$, and red contours
continue this sequence from $\nu \geq 4.5$.
\label{fig:fig13}}
\end{center}
\end{figure*}

\begin{figure*}
 \centering
  \begin{minipage}[t]{8.5cm}
\begin{center}
\includegraphics[trim=0 0 0 0,angle=0, width=1.0\textwidth]{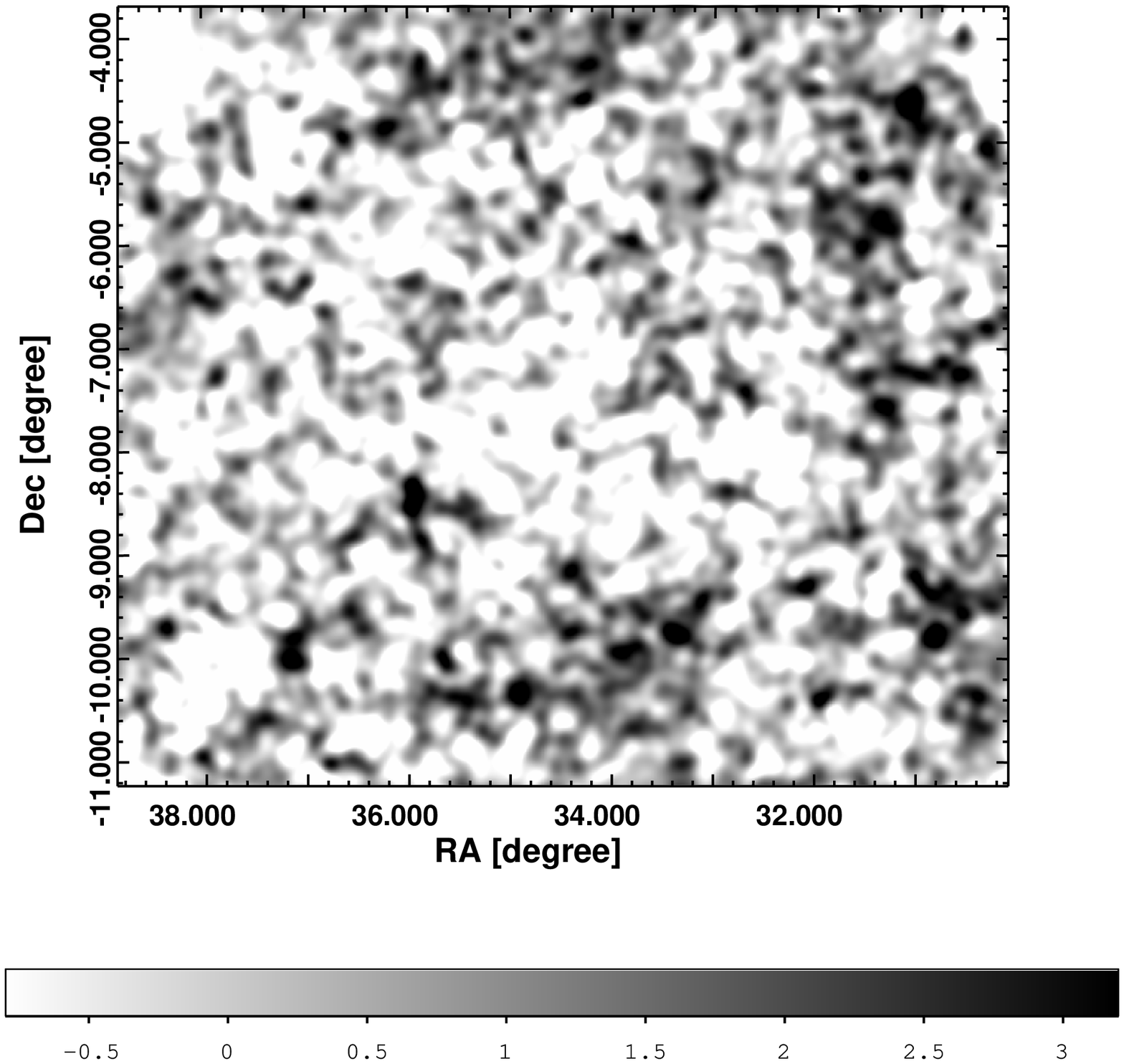}
\end{center}
 \end{minipage}
\qquad
 \begin{minipage}[t]{8.5cm}
\begin{center}
\includegraphics[trim=0 0 0 0,angle=0, width=1.0\textwidth]{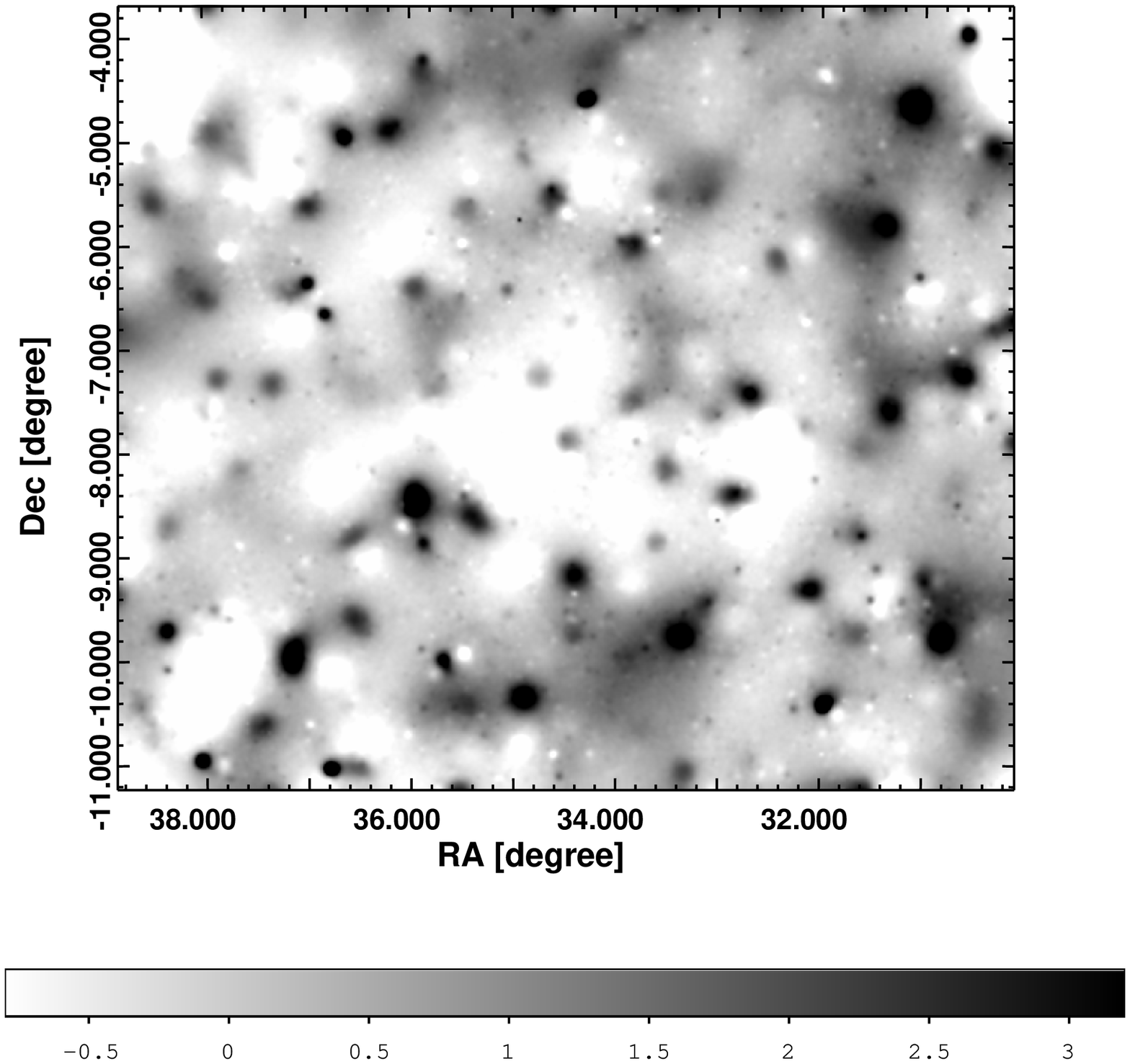}
\end{center}
\end{minipage}
\caption{Reconstructed ``dark matter mass'' convergence map for the entire $64$~$\rm deg^2$ CFHTLS-Wide W1 field
showing the same data as Figure~\ref{fig:fig13}, but smoothed with a Gaussian filter of width
$\theta_G=6\arcmin$ (left) and multi-scale wavelet filtering (right) to highlight the large-scale features.
\label{fig:fig14}}
\end{figure*}

\begin{figure}
\begin{center}
\includegraphics[width=0.75\columnwidth]{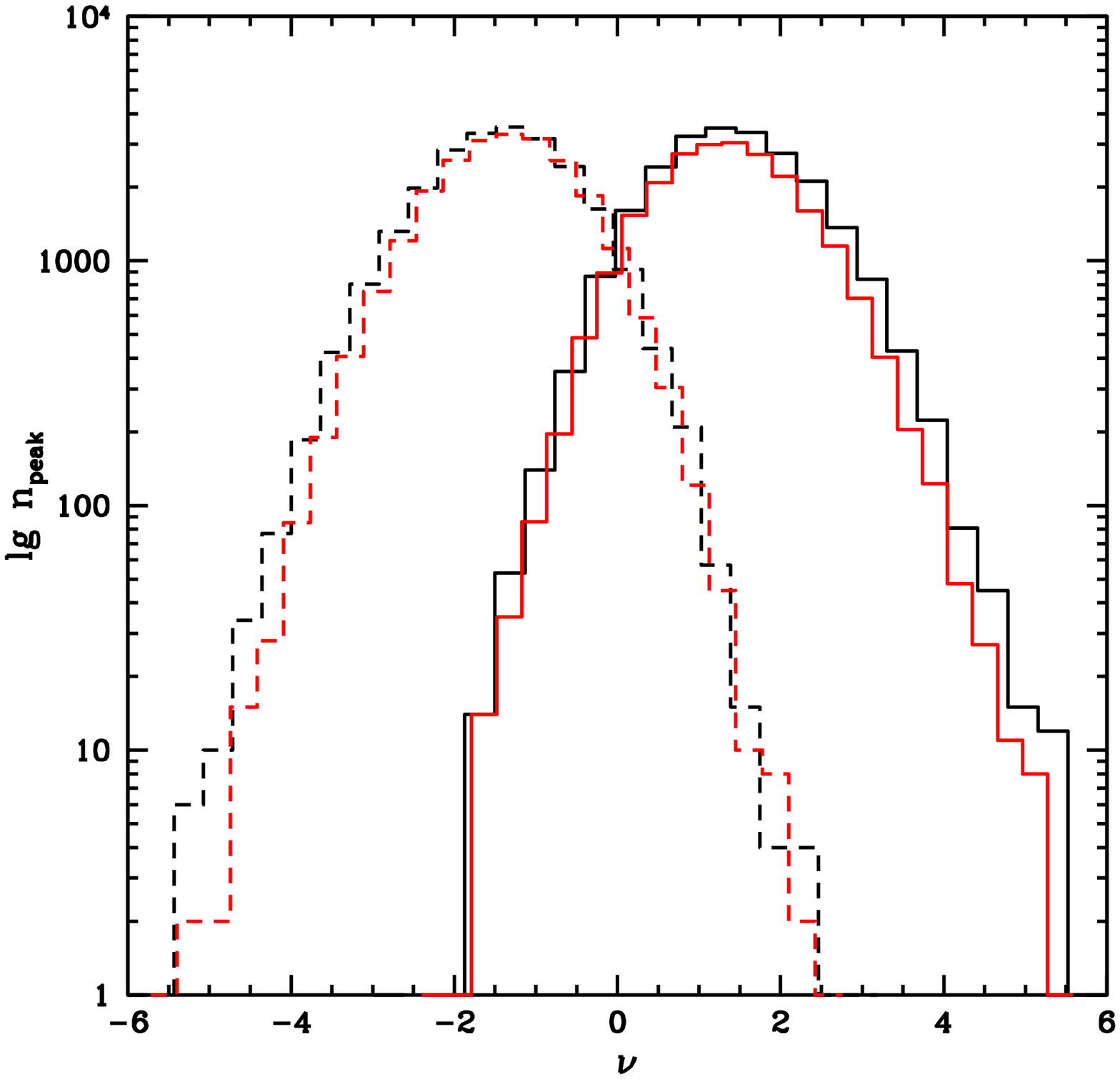}
\includegraphics[width=0.75\columnwidth]{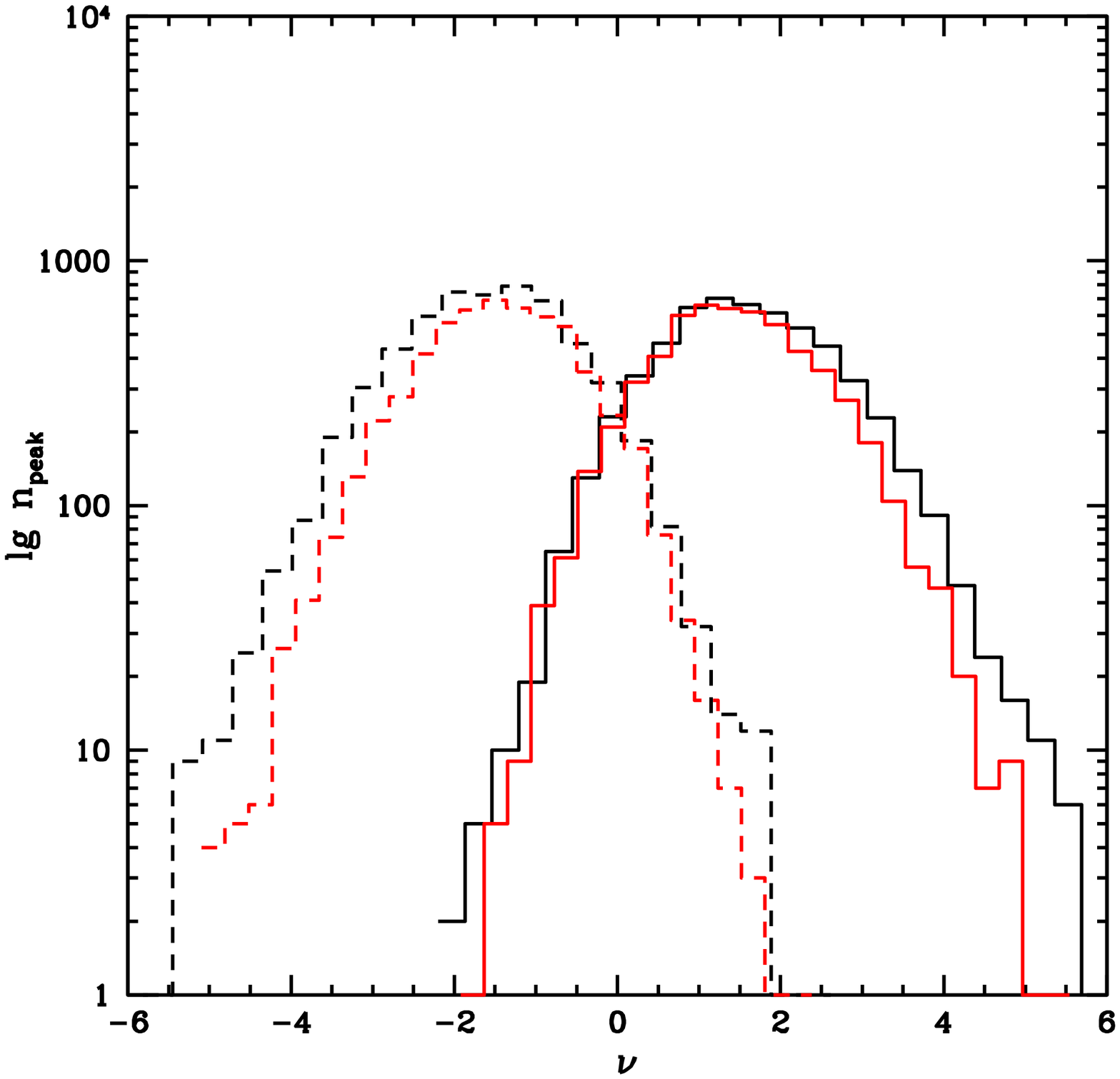}
\caption{Numbers of local maxima (solid line) and minima (dashed line) in our $E$-mode (black) and $B$-mode (red)
convergence map of the CFHTLS-Wide W1 field,
with smoothing scales $\theta_G=1\arcmin$ (top-panel) and $\theta_G=2\arcmin$ (bottom-panel).
Local maxima can still have a slightly negative peak height if they occur
along the same line of sight as a negative noise fluctuation (or a huge void),
and local minima can similarly have a slightly positive peak height.
To eliminate spurious noise peaks, we shall mainly consider maxima or minima with $|\nu|>3.5$.
In this regime, there is an excess of local maxima over local minima, and an excess of $E$-mode peaks over $B$-mode peaks (see Table~\ref{tab:tab2}). \label{fig:fig15}}
\end{center}
\end{figure}

\begin{figure}
\begin{center}
\includegraphics[width=0.75\columnwidth]{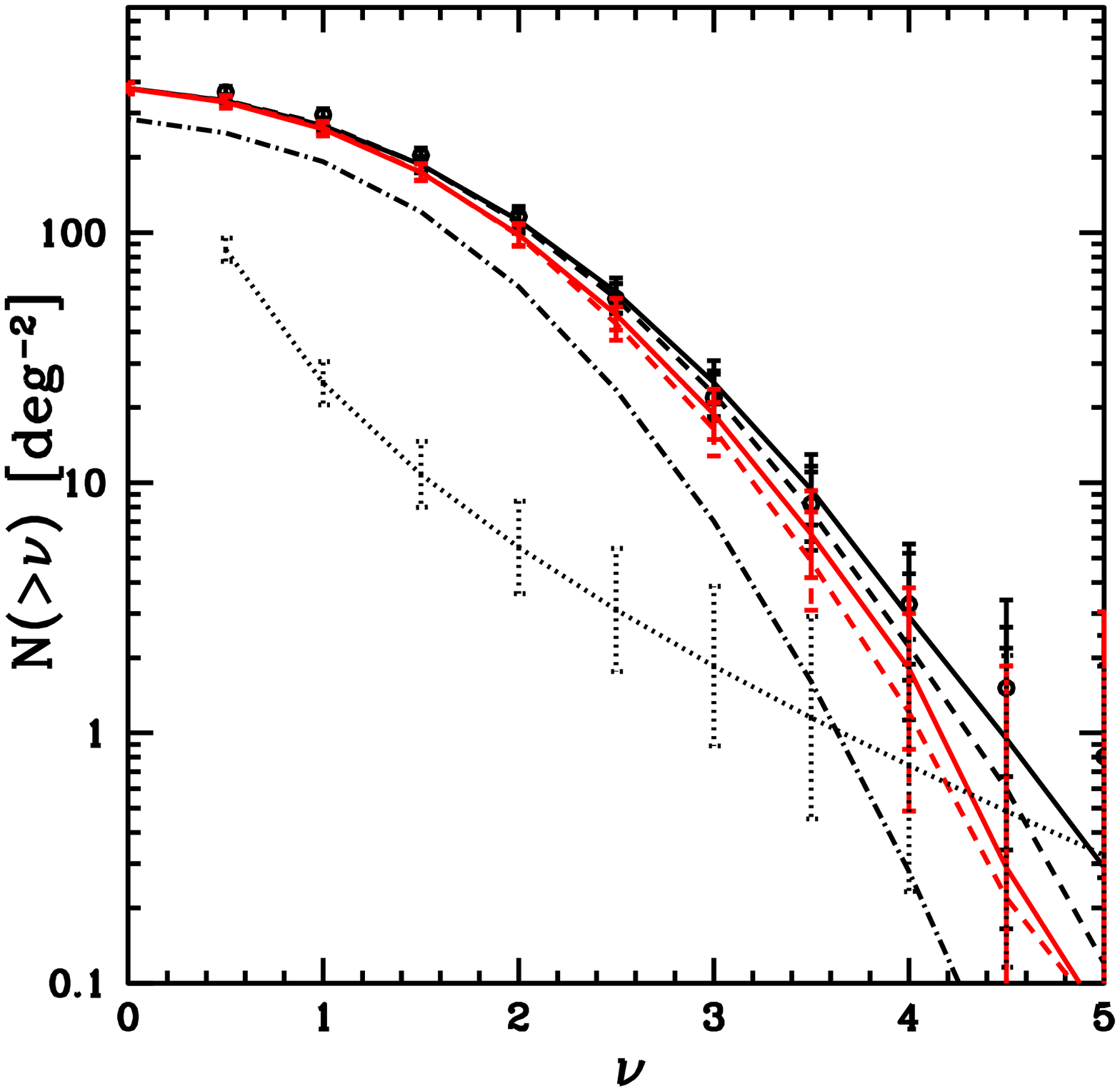}
\includegraphics[width=0.75\columnwidth]{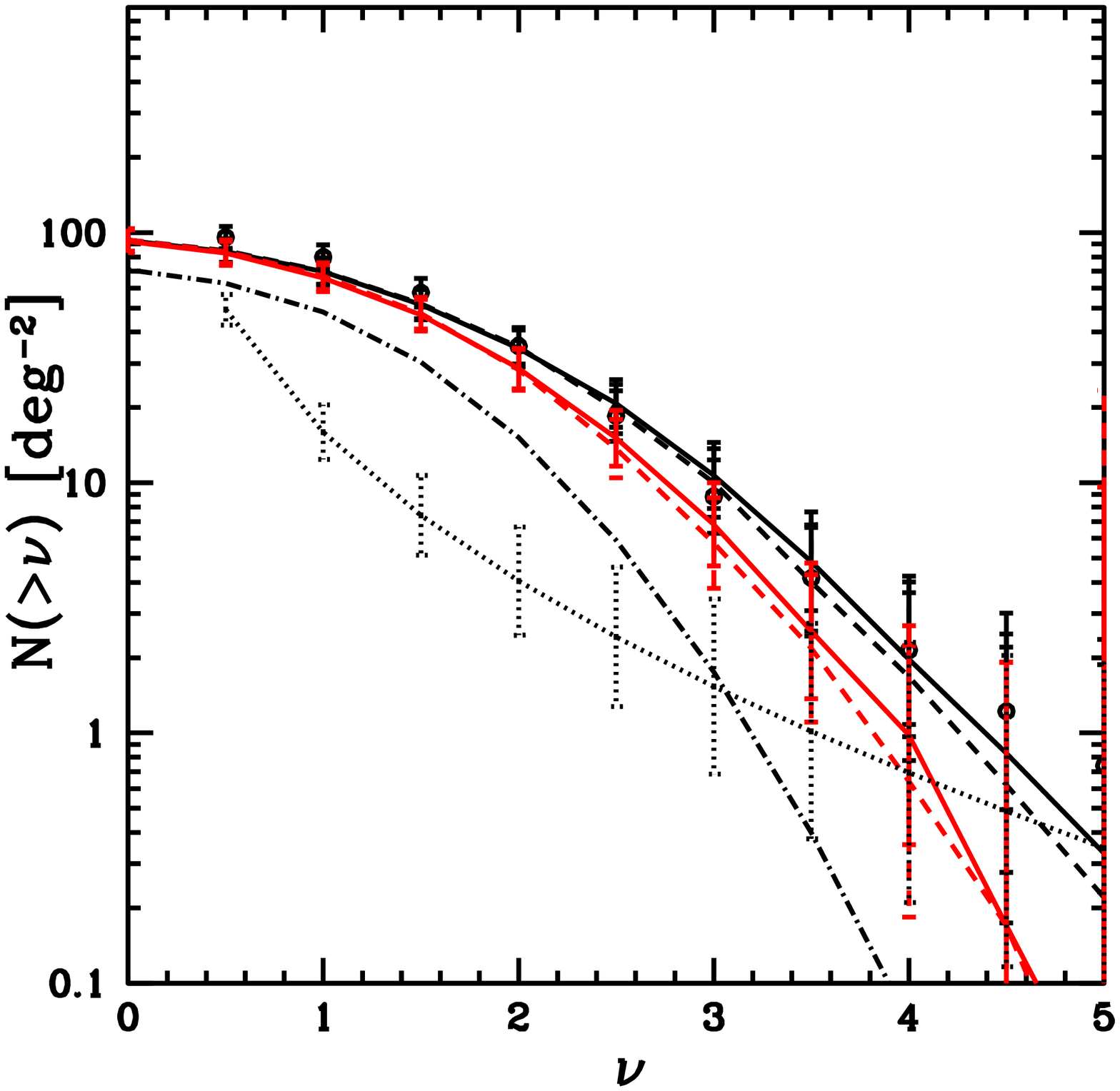}
\caption{Cumulative density of local maxima $N(>\nu)$ (solid line)
and corresponding density of local minima $N(< -\nu)$ (dashed line)
in the CFHTLS-Wide W1 $E$-mode (black) and $B$-mode (red) convergence map with
smoothing scale $\theta_G=1\arcmin$ (top-panel) and $\theta_G=2\arcmin$ (bottom-panel).
Error bars are simply $1\sigma$ uncertainties assuming Poisson shot noise.
The dot-dashed line shows the prediction from Gaussian random field theory (van Waerbeke 2000).
Circles with additional error bars and the dotted curves show an analytical prediction in a $\Lambda$CDM universe (Fan et al.\ 2010), with and without the influence of random noise.
The excess of positive maxima over negative minima demonstrates
the non-Gaussianity of the convergence field.
\label{fig:fig16}}
\end{center}
\end{figure}

\begin{figure}
\centering
\begin{center}
\includegraphics[width=\columnwidth]{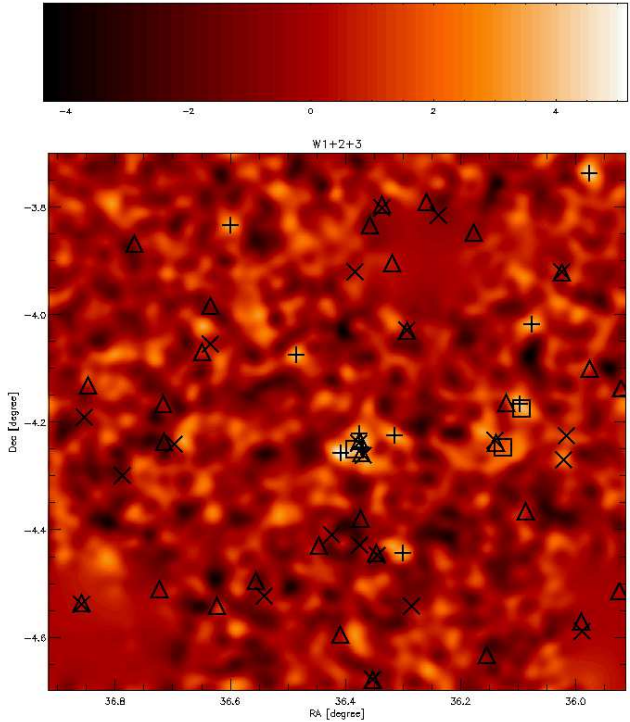}\\[0.1cm]
\end{center}
\caption{Reconstructed lensing convergence signal-to noise map for the W1+2+3 pointing,
plus overlays showing optically and X-ray selected cluster counterparts.
The smoothing scale of the background map is $\theta_G=1\arcmin$.
Symbols indicate the positions of\\
$+$: lensing peaks detected with $\nu > 3.5$ in the $\theta_G=1\arcmin$ map,\\
$\square$:  lensing peaks detected with $\nu > 3.5$ in the $\theta_G=2\arcmin$ map,\\
$\triangle$: optically-detected clusters in the K2 catalog, and\\
$\times$: X-ray selected clusters found in XMM-LSS (Adami et al.\ 2011). \label{fig:fig17}}
\end{figure}

\begin{figure*}
\centering
\begin{minipage}[t]{8.5cm}
\begin{center}
\includegraphics[trim=0 0 0 0,width=1.0\textwidth]{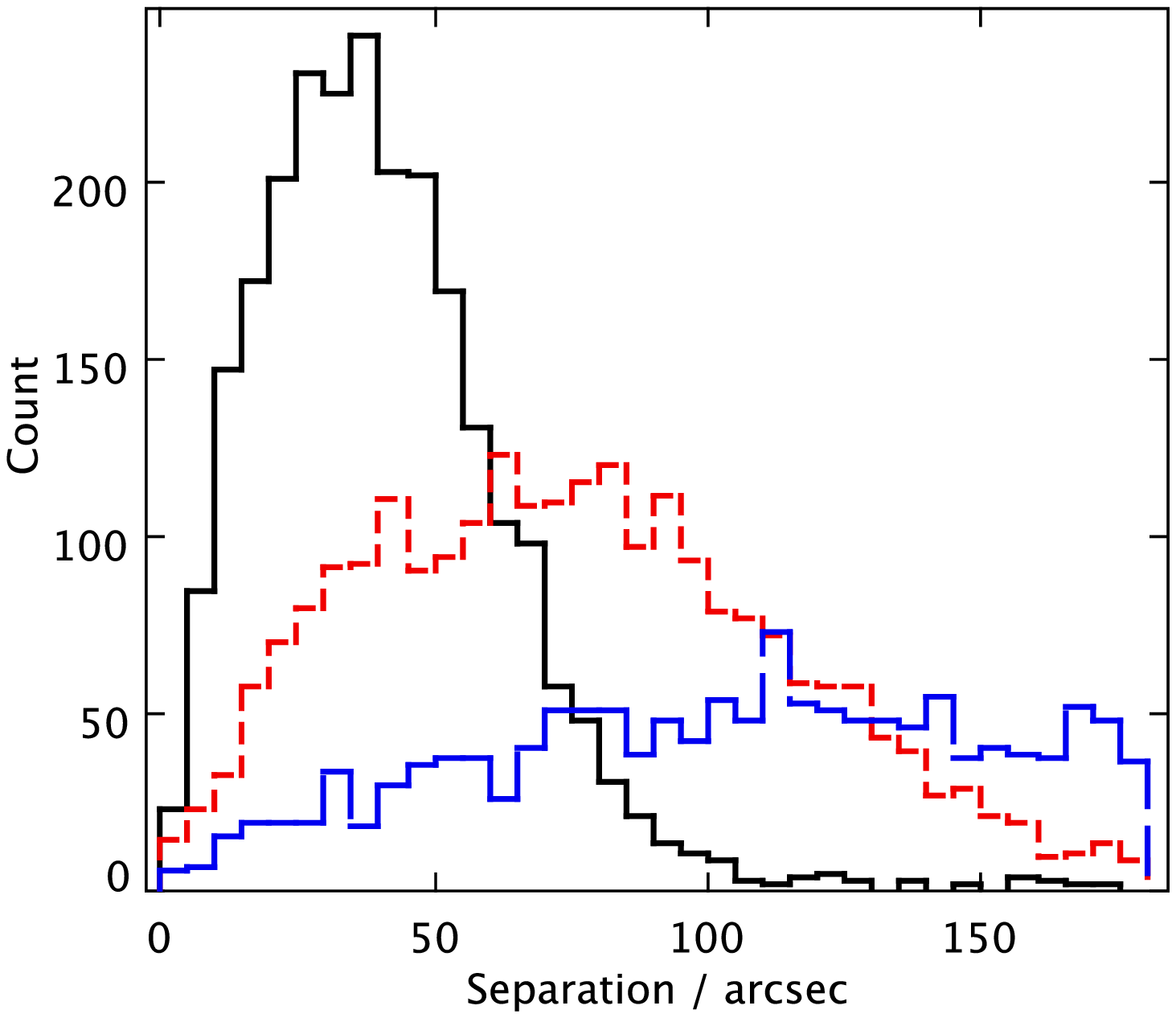}\\[0.1cm]
\end{center}
\end{minipage}
\qquad
\begin{minipage}[t]{8.5cm}
\begin{center}
\includegraphics[trim=0 0 0 0,width=1.0\textwidth]{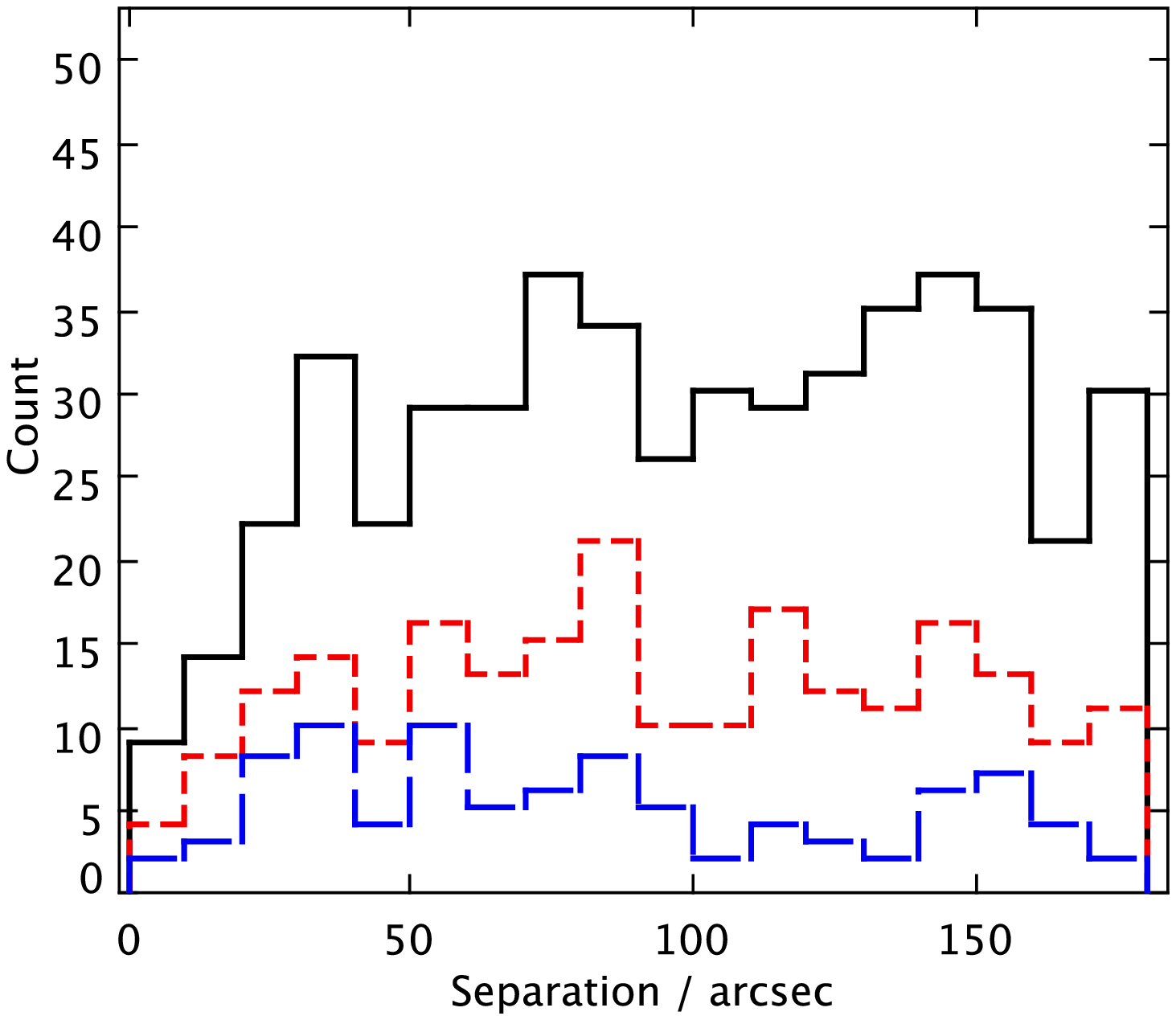}\\[0.1cm]
\end{center}
\end{minipage}
\caption{Distribution of the offsets between matched pairs of weak lensing peaks and K2-detected clusters.
Left: offsets for all peaks in maps with smoothing scales of  $0\arcmin.5$ (solid black),
$1\arcmin$ (dotted red), and $2\arcmin$ (dashed blue).
Right: offsets for peaks with $\nu>3$ (solid
black histograms), $\nu>3.5$ (dotted red histograms) and $\nu>4$ (dashed
blue histograms), all with the smoothing scale $\theta= 1~\rm arcmin$. \label{fig:fig18}}
\end{figure*}

\begin{figure}
\begin{center}
\protect
\includegraphics[width=0.95\columnwidth]{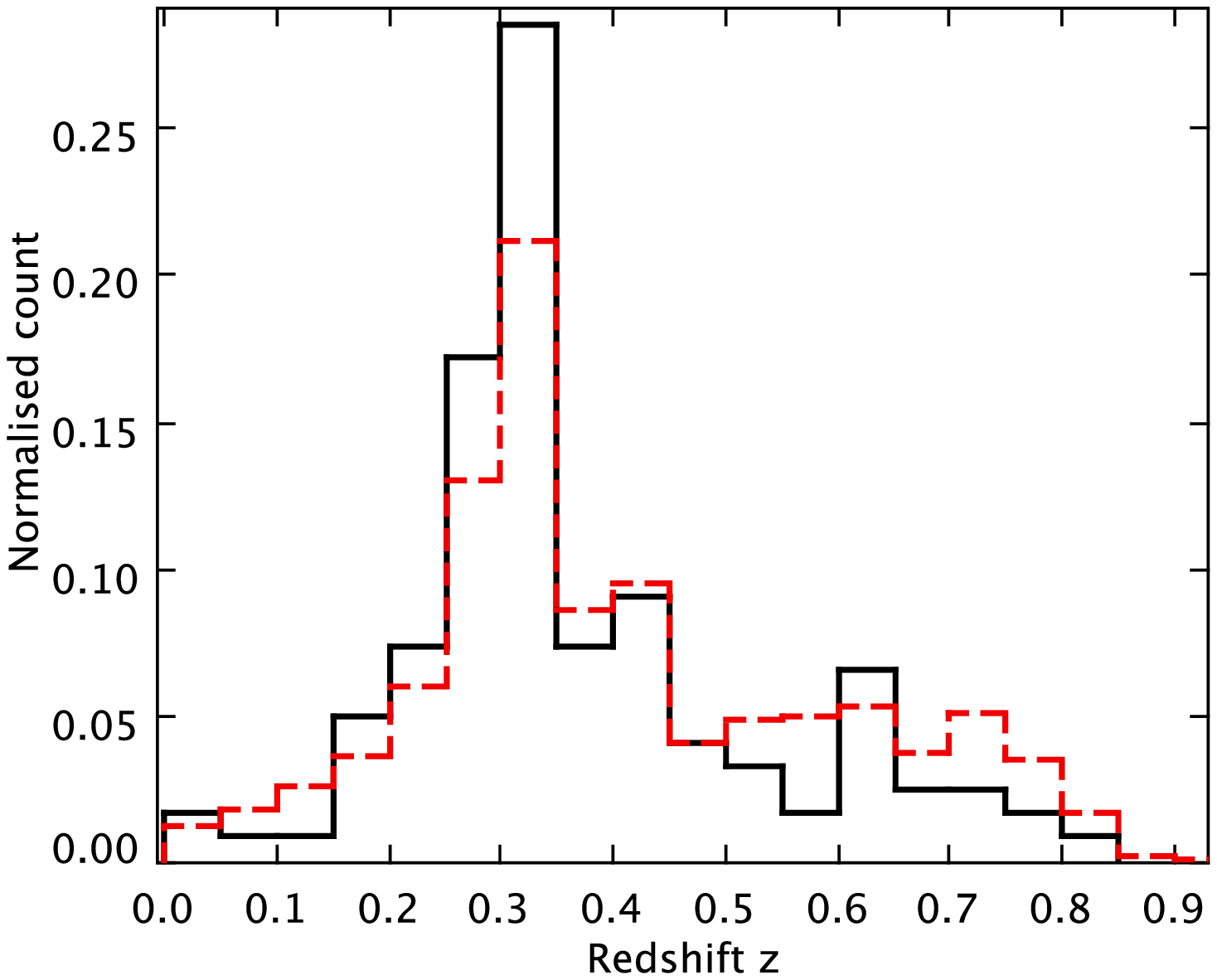}
\caption{Cluster redshift distribution for the matched clusters of weak lensing
and K2 (black solid histogram) and the total K2 detected clusters
with K2 detection significance $(r-i)>3$ (red dashed histogram).\label{fig:fig19}}
\end{center}
\end{figure}

Because gravitational lensing is achromatic while systematics are typically not,
we can also assess the robustness of our measurements by
comparing shears measured from independent imaging acquired in multiple bands.
The first attempt at comparing multicolor shear measurements was
made by Kaiser et al.\ (2000) using the CFHT12K camera. The $I$ and $V$
bands showed significantly different signals that were inconsistent
with the change in redshift distribution between the two filters.
After a great deal of algorithmic progress, Semboloni et al.\ (2006) obtained
consistent shear measurements from $i'$-band and $r'$-band CFHTLS-Deep data.
Gavazzi \& Soucail (2007) extracted consistent shear from the $g'$, $r'$, $z'$ and $i'$ bands of CFHTLS-Deep.
Gavazzi et al.\ (2009) also measured consistent values of the PSF-corrected
ellipticities of central Coma cluster galaxies in MegaCam-$u^*$ and CFH12k-$I$ bands.
Our analysis pipeline measures shears in the independent CFHTLS $r'$-band and $i'$-band
imaging that are consistent within the $\sim 0.1$ rms noise (Figure~\ref{fig:fig11}).
To maximize the total number of galaxies in our shear catalog, we therefore combine
$r'$-band and $i'$-band measurements. Only the unique $r'$-band galaxies are added to the
$i'$-band catalog. The combined catalog includes $\sim 2.66$~million galaxies, with
$n_g \sim 14.5~{\rm arcmin}^{-2}$ for lensing measurements.
%$n_g \sim 12.8~{\rm arcmin}^{-2}$ for lensing measurements.

We shall continue testing for systematics at each stage of our analysis, by checking that the shear signal behaves as expected, and is consistent with external data sets.
An important example of this is the (nonphysical) $B$-mode signal, which we shall compute wherever
we reconstruct the (physical) $E$-mode.
Pure gravitational fields produce zero $B$-mode for isolated clusters and only tiny $B$-modes through coupling
between multiple systems along adjacent lines of sight (Schneider et al.\ 2002).
The $B$-mode signal corresponds to the imaginary component of $P_{\kappa}(\ell)$; it can be conveniently
measured by rotating all galaxy shears through $45^\circ$ then remeasuring the $E$-mode signal
(Crittenden et al.\ 2002).

\section{Mass reconstructions}\label{sec:mass2d}

\subsection{Kaiser-Squires Inversion and Masking}

The shear field $\gamma_i(\mathbf{\theta})$ is sparsely and noisily sampled by
measurements of the shapes of galaxies at positions $\mathbf{\theta}$.
The smooth, underlying shear field $\gamma_i(\mathbf{\theta})$ can be written in terms of
the lensing potential $\phi(\mathbf{\theta})$ as
\begin{equation}
\gamma_1=\frac{1}{2}(\partial_1^2-\partial_2^2)\phi, \label{eqn:gamma1phi}
\end{equation}
\begin{equation}
\gamma_2=\partial_1 \partial_2 \phi, \label{eqn:gamma2phi}
\end{equation}
where the partial derivatives $\partial_i$ are with respect to $\theta_i$.
The convergence field $\kappa(\mathbf{\theta})$, which is proportional to the mass projected along a line of sight,
can also be expressed in terms of the lensing potential as
\begin{equation}
\kappa=\frac{1}{2}(\partial_1^2+\partial_2^2)\phi.
\end{equation}

We shall reconstruct the convergence field from our shear measurements via the Kaiser \& Squires (1993) (KS93) method.
This is obtained by inverting Equations~(\ref{eqn:gamma1phi}) and (\ref{eqn:gamma2phi}) in Fourier space:
\begin{equation}
\hat \gamma_i=\hat P_i \hat \kappa
\end{equation}
for $i=1,2$, where the hat symbol denotes Fourier transforms, we define $k^2=k_1^2+k_2^2$ and
\begin{equation}
\hat P_1(k)=\frac{k_1^2-k_2^2}{k^2},
\end{equation}
\begin{equation}
\hat P_2(k)=\frac{2 k_1 k_2}{k^2}.
\end{equation}
This inversion is non-local, so we deal with masked regions of the shear field by masking out the
same area in the convergence field, plus a $1\arcmin.5$ border.
We shall ignore any signal within these regions, and set the convergence to zero in relevant figures.

For the finite density of source galaxies resolved by CFHT, the scatter of their intrinsic ellipticities
means that a raw, unsmoothed convergence map $\kappa(\mathbf{\theta})$ will be very noisy.
Following Miyazaki et al.\ (2002), we smooth the convergence map by convolving it (while still in Fourier space)
with a Gaussian window function,
\begin{equation}
W_G(\theta)=\frac{1}{\pi \theta_G^2} {\rm exp} \left( -\frac{\theta^2}{\theta_G^2} \right ),
\end{equation}
As shown by van Waerbeke (2000), if different galaxies'
intrinsic ellipticities are uncorrelated, the statistical properties of the resulting noise field
can be described by Gaussian random field theory (Bardeen et al.\ 1986; Bond \& Efstathiou 1987)
on scales where the discreteness effect of source galaxies can be ignored.
The Gaussian field is uniquely specified by the variance of the noise, which is in turn
controlled by the number of galaxies within a smoothing aperture
(Kaiser \& Squires 1993; Van Waerbeke 2000)
\begin{equation}
\sigma_{\rm noise}^2=\frac{\sigma_e^2}{2} \frac{1}{2\pi \theta_G^2 n_g},
\end{equation}
where $\sigma_e$ is the rms amplitude of the intrinsic ellipticity
distribution and $n_g$ is the density of source galaxies. We
define the signal-to-noise ratio for weak lensing detections by
\begin{equation}
\nu\equiv\frac{\kappa}{\sigma_{\rm noise}}.
\end{equation}
To define the noise level in theoretical calculations of $\nu$, we adopt a constant effective density of galaxies
equal to the mean within our survey. For observational calculations of $\nu$, we use the mean galaxy density in each
pointing --- but do not consider the non-uniformity of the density within each field due to masks or galaxy clustering.

It turns out that a simple Gaussian filter of width $\theta_G\approx1\arcmin$
is close to the optimal linear filter for cluster detection, and this choice has been
extensively studied in simulations (White et al.\ 2002; Hamana et al.\ 2004;
Tang \& Fan 2005). Because of our relatively low source galaxy density,
the galaxies' random intrinsic shapes will produce spurious noise peaks,
degrading the completeness and purity of our cluster detection.
To reduce contamination, we repeat our mass reconstruction using two
smoothing scales $\theta_G=1\arcmin$ and $\theta_G=2\arcmin$.
The map with greater smoothing will be less noisy; to help remove
spurious peaks from the higher resolution map, we consider only those
peaks detected above a signal-to-noise threshold in both maps.

Figure~\ref{fig:fig12} shows the reconstructed convergence field corresponding to foreground mass in the W1+2+3 pointing.
The left panel shows the $E$-mode reconstruction with KS93 method after smoothing by a $1 \rm arcmin$ Gaussian kernel.
This contains several high signal-to-noise ratio peaks, while the associated $B$-mode systematics measurement in the
right panel
is statistically consistent with zero, with fewer peaks.
As weak lensing produces only curl-free or $E$-mode distortions,
a detection (significant above statistical noise) of curl or $B$-mode signal
would have indicated contamination from residual systematics, e.g.\ imperfect PSF correction.

\subsection{Large-scale Lensing Mass Map}

A reconstructed ``dark matter mass'' convergence map for the entire $64~{\rm deg}^2$ CFHTLS-Wide W1 field
is presented in Figure~\ref{fig:fig13}. We detect $301$ peaks with $\nu>3.5$ in maps with both smoothing scales
$\theta_G=1\arcmin$ and $\theta_G=2\arcmin$. The same information is reproduced in Figure~\ref{fig:fig14}, with
$\theta_G=6\arcmin$ and after multi-scale entropy restoration filtering (MRLens; Starck et al.\ 2006), to
better display large-scale features. The MRLens filtering effectively suppresses noise peaks, but results in
non-Gaussian noise that complicates the peak selection (Jiao et al.\ 2011), so we shall not use it further.

\begin{table}
\centering
\caption{The Number of Local Maxima and Minima in the Convergence Map of the CFHTLS-Wide W1 Field,
as a Function of Smoothing Scale.}
\begin{tabular}{ccccc}
\hline
\hline
Smoothing &  $E$-mode & $E$-mode &  $B$-mode & $B$-mode \\
Scale $\theta_G$ &  $\nu>3.5$ & $\nu<-3.5$ &  $\nu>3.5$ & $\nu<-3.5$ \\
\hline
$0\arcmin.5$ & $1512$ & $1270$ & $1244$ & $1033$ \\
$1\arcmin.0$ & $543$ & $445$ & $361$ & $282$ \\
$2\arcmin.0$ & $281$ & $233$ & $148$ & $126$ \\
\hline
\end{tabular}
\label{tab:tab2}
\end{table}

To assess the reliability of this map, we shall first investigate the statistical properties of local maxima and minima.
Figure~\ref{fig:fig15} shows the distribution of peak heights, as a function of detection signal-to-noise.
The bimodal distribution in both the $B$-mode and $E$-mode signals is dominated by
positive and negative noise fluctuations,
but an asymmetric excess in the $E$-mode signal is apparent at both $\nu>3.5$ and,
at lower significance, $\nu<-3.5$.
The amplitude, slope and non-Gaussianity of this excess are all powerful discriminators between values of
parameters in cosmological models (Pires et al.\ 2009).
Positive peaks correspond mainly to dark matter halos around galaxy clusters.
Local minima could correspond to voids (Jain \& van Waerbeke 2000; Miyazaki et al.\ 2002),
but the large size of voids is ill-matched to our $\theta_G=1\arcmin$ filter width,
and their density contrast can never be greater than unity, so this aspect of our data is likely just noise.

Figure~\ref{fig:fig16} recasts the peak distribution into a cumulative density of positive maxima or negative minima.
As expected, we find a non-Gaussian mass distribution with more highly significant positive maxima
(corresponding to mass overdensities) than highly significant negative minima (see Table~\ref{tab:tab2}).
Analytic predictions of peak counts are also overlaid.
Following van Waerbeke (2000), dot-dashed lines show the expected density of pure noise peaks,
and dotted lines show the expected number of true dark matter halos.
Predictions from Fan et al.\ (2010), which also take into account the effect of noise on the heights of true peaks and the clustering of noise peaks near dark matter halos, are shown as circles with error bars.
In these theoretical calculations, we model the population of background galaxies as having an
intrinsic ellipticity dispersion $\sigma_e=0.278$, density $n_g=14.5~\rm arcmin^{-2}$ and the
redshift distribution from  Fu et al.\ (2008) in a $\Lambda$CDM universe.
At $\nu>4.5$, it appears that theory may begin to predict more peaks than are observed. However, these are very small number statistics, and our observations are consistent with analytical predictions within Poisson noise. At very low $\nu$, the number of peaks is washed out and actually decreases when noise is superimposed, because it is impossible to extract very low-$\nu$ peaks. However, if larger and deeper surveys still find fewer low-$\nu$ or high-$\nu$ peaks than expected, and systematic effects such as the consequences of masked regions are more fully understood, it may indicate a cosmology with, e.g., lower $\Omega_m$ and $\sigma_8$ than the values used for our predictions.

Our main conclusions about the distribution of convergence peaks are as follows.
\begin{itemize}
\item
Peak counts detected in CFHTLS-Wide are consistent with predictions
from a $\Lambda$CDM cosmological model, once noise effects are properly included
(van Waerbeke 2000; Fan et al.\ 2010).
\item
The convergence field is non-Gaussian. The excess of local maxima with $\nu>3.5$
compared to local minima with $\nu<-3.5$ is also consistent with models (Miyazaki et al.\ 2002;
Gavazzi \& Soucail 2007).
\item
Noise peaks dominate the expected peak counts due to cosmological weak lensing
below $|\nu|\simlt3$.
\end{itemize}

We expect that weak lensing peak counts will become reliably employed to constrain
cosmological parameters in future lensing analyses.
Although it is difficult to remove the contributions of noise from intrinsic galaxy shapes and
the projection of large scale structures, these effects can be analytically predicted.
Pushing these predictions into the low signal-to-noise regime might also allow constraints to
use a much higher number density of less massive peaks, tightening predictions by reducing Poisson noise.

\section{Optical/X-ray counterparts}\label{sec:counterparts}

We shall now compare our weak lensing peak detections with catalogs of overdensities detected via optical or X-ray emission.
We adopt the {\em K2} optical cluster catalog constructed using photometric redshifts
from the same CFHTLS-Wide W1 imaging (Thanjavur et al.\ 2009),
and the XMM-LSS X-ray selected cluster catalog (Adami et al.\ 2011), which partially overlaps our survey field.

\begin{table*}
\centering
\caption{The Detection Purity of Various Weak Lensing Cluster Surveys Described in the Literature.}
\begin{tabular}{ccccc}
\hline
\hline
   &  \rm Weak Lensing Cluster Sample & \rm Comparison Sample & Purity\\
\hline
Hamana et al. \ (2004) & $n_\mathrm{gal}=30$~arcmin$^{-2}$, $\nu>4$ & Halos (simulations) & $60\%$ \\
Dietrich et al. \ (2007) & $n_\mathrm{gal}=18$~arcmin$^{-2}$, & Halos (simulations) & $75\%$\\
Miyazaki et al. \ (2007) & $\nu>3.69$ & X-ray clusters (XMM-LSS) & $80\%$ \\
Gavazzi \& Soucail \ (2007) & $n_\mathrm{gal} \sim 35$~arcmin$^{-2}$, $\nu>3.5$ & Photometric clusters & $\sim 65\%$ \\
Schirmer et al. \ (2007) & $\nu>4$ & BCG & $\sim 45\%$\\
Geller et al. \ (2010) & $\nu>3.5$ & Spectroscopic clusters (SHELS)
 & $\sim 33\%$\\
\hline
\end{tabular}
\vspace{5mm}
\label{tab:tabpurity}
\end{table*}

We define the purity $f_p$ of our blind weak lensing cluster search
as the fraction of peaks above a given detection threshold $\nu_{\rm th}$
that are associated with an optically detected cluster
%relative to all the peaks with height greater than that threshold
\begin{equation}
f_p=\frac{N_{\rm matched}(\nu>\nu_{\rm th})}{N(\nu>\nu_{\rm th})}~.
\end{equation}
The expected purity depends upon the survey depth (galaxy density), systematics, and the extent of multiband or
spectroscopic follow-up.
Using numerical simulations with $n_\mathrm{gal}=30$~arcmin$^{-2}$, Hamana et al.\ (2004) predict a purity of more than $60\%$ for convergence peaks with $\nu>4$.
The Bonn Lensing, Optical, and X-ray selected galaxy clusters (BLOX) simulations by Dietrich et al.\ (2007) with $n_\mathrm{gal}=18$~arcmin$^{-2}$ also predict that $75\%$ of matches between convergence map peaks and massive halos are within $2\arcmin.15$.
In practice, Miyazaki et al.\ (2007) achieve $80\%$ purity for the $\sim100$ peaks in Subaru convergence maps with $\nu>3.69$.
Gavazzi \& Soucail (2007) obtain $\sim 65\%$ purity from $14$ peaks in CFHTLS-Deep with $\nu>3.5$.
Schirmer et al.\ (2007) obtain $\sim45\%$ purity for the $158$ possible mass concentrations identified in the
Garching-Bonn Deep Survey (GaBoDS) at $\nu>4$,
consistent with an earlier evaluation of a subsample of the survey (Maturi et al.\ 2007).
Geller et al.\ (2010) find only $\sim 33\%$ purity by combining the Deep Lens Survey with the Smithsonian Hectospec Lensing Survey (SHELS) (see Table~\ref{tab:tabpurity}).

Figure~\ref{fig:fig17} shows a subset of our cluster search in the representative W1+2+3 pointing, which also overlaps
with the XMM-LSS survey (Pacaud et al.\ 2007; Adami et al.\ 2011).
The observed weak lensing peak positions may not coincide exactly with cluster centers defined from optical emission,
because of a combination of noise, substructure, and
physical processes associated with cluster mergers (Fan et al.\ 2010; Hamana et al.\ 2004).
We therefore search for matched pair candidates in K2 within a $3\arcmin.0$ radius of peaks
that appear in both the $\theta_G=1\arcmin$ and $\theta_G=2\arcmin$ lensing map.
This search radius is chosen to be larger than the smoothing scale,
but smaller than the angular virial radius of a massive cluster at $0.1<z<0.9$ (Hamana et al.\ 2004).
If more than one pair exists within $3\arcmin.0$, we adopt the closest match as the primary candidate.

\begin{table}
\centering
\caption{Cluster Search Purity as a Function of Peak Height Threshold $\nu_{\rm th}$.}
\begin{tabular}{cccc}
\hline
\hline
$\nu_{\rm th}$ &  $N(\nu>\nu_{\rm th})$ & $N_{\rm matched}(\nu>\nu_{\rm th}, \rm K2)$ & $f_p$ \\
\hline
3.5 & 301 & 126 & 42\% \\
4.0 & 125 & 67 & 54\% \\
4.5 &  51 & 30 & 59\% \\
\hline
\end{tabular}
\vspace{5mm}
\label{tab:tab3}
\end{table}

We obtain $126$ matches between weak lensing peaks with $\nu>3.5$ and optical K2 clusters: corresponding to
a purity of $42\%$. This is lower than in CFHTLS-Deep (Gavazzi \& Soucail 2007) because of the much lower source galaxy density.
The purity is listed as a function of detection threshold $\nu_{\rm th}$ in Table~\ref{tab:tab3}, and
the complete catalog of matched pairs is presented in Table~\ref{tab:matchpeaks}.
Figure~\ref{fig:fig18} shows the separations between the matched weak lensing peaks and K2 centers.
The left panel shows the offsets for peaks with $\nu>0$ in maps with various smoothing scales;
the separation between peaks typically matches the smoothing scale.
The right panel shows the offsets for only the reliably-detected peaks with high $\nu$;
the finite number of clusters is too low to draw solid conclusions about the typical separation.
Figure~\ref{fig:fig19} shows the redshift distribution of matched clusters, relative to the overall {\em K2} sample.
Our lensing selection preferentially detects clusters at $0.2<z<0.4$, and becomes inefficient above $z>0.5$.

We also compare our $\nu>3.5$ lensing peaks with X-ray observations from the XMM-LSS survey.
Adami et al.\ (2011) present $66$ spectroscopically confirmed clusters $(0.05<z<1.5)$ within the $6~\rm deg^2$ XMM-LSS survey.
This partially overlaps with the CFHTLS-W1 field ($53$ X-ray clusters are within W1).
In this overlap region, we find $31$ lensing peaks, $7$ of which are within $5\arcmin.0$ of X-ray clusters.
Note that we increase the distance threshold for matches because of additional noise in the X-ray centers and
the common phenomenon of separation between the gravitational field and X-ray gas
--- Shan et al.\ (2010) found that $45\%$ of a sample of $38$ clusters had X-ray offsets $>10 \arcsec$.
Indeed, in our new sample, we also find that the offsets between weak lensing and X-ray centers are always
comparable to or much larger than the offset between weak lensing and optical centers.

Miyazaki et al.\ (2007) also perform a weak lensing analysis of $0.5~\mathrm{deg}^2$ of the XMM-LSS survey.
They find $15$ lensing peaks with optical counterparts, of which $10$ match X-ray selected clusters (Adami et al.\ 2011).
Many of the X-ray clusters are simply at too high a redshift to be detected by the CFHTLS lensing data.
However, three of these clusters are detected in both our analysis (c77, c92 and c93 in Table~\ref{tab:matchxray}) and that of Miyazaki et al.\ (2007).

\section{Tomographic analysis of lensing peaks}\label{sec:mass3d}

Since multi-band photometric redshift estimates are available for $76\%$ of the source galaxies,
we shall now perform a 3D, tomographic shear analysis around all $\nu>3.5$ cluster candidates with an optical or X-ray counterpart.
This process yields an estimate of the cluster redshift (and mass) independently of its visible emission.
It also further cleans the cluster catalog of spurious peaks created by either noise
or the projection of multiple small systems along a single line of sight.
%Behind a real lens cluster at redshift $z_l$, the shear signal increase as a
%function of the distance to the source galaxy
%\begin{equation}
%\Sigma_c(z_s;z_l)=\frac{c^2}{4\pi G}\frac{D_s}{D_lD_{ls}} ~,
%\end{equation}
%where $D$ quantities are angular diameter distances to the source, to the lens, and from the lens to the source.
%Any shear signal not showing this characteristic rise with redshift is inconsistent with the single lens hypothesis.

We fit the 3D shear signal around each cluster candidate to both a
singular isothermal sphere (SIS) and a Navarro et al. (1996; NFW) model,
with the cluster lens as an additional free parameter.
We assume that the center of each cluster candidate is at the position of the brightest central galaxy (BCG).
This may not be optimal (Johnston et al. 2007), but it is much more precisely known than the peak of the lensing
signal (Fan et al.\ 2010).
For each source galaxy, we adopt the best-fit redshift from the cleaned Arnouts et al.\ (2010) catalog.
Following Gavazzi \& Soucail (2007), we measure the mean tangential shear in a $1\arcmin$--$5\arcmin$ annulus
from the cluster center, excluding the core to minimize dilution of the signal from any cluster member galaxies
with incorrect photometric redshifts.

For an SIS model, the component of shear tangential to the cluster center is
\begin{equation}
\gamma_t(\theta,z_s;z_l)=\frac{D_{ls}}{D_s} \frac{4\pi\sigma^2_v}{c^2}\frac{1\arcsec}{2\theta}~,
\end{equation}
where $\theta$ is the angular distance from the center.
We simultaneously fit the unknown lens redshift and characteristic cluster velocity dispersion $\sigma_v$ by minimizing
\begin{equation}
\chi_{\rm SIS}^2(z_l,\sigma_v)=\sum_i \frac{\big(\gamma_{t,i}-\gamma_{{\rm SIS},i} (z_l,\sigma_{\rm tomo})\big)^2} {\sigma_{e,i}^2} ~,
\end{equation}
where $w_i=w(z_l,z_{s,i})$ and $\sigma_{e,i}^2$ is given by Equation.~(2).

The full NFW model has two free parameters, but we assume the Bullock et al.\ (2001) relation between concentration $c$
and virial mass $M_{\rm vir}$ seen in numerical simulations
\begin{equation}
c_{\rm NFW}(M_{\rm vir},z)=\frac{c_*}{1+z} \left( \frac{M_{\rm vir}}{10^{14} h^{-1}M_{\odot}} \right)^{-0.13},
\end{equation}
where $c_*=8$.
As for the SIS model, there is then only one free parameter, and we fit the shear field for lens redshift and halo mass
by minimizing
\begin{equation}
\chi_{\rm NFW}^2(z_l,M_{\rm vir})=\sum_i \frac{\big(\gamma_{t,i}-\gamma_{\rm NFW}(z_l,M_{\rm vir})\big)^2}{\sigma_{e,i}^2}.
\end{equation}
To aid comparison with other work, for each cluster we calculate both the virial mass $M_{\rm vir}$  and the
mass $M_{\rm 200}$ enclosed within a radius $r_{\rm 200}$ in which the mean
density of the halo is $200$ times the critical density at the redshift of the cluster.

\begin{figure}
\begin{center}
\includegraphics[width=0.75\columnwidth]{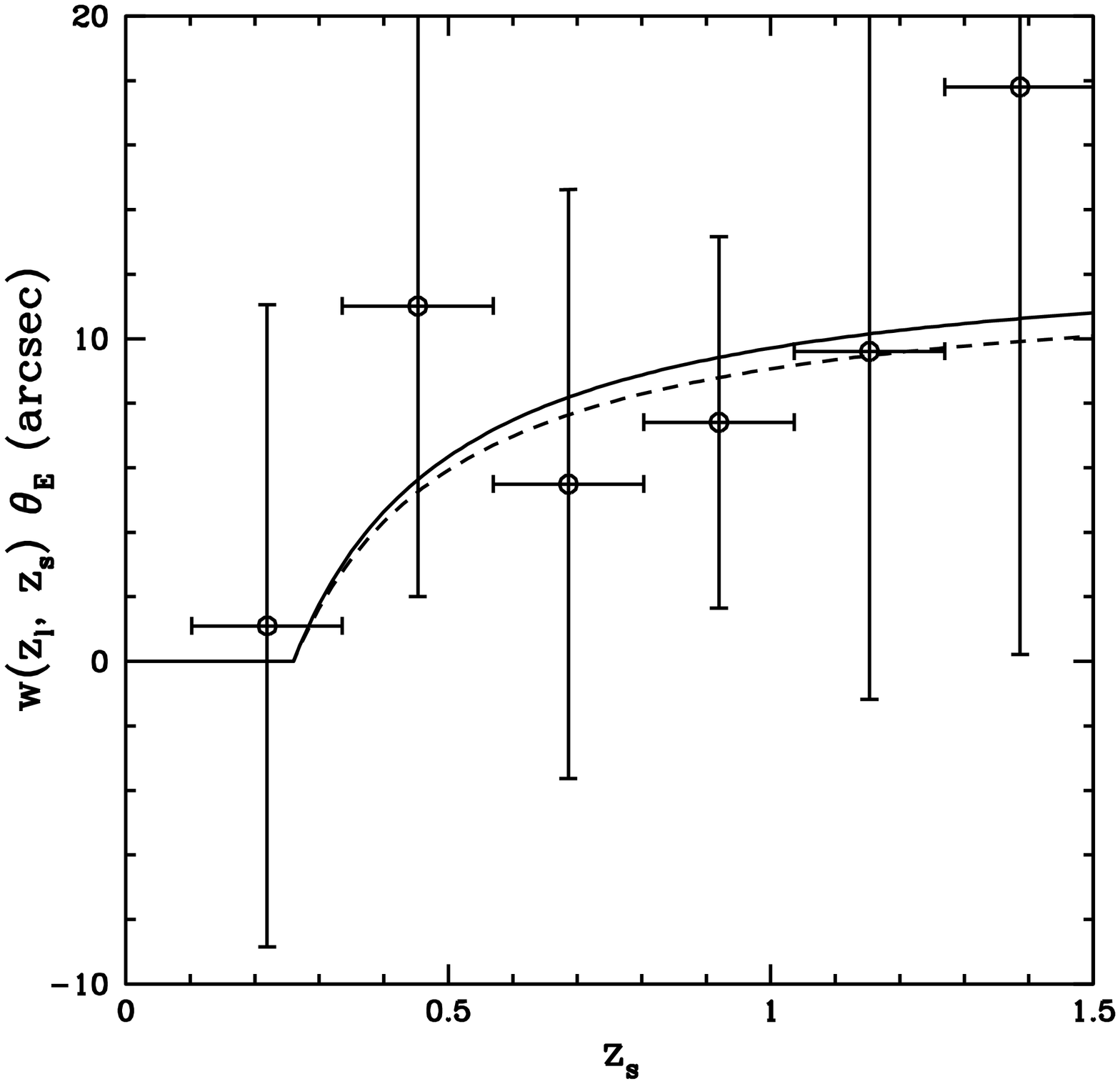}
\includegraphics[width=0.75\columnwidth]{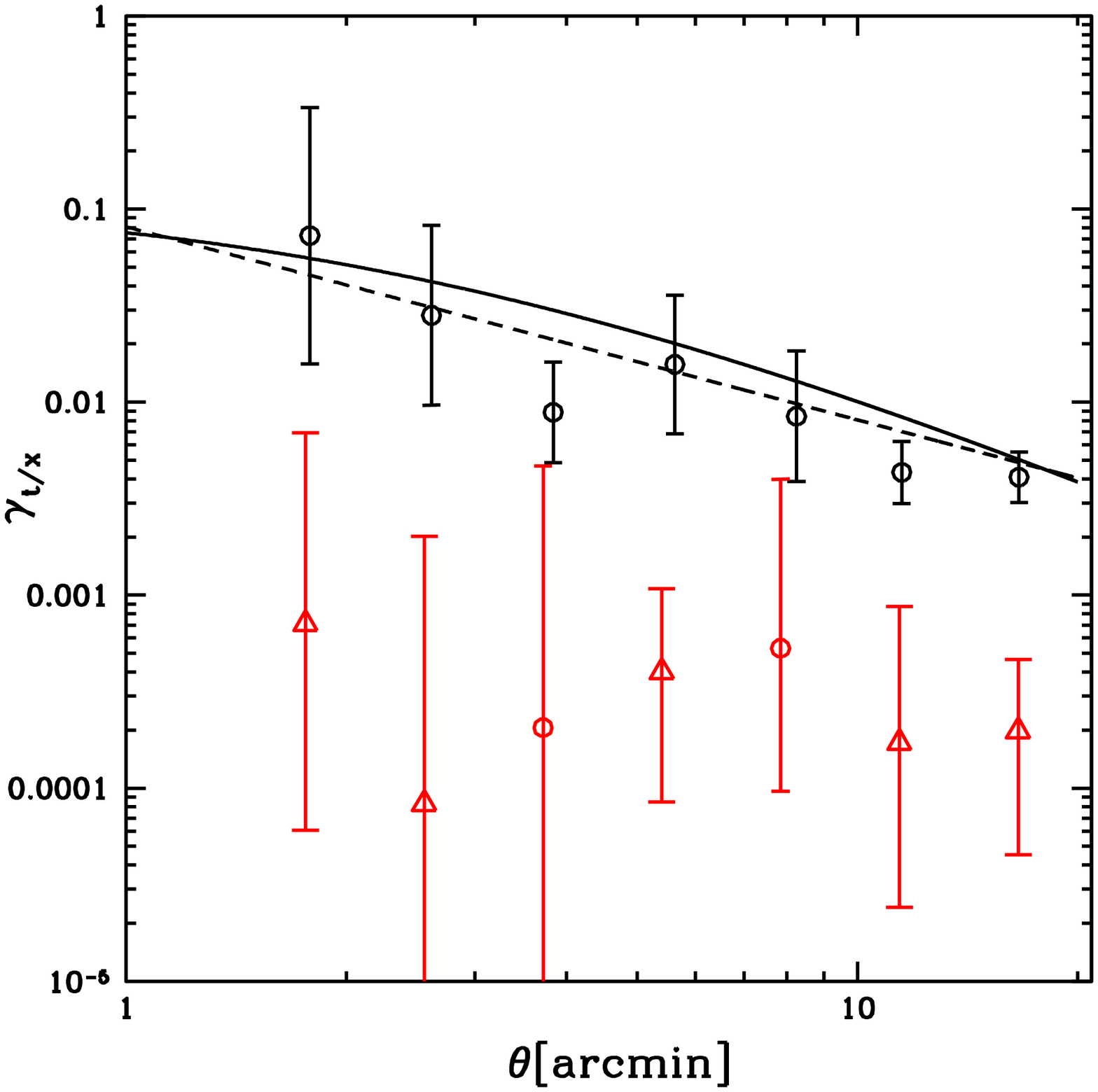}
\includegraphics[angle=0.0,trim=0 0 0 0,width=0.75\columnwidth]{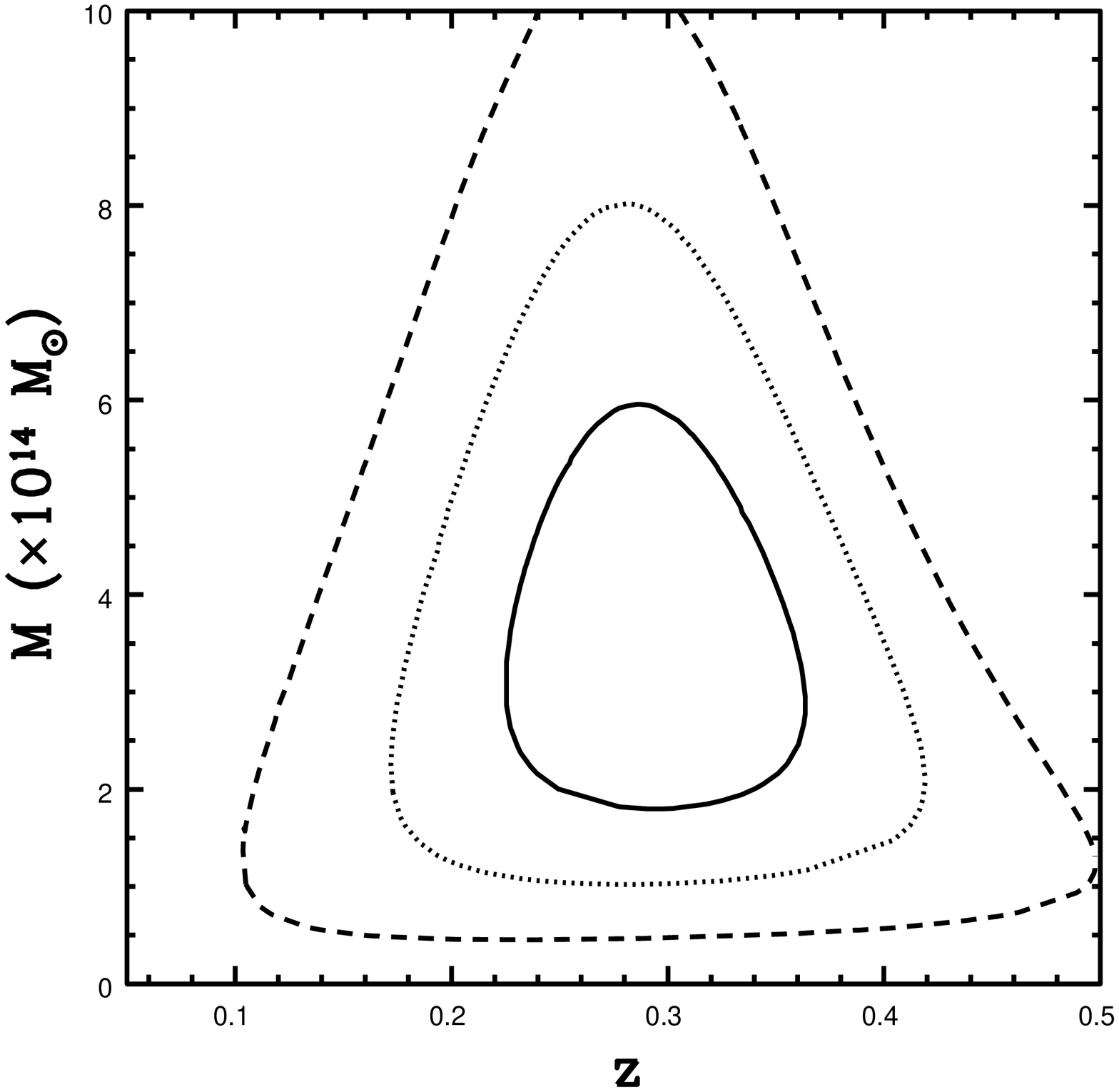}
\caption{Results of a tomographic weak lensing analysis around one example peak (c3 in our catalog), which has
an optical counterpart at redshift $z=0.28$.
The solid and dashed lines are the best-fit SIS and NFW models.
Top: projection in the redshift direction showing the characteristic increase with redshift of a real signal.
The best-fit lens redshift is $0.26^{+0.11}_{-0.10}$ for an SIS model and $0.26^{+0.12}_{-0.12}$ for an NFW model.
Middle: radial profile projected onto the plane of the sky.
Black circles show the $E$-mode tangential shear signal.
Red circles (triangles) show positive (negative) values of the $B$-mode systematic signal,
which oscillates about zero.
Bottom: joint redshift-mass constraints from the best fit NFW model. Contours show 68\%, 95\% and 99\% confidence limits.
\label{fig:fig20}}
\end{center}
\end{figure}

\begin{figure*}
\begin{center}
\includegraphics[angle=0.0,width=2.0\columnwidth]{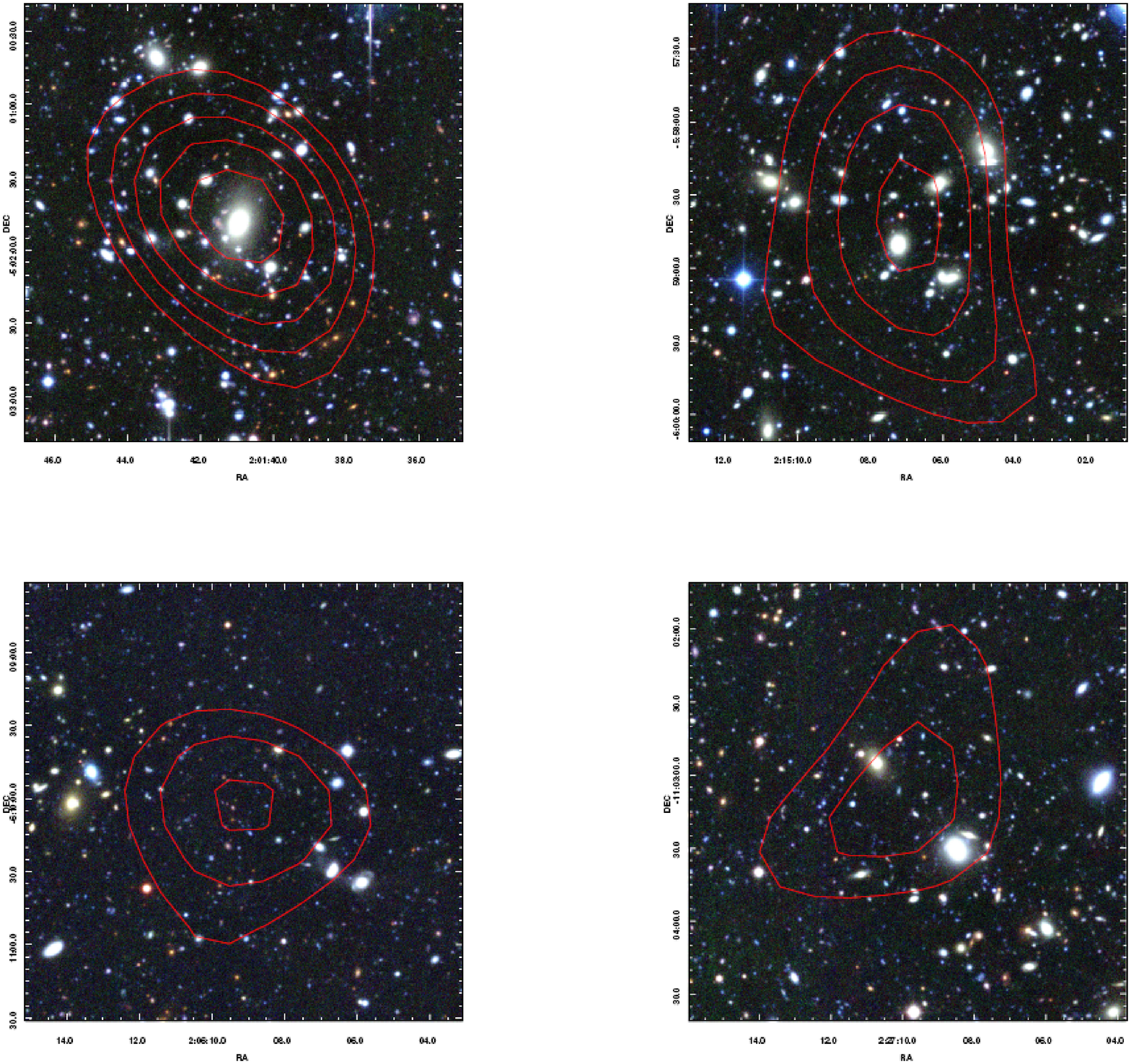}
\caption{Composite $3\times 3$~arcmin$^2$ CFHTLS $g'$, $r'$, $i'$ color images
for four clusters detected in both weak lensing and the {\em K2} optical catalog.
Contours show signal-to-noise in reconstructed convergence, starting at $\nu=3.0$ and increasing
in steps of $0.5$ and the images are centered on the convergence peak.
Top-left: cluster c3 with $\nu=5.395$.
Top-right: cluster c49 with $\nu=4.728$.
Bottom-left: cluster c21 with $\nu=4.142$.
Bottom-right: cluster c103 with $\nu=3.725$.
Candidate c21 is included here as an example of a 2D lensing peak that is probably spurious: it does not appear obviously associated with an overdensity of galaxies in the optical imaging, and is not well-fit by a 3D shear signal.
\label{fig:fig21}}
\end{center}
\end{figure*}

Figure~\ref{fig:fig20} illustrates our 3D tomographic results on one cluster (identification c3).
The mean tangential shear is consistent with zero for $z_s \le 0.26$.
The subsequent increase with redshift is clear and allows for an unambiguous identification of the lens redshift.
Error bars are derived from the scatter in observed ellipticities (intrinsic+measurement error), as determined by Equation~(\ref{eqn:weight}).
The observed tangential shear profile is consistent with either an SIS or NFW model.
Due to our low density of source galaxies and our exclusion of galaxies within the central $1\arcmin$,
we have insufficient signal-to-noise for individual clusters to distinguish between the two models,
which differ most noticeably near the core.
The amplitude of the systematic $B$-mode signal, computed by rotating all shear estimates by $45^{\circ}$,
is always at least one order of magnitude smaller than the tangential shear, and it oscillates about zero.

Figure~\ref{fig:fig21} shows the signal-to-noise contours of the convergence signal reconstructed around four clusters
from the 3D shear signal.

\begin{figure}
\begin{center}
\includegraphics[angle=0.0,width=0.95\columnwidth]{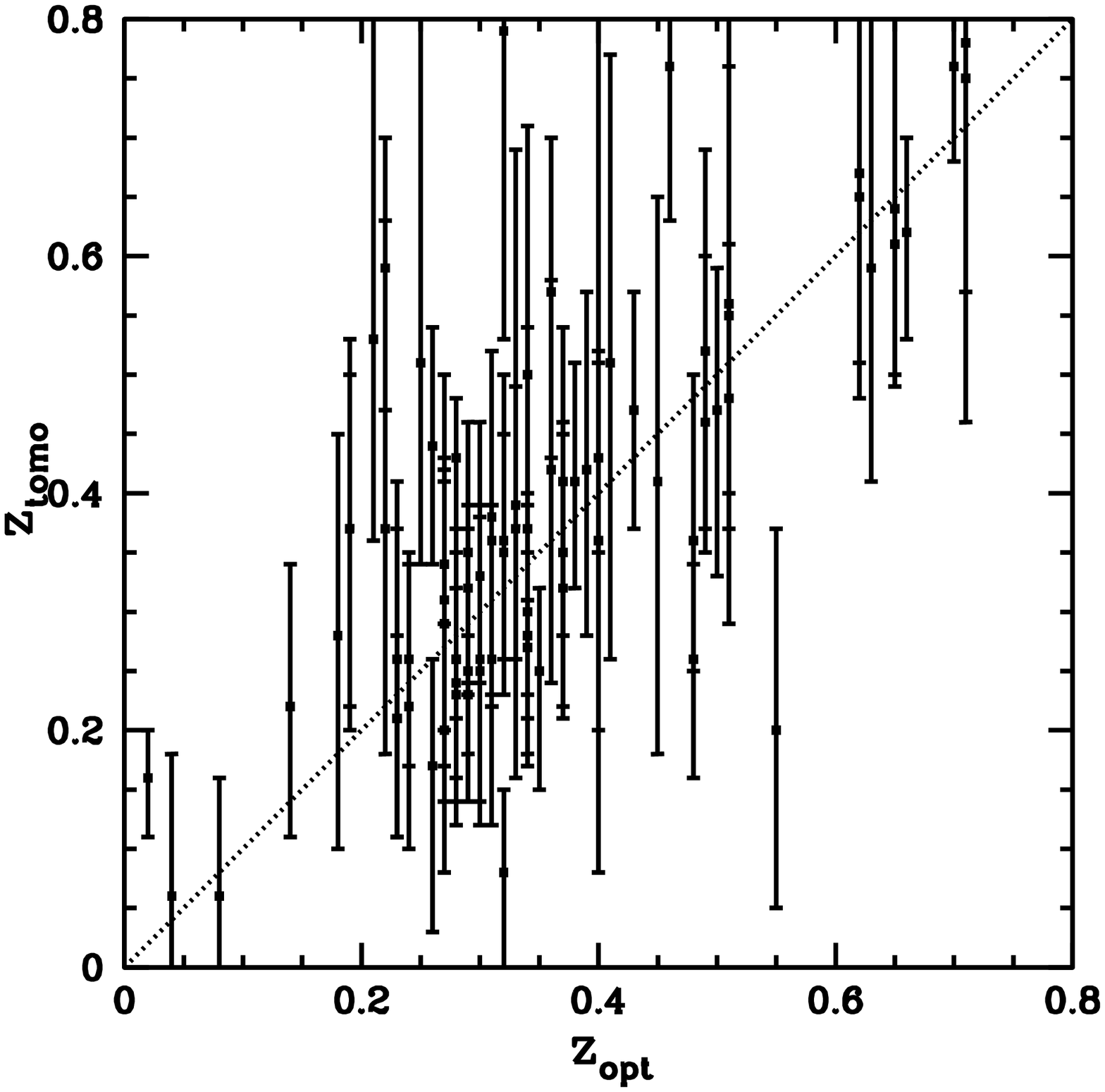}
\caption{Comparison between cluster redshifts derived directly from
the photometric redshifts of cluster member galaxies $z_{\rm opt}$
and from weak lensing tomography $z_{\rm tomo}$.
This figure includes all $85$ $\nu > 3.5$ weak lensing peaks with optical counterparts,
whose 3D shear signal is consistent at $\chi^2_\mathrm{red}<3$ with the expected
increase as a function of redshift. \label{fig:fig22}}
\end{center}
\end{figure}

Our tomographic analysis confirms the identification of $85$ clusters with $\chi^2_\mathrm{red}=\chi^2/{\rm dof}<3$.
The mean redshift and velocity dispersion of these clusters are $\langle z_c\rangle=0.36$ and
$\langle\sigma_v\rangle=658.8 \rm km~s^{-1}$. Their full properties are listed in Table~\ref{tab:matchpeaks}.
Reassuringly, we find that the inferred lens redshifts are effectively identical for either SIS or NFW profile fits,
and are consistent (although noisy) with the (independent) photometric redshifts of the cluster member galaxies.
Figure~\ref{fig:fig22} compares the redshift estimates from tomographic gravitational lensing and optical spectroscopy of member galaxies for these $85$ clusters.

At the faint end of our source galaxy sample, photo-$z$ estimates will be unreliable because of noise in the photometry and degeneracies in
the broad-band colors of galaxies with different spectral energy distributions at different redshifts. This will show up as a ``double peak'' in the posterior
probability of the redshift distribution. Using only the best-fit peak might be randomly picking whichever of these peaks
is higher because of noise. This often biases lensing analyses because the expected lensing signal may be much higher at one redshift than the other. To check for such an effect, we redo the tomographic analysis without any photo-$z$s that have a
double peak (Arnouts et al.\ 2010). For cluster
c3 in our catalog, we get very similar fit results: $z_{\rm SIS}=0.26^{+0.13}_{-0.12}$ and $\sigma_v=724.0^{+128.3}_{-131.5}$
for the SIS model, and $z_{\rm NFW}=0.26^{+0.09}_{-0.11}$ and $m_{\rm NFW}=4.14^{+0.81}_{-0.76}$ for the NFW model.
This suggests that the fit is not very sensitive to that population of galaxies with ``double peaks'' in photometric redshifts.

In cases where the data are poorly fit ($\chi^2_\mathrm{red}>3$) by a 3D lensing signal, the inferred velocity dispersion or virial mass are typically very small (often consistent with zero).
These systems are probably spurious peaks due to noise or projection effects.
In addition, two peaks (c71 and c104) have an unphysically high velocity dispersion $\sigma_{\rm tomo}>1300~\rm km~s^{-1}$.
After careful examination of the image data, we found that they lie near two strongly saturated stars
whose flux extends beyond the masked regions, possibly degrading galaxy shape measurements.

One important goal of cluster lensing is to measure the total mass of systems,
which can also be estimated from the velocity dispersion of its member galaxies or
from its X-ray emission, under various assumptions
about the state of the intra-cluster medium and hydrostatic equilibrium
(e.g.,\ Bahcall et al.\ 1995; Carlberg et al.\ 1996;  Carlstrom et al.\ 2002).

\begin{figure}
\begin{center}
\includegraphics[width=0.95\columnwidth]{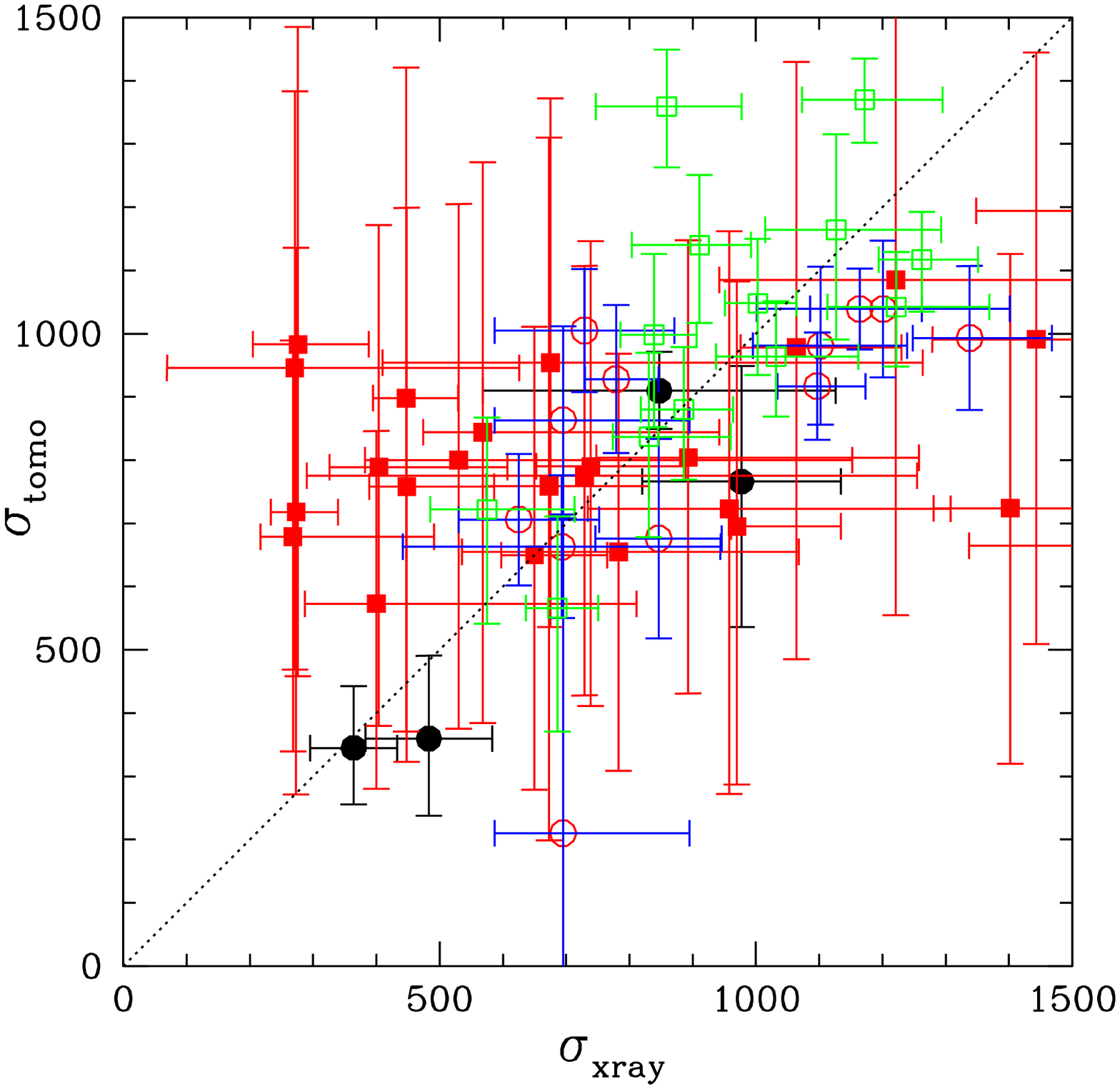}
\caption{Scaling relation between X-ray and weak lensing tomographic measurements of the velocity dispersion of clusters.
Black filled circles show our data for four lensing-selected clusters.
Red filled squares show the lensing-selected clusters of Hamana et al.\ (2009).
Blue open circles and green open squares denote the X-ray selected clusters from Cypriano et al.\ (2004) and
Hoekstra (2007) respectively.
\label{fig:fig23}}
\end{center}
\end{figure}

Figure~\ref{fig:fig23} shows the relation between one mass proxy obtained from X-ray observations
and another obtained from weak lensing, for the $4$ of $7$ clusters in our sample with an X-ray counterpart
and an SIS velocity dispersion parameter well constrained by lensing.
These are overlaid on the results of another lensing-selected cluster sample (Hamana et al.\ 2009)
and two X-ray selected cluster samples (Cypriano et al.\ 2004; Hoekstra 2007).
Even combining these samples, the scatter is large.
However, the consistency of the scatter between the four samples suggests that neither detection
method produces a strong selection bias.
Note that, since the density (and shear) profile of real cluster is not necessarily a single power law,
the best-fit SIS model may depend on details of the fitting method, and the range over which data are fit.
Our results therefore are somewhat method-dependent.
A corollary of this issue is that it might also be possible to minimize scatter in the
$\sigma_{\rm xray}$-$\sigma_{\rm tomo}$ relation by optimizing the tomographic lens fitting method.

\section{Conclusions}\label{sec:conc}

We have presented a tomographic weak gravitational lensing analysis of
the completed $64 ~\rm deg^2$ CFHTLS-Wide W1 field,
demonstrating some of the power of lensing to probe mass in galaxy clusters.
We measured the shapes of distant galaxies using the KSB shape measurement method,
which we verified against shape measurements from high-resolution {\em HST} imaging of an
overlapping sky area.
We also obtained consistent shape measurements using two independent imaging bands.
The level of residual shape measurement systematics is an order of magnitude lower than
the $1\%$--$10\%$ shear signal expected in galaxy clusters, so this analysis is acceptable
for cluster studies.

We have reconstructed the largest contiguous ``dark matter map'' convergence field to date,
using two different smoothing scales to help remove spurious noise peaks.
From this map, we performed the largest lensing-selected blind cluster search to date, finding
$301$ local maxima in the lensing map with detection significance $\nu>3.5$.
Once sources of noise are properly modeled from the intrinsic shapes of galaxies and substructure,
this is consistent with predictions from a $\Lambda$CDM cosmology.
Note that our theoretical calculations (Fan et al.\ 2010) do not consider projection effects of
structures along the line of sight.
%, because this is expected to be insignificant for relatively high-$\nu$ peaks
%(Dietrich \& Hartlap 2010).}

To identify counterparts of the weak lensing peaks, we match our cluster candidates to the K2 galaxy cluster catalog,
created using photometric redshift estimates across the CFHTLS-Wide W1 field (Thanjavur et al.\ 2009).
Of the $301$ peaks with $\nu>3.5$, $126$ have a corresponding optically-detected BCG within $3\arcmin.0$.
In the (much smaller) survey area that overlaps the XMM-LSS survey, we also find matches for seven lensing peaks with
X-ray selected clusters. Thus, many of the candidate peaks are indeed likely just noise.

Tomographic weak lensing techniques dramatically improve standard 2D algorithms.
In a full 3D lensing analysis of the $\nu>3.5$ peaks, we further distinguish real clusters from noise fluctuations,
and confirm (at $\chi^2_{\rm red}<3$) the identification of $85$ clusters.
Importantly, we obtain independent measurements of the cluster lens redshifts,
which are consistent with the redshifts of their previously-identified optical counterparts.
For each cluster, we fit NFW and SIS radial profiles to the lensing data to measure the mass
or velocity dispersion $\sigma_{\rm tomo}$.
The clusters' mean redshift and velocity dispersion is $\langle z_c\rangle=0.36$ and $\langle\sigma_c\rangle=658.8~\rm km~s^{-1}$.
Weak lensing measurements of the total mass in the four of our clusters with X-ray counterparts
are also in reasonable agreement with mass estimates obtained from X-ray emission.
Future surveys, such as DES, LSST, KDUST and EUCLID, will be able to apply these techniques to
map clusters in much larger volumes.
Such large catalogs will also be able to tightly constrain cosmological models (Takada \& Bridle 2007; Dietrich \&
Hartlap 2010).

\section*{Acknowledgments}
The authors thank Bernard Fort, Liping Fu, Bo Qin, Catherine Heymans, and Ludovic Van Waerbeke
for useful discussions. H.Y.S. acknowledges support from Sino French laboratories FCPPL and Origins and CPPM hospitality
during stays in France.
H.Y.S. acknowledges the support from NSFC of China under grants 11103011 and China Postdoctoral Science Foundation.
J.P.K. acknowledges supports from CNRS as well as
PNCG and CNES. Z.H.F. acknowledges the support from NSFC of China under grants 10773001, 11033005,
and 973 program 2007CB815401. M.L. acknowledges the Centre National de la Recherche Scientifique (CNRS) for its
support. The Dark Cosmology Centre is funded by the Danish National Research Foundation.

This work is based on observations obtained with MegaPrime/MegaCam, a joint project of CFHT and CEA/DAPNIA, at the
Canada-France-Hawaii Telescope (CFHT) which is operated by the National Research Council (NRC) of Canada,
the Institut National des Science de l'Univers of the Centre National de la Recherche Scientifique (CNRS)
of France, and the University of Hawaii. This work is based in part on data products produced at TERAPIX
and the Canadian Astronomy Data Centre as part of the Canada-France-Hawaii Telescope Legacy Survey, a
collaborative project of NRC and CNRS.

This work also uses observations obtained with the {\em Hubble Space Telescope}.
The HST COSMOS Treasury program was supported through NASA grant
HST-GO-09822. We thank Tony Roman, Denise Taylor, and David
Soderblom for their assistance in planning and scheduling the extensive
COSMOS observations. We thank the NASA IPAC/IRSA staff (Anastasia Laity,
Anastasia Alexov, Bruce Berriman, and John Good) for providing online
archive and server capabilities for the COSMOS data sets. It is also our
pleasure to gratefully acknowledge the contributions of the entire COSMOS
collaboration, consisting of more than 70 scientists. More information on
the COSMOS survey is available at {\tt \url{http://www.astro.caltech.edu/~cosmos}}.

\newpage
\begin{turnpage}
\begin{center}
\tabletypesize{\tiny}
\begin{longtable}{cccccccccccccccc}
\caption{\rm Catalog of the Matching Groups/Clusters between $\nu>3.5$
Convergence Peaks and K2-detected Groups/Clusters in CFHTLS-Wide W1 Fields. K2 redshift denotes the median
redshift of bright ($i \le 20$) cluster members} \label{tab:matchpeaks} \\
\hline
ID & $z_{\rm K2}$ & $\alpha_{\rm K2}$ & $\delta_{\rm K2}$ & $\rm sig_{r-i}$ & $\alpha$ & $\delta$ & d & $z_{\rm SIS}$ & $\sigma_v$ & $\chi^2_{\rm SIS}$ & $z_{\rm NFW}$ & $m_{\rm NFW}$ & $m_{200}$ & $\chi^2_{\rm NFW}$ & $\nu$ \\ \hline
 & & & & & & & arcsec & & $\rm km/s$ & & & $10^{14}M_{\odot}/h$ & $10^{14}M_{\odot}/h$ & & \\ \hline
c1-w1 & 0.18 & 30.990 & -4.223 & 3.19 & 30.976 & -4.231 & 59.2 & $0.28^{+0.17}_{-0.10}$ & $829.7^{+225.1}_{-308.9}$ & 0.42 & $0.29^{+0.16}_{-0.14}$ & $2.88^{+6.22}_{-2.15}$ & $2.50^{+5.39}_{-1.86}$ & 0.41 & 4.458 \\
c2-w1 & 0.48 & 30.625 & -3.966 & 4.09 & 30.619 & -3.970 & 29.6 & $0.26^{+0.08}_{-0.10}$ & $945.5^{+173.8}_{-212.5}$ & 1.52 & $0.26^{+0.07}_{-0.08}$ & $11.0^{+17.8}_{-7.20}$ & $9.32^{+15.1}_{-6.10}$ & 1.50 & 4.113 \\
c3-w1 & 0.28 & 30.421 & -5.030 & 7.45 & 30.420 & -5.030 & 1.62 & $0.26^{+0.11}_{-0.10}$ & $701.6^{+134.2}_{-132.8}$ & 0.85 & $0.26^{+0.12}_{-0.12}$ & $3.51^{+0.63}_{-0.66}$ & $3.02^{+0.54}_{-0.57}$ & 0.88 & 5.395 \\
c4-w1 & 0.72 & 30.296 & -6.089 & 9.85 & 30.260 & -6.088 & 128.5 & $0.69^{+0.21}_{-0.22}$ & $\sim 0$ & 4.33 & $0.69^{+0.23}_{-0.20}$ & $\sim 0$ & $\sim 0$ & 4.25 & 3.732 \\
c5-w1 & 0.38 & 30.860 & -5.925 & 4.72 & 30.867 & -5.945 & 74.3 & $0.41^{+0.10}_{-0.09}$ & $503.1^{+102.5}_{-104.1}$ & 1.37 & $0.40^{+0.10}_{-0.11}$ & $0.78^{+0.16}_{-0.17}$ & $0.70^{+0.14}_{-0.15}$ & 1.37 & 3.510 \\
c6-w1 & 0.27 & 30.840 & -7.331 & 3.68 & 30.836 & -7.341 & 38.2 & $0.20^{+0.09}_{-0.12}$ & $559.8^{+131.8}_{-139.6}$ & 0.68 & $0.22^{+0.11}_{-0.10}$ & $0.79^{+0.18}_{-0.17}$ & $0.69^{+0.16}_{-0.15}$ & 0.51 & 4.184 \\
c7-w1 & 0.33 & 31.077 & -7.213 & 3.20 & 31.080 & -7.216 & 15.5 & $0.39^{+0.30}_{-0.23}$ & $646.1^{+183.9}_{-222.3}$ & 1.05 & $0.40^{+0.25}_{-0.35}$ & $1.38^{+2.09}_{-1.18}$ & $1.23^{+1.86}_{-1.05}$ & 1.03 & 3.677 \\
c8-w1 & -99.0 & 30.779 & -7.142 & 4.28 & 30.785 & -7.141 & 20.1 & $0.76^{+0.15}_{-0.19}$ & $162.3^{+137.0}_{-125.4}$ & 0.78 & $0.80^{+0.16}_{-0.17}$ & $0.11^{+0.08}_{-0.09}$ & $0.10^{+0.07}_{-0.08}$ & 0.73 & 4.586 \\
c9-w1 & 0.81 & 30.884 & -6.736 & 5.78 & 30.882 & -6.749 & 46.8 & $0.77^{+0.25}_{-0.26}$ & $892.1^{+274.1}_{-291.7}$ & 1.25 & $0.78^{+0.24}_{-0.26}$ & $4.07^{+0.98}_{-1.04}$ & $3.82^{+0.92}_{-0.97}$ & 1.21 & 3.722 \\
c10-w1 & 0.36 & 30.628 & -6.524 & 5.11 & 30.649 & -6.536 & 85.3 & $0.41^{+0.13}_{-0.12}$ & $\sim 0$ & 9.34 & $0.41^{+0.12}_{-0.11}$ & $\sim 0$ & $\sim 0$ & 8.30 & 3.897 \\
c11-w1 & 0.31 & 30.364 & -8.283 & 4.20 & 30.369 & -8.275 & 35.2 & $0.35^{+0.16}_{-0.13}$ & $\sim 0$ & 5.22 & $0.35^{+0.15}_{-0.17}$ & $\sim 0$ & $\sim 0$ & 4.28 & 3.710 \\
c12-w1 & 0.14 & 30.325 & -7.651 & 4.43 & 30.346 & -7.659 & 81.9 & $0.22^{+0.12}_{-0.11}$ & $396.6^{+128.6}_{-102.1}$ & 1.96 & $0.42^{+0.15}_{-0.14}$ & $0.28^{+0.07}_{-0.06}$ & $0.25^{+0.06}_{-0.05}$ & 1.59 & 4.728 \\
c13-w1 & 0.23 & 30.708 & -9.337 & 4.83 & 30.704 & -9.356 & 69.7 & $0.21^{+0.16}_{-0.10}$ & $473.8^{+221.3}_{-239.5}$ & 2.15 & $0.21^{+0.13}_{-0.11}$ & $0.16^{+0.08}_{-0.09}$ & $0.14^{+0.07}_{-0.08}$ & 2.01 & 4.084 \\
c14-w1 & 0.40 & 30.534 & -8.437 & 7.52 & 30.537 & -8.417 & 71.4 & $0.43^{+0.09}_{-0.08}$ & $691.1^{+102.1}_{-97.8}$ & 0.14 & $0.43^{+0.09}_{-0.11}$ & $1.50^{+0.19}_{-0.17}$ & $1.34^{+0.17}_{-0.15}$ & 0.14 & 4.045 \\
c15-w1 & 0.32 & 30.870 & -9.817 & 18.02 & 30.881 & -9.838 & 86.2 & $0.36^{+0.14}_{-0.13}$ & $483.9^{+120.9}_{-112.7}$ & 0.68 & $0.36^{+0.13}_{-0.15}$ & $0.70^{+0.16}_{-0.18}$ & $0.62^{+0.14}_{-0.16}$ & 0.59 & 4.835 \\
c16-w1 & 0.66 & 30.452 & -10.853 & 5.81 & 30.463 & -10.840 & 59.6 & $0.62^{+0.08}_{-0.09}$ & $747.7^{+114.3}_{-117.2}$ & 1.15 & $0.62^{+0.10}_{-0.08}$ & $2.24^{+0.33}_{-0.35}$ & $2.06^{+0.30}_{-0.32}$ & 1.07 & 3.511 \\
c17-w1 & 0.32 & 32.021 & -4.583 & 3.85 & 32.045 & -4.542 & 171.4 & $0.35^{+0.10}_{-0.09}$ & $931.9^{+218.7}_{-239.2}$ & 1.12 & $0.35^{+0.11}_{-0.14}$ & $5.52^{+5.32}_{-3.20}$ & $4.81^{+4.63}_{-2.79}$ & 0.93 & 3.628 \\
c18-w1 & 0.25 & 31.797 & -4.002 & 3.04 & 31.772 & -3.963 & 165.9 & $0.51^{+0.45}_{-0.17}$ & $874.1^{+175.9}_{-340.5}$ & 0.69 & $0.50^{+0.21}_{-0.14}$ & $5.09^{+6.41}_{-3.42}$ & $4.56^{+5.75}_{-3.07}$ & 0.61 & 5.011 \\
c19-w1 & 0.29 & 31.793 & -3.894 & 4.37 & 31.792 & -3.854 & 142.4 & $0.39^{+0.19}_{-0.18}$ & $\sim 0$ & 6.94 & $0.38^{+0.20}_{-0.17}$ & $\sim 0$ & $\sim 0$ & 5.82 & 4.540 \\
c20-w1 & 0.19 & 31.283 & -4.972 & 3.17 & 31.308 & -4.993 & 118.5 & $0.12^{+0.12}_{-0.13}$ & $\sim 0$ & 5.12 & $0.11^{+0.11}_{-0.12}$ & $\sim 0$ & $\sim 0$ & 4.64 & 3.899 \\
c21-w1 & 0.42 & 31.558 & -6.167 & 4.34 & 31.538 & -6.167 & 70.2 & $0.32^{+0.14}_{-0.16}$ & $\sim 0$ & 3.73 & $0.31^{+0.13}_{-0.11}$ & $\sim 0$ & $\sim 0$ & 4.16 & 4.142 \\
c22-w1 & 0.41 & 31.272 & -5.923 & 4.63 & 31.254 & -5.943 & 95.4 & $0.51^{+0.26}_{-0.25}$ & $273.0^{+164.5}_{-157.3}$ & 1.51 & $0.49^{+0.15}_{-0.18}$ & $0.24^{+0.16}_{-0.15}$ & $0.22^{+0.15}_{-0.14}$ & 1.22 & 3.860 \\
c23-w1 & 0.35 & 31.667 & -7.384 & 4.89 & 31.665 & -7.400 & 57.0 & $0.26^{+0.21}_{-0.22}$ & $\sim 0$ & 3.14 & $0.26^{+0.19}_{-0.20}$ & $\sim 0$ & $\sim 0$ & 3.30 & 4.211 \\
c24-w1 & 0.40 & 31.358 & -7.588 & 10.03 & 31.353 & -7.589 & 20.6 & $0.88^{+0.55}_{-0.80}$ & $659.3^{+141.3}_{-181.0}$ & 0.82 & $0.88^{+0.11}_{-0.12}$ & $1.16^{+0.70}_{-0.45}$ & $1.11^{+0.67}_{-0.43}$ & 0.71 & 5.311 \\
c25-w1 & 0.42 & 31.927 & -7.502 & 3.33 & 31.953 & -7.489 & 101.7 & $0.32^{+0.31}_{-0.28}$ & $\sim 0$ & 3.48 & $0.32^{+0.29}_{-0.27}$ & $\sim 0$ & $\sim 0$ & 3.48 & 3.860 \\
c26-w1 & 0.24 & 31.320 & -7.680 & 3.00 & 31.326 & -7.664 & 63.5 & $0.22^{+0.13}_{-0.12}$ & $417.4^{+138.7}_{-131.5}$ & 0.86 & $0.22^{+0.12}_{-0.12}$ & $0.49^{+0.17}_{-0.16}$ & $0.43^{+0.15}_{-0.14}$ & 0.66 & 4.430 \\
c27-w1 & 0.26 & 31.595 & -8.807 & 4.22 & 31.617 & -8.806 & 75.8 & $0.17^{+0.09}_{-0.14}$ & $646.2^{+103.0}_{-135.1}$ & 1.07 & $0.19^{+0.11}_{-0.16}$ & $1.95^{+1.54}_{-1.27}$ & $1.67^{+1.32}_{-1.09}$ & 1.12 & 4.423 \\
c28-w1 & 0.08 & 31.769 & -8.495 & 3.43 & 31.724 & -8.512 & 171.6 & $0.06^{+0.10}_{-0.11}$ & $426.0^{+153.1}_{-166.8}$ & 2.41 & $0.08^{+0.12}_{-0.10}$ & $0.65^{+0.21}_{-0.18}$ & $0.55^{+0.18}_{-0.15}$ & 1.54 & 3.651 \\
c29-w1 & 0.27 & 31.230 & -10.400 & 14.90 & 31.230 & -10.407 & 24.4 & $0.21^{+0.16}_{-0.18}$ & $\sim 0$ & 5.94 & $0.22^{+0.13}_{-0.15}$ & $\sim 0$ & $\sim 0$ & 4.60 & 4.104 \\
c30-w1 & 0.19 & 32.773 & -4.129 & 3.19 & 32.758 & -4.151 & 96.1 & $0.37^{+0.13}_{-0.15}$ & $455.5^{+132.6}_{-138.9}$ & 0.34 & $0.40^{+0.14}_{-0.15}$ & $0.48^{+0.11}_{-0.13}$ & $0.43^{+0.10}_{-0.12}$ & 0.43 & 4.674 \\
c31-w1 & 0.34 & 32.634 & -4.123 & 6.94 & 32.653 & -4.132 & 76.5 & $0.28^{+0.12}_{-0.11}$ & $1134.9^{+243.1}_{-237.0}$ & 1.38 & $0.28^{+0.12}_{-0.12}$ & $12.3^{+2.41}_{-2.35}$ & $10.5^{+2.05}_{-2.00}$ & 1.17 & 4.899 \\
c32-w1 & 0.28 & 32.981 & -4.688 & 3.95 & 32.957 & -4.652 & 157.7 & $0.22^{+0.17}_{-0.14}$ & $\sim 0$ & 3.59 & $0.23^{+0.16}_{-0.17}$ & $\sim 0$ & $\sim 0$ & 3.15 & 3.737 \\
c33-w1 & 0.23 & 32.435 & -5.730 & 3.36 & 32.456 & -5.726 & 77.9 & $0.21^{+0.07}_{-0.10}$ & $599.6^{+97.8}_{-117.5}$ & 1.04 & $0.22^{+0.11}_{-0.14}$ & $1.09^{+0.12}_{-0.13}$ & $0.94^{+0.10}_{-0.11}$ & 1.11 & 3.576 \\
c34-w1 & 0.48 & 32.669 & -7.455 & 5.61 & 32.675 & -7.449 & 31.3 & $0.36^{+0.14}_{-0.11}$ & $703.7^{+150.7}_{-193.7}$ & 0.73 & $0.39^{+0.25}_{-0.17}$ & $1.71^{+2.27}_{-1.28}$ & $1.52^{+2.02}_{-1.14}$ & 0.68 & 4.474 \\
c35-w1 & 0.45 & 32.986 & -7.123 & 3.48 & 32.941 & -7.132 & 163.9 & $0.41^{+0.24}_{-0.23}$ & $1071.5^{+262.2}_{-239.6}$ & 1.89 & $0.42^{+0.17}_{-0.18}$ & $6.89^{+1.07}_{-1.13}$ & $6.10^{+0.95}_{-1.00}$ & 1.81 & 3.911 \\
c36-w1 & 0.40 & 32.865 & -6.694 & 3.31 & 32.839 & -6.726 & 145.2 & $0.36^{+0.15}_{-0.16}$ & $580.5^{+121.5}_{-124.3}$ & 1.51 & $0.36^{+0.16}_{-0.17}$ & $1.27^{+0.23}_{-0.25}$ & $1.13^{+0.20}_{-0.22}$ & 1.28 & 4.430 \\
c37-w1 & 0.62 & 32.599 & -6.543 & 6.99 & 32.575 & -6.541 & 85.3 & $0.51^{+0.29}_{-0.27}$ & $\sim 0$ & 4.12 & $0.51^{+0.28}_{-0.27}$ & $\sim 0$ & $\sim 0$ & 4.54 & 3.504 \\
c38-w1 & 0.33 & 32.837 & -8.396 & 8.91 & 32.824 & -8.403 & 55.2 & $0.39^{+0.10}_{-0.10}$ & $\sim 0$ & 3.28 & $0.41^{+0.11}_{-0.10}$ & $\sim 0$ & $\sim 0$ & 3.17 & 3.774 \\
c39-w1 & 0.34 & 32.135 & -7.730 & 3.54 & 32.167 & -7.750 & 133.5 & $0.37^{+0.17}_{-0.19}$ & $682.9^{+225.9}_{-241.1}$ & 1.34 & $0.36^{+0.16}_{-0.18}$ & $2.50^{+2.01}_{-1.67}$ & $2.19^{+1.76}_{-1.46}$ & 1.13 & 3.632 \\
c40-w1 & 0.46 & 33.012 & -7.676 & 9.58 & 33.020 & -7.659 & 69.4 & $0.76^{+0.13}_{-0.13}$ & $1101.3^{+261.1}_{-263.8}$ & 1.21 & $0.83^{+0.13}_{-0.15}$ & $4.49^{+1.18}_{-1.26}$ & $4.23^{+1.11}_{-1.19}$ & 1.05 & 4.486 \\
c41-w1 & 0.51 & 32.658 & -7.471 & 6.09 & 32.655 & -7.445 & 93.5 & $0.55^{+0.21}_{-0.18}$ & $\sim 0$ & 4.11 & $0.55^{+0.17}_{-0.16}$ & $\sim 0$ & $\sim 0$ & 3.65 & 4.157 \\
c42-w1 & 0.71 & 32.313 & -9.244 & 15.30 & 32.324 & -9.237 & 44.9 & $0.78^{+2.12}_{-0.32}$ & $530.9^{+138.9}_{-208.4}$ & 1.68 & $0.77^{+0.13}_{-0.12}$ & $1.10^{+0.60}_{-0.40}$ & $1.04^{+0.57}_{-0.38}$ & 1.31 & 3.792 \\
c43-w1 & 0.65 & 32.795 & -9.139 & 6.56 & 32.815 & -9.157 & 95.8 & $0.61^{+0.20}_{-0.12}$ & $678.1^{+112.3}_{-143.9}$ & 0.83 & $0.66^{+0.11}_{-0.09}$ & $1.80^{+1.15}_{-0.95}$ & $1.67^{+1.07}_{-0.88}$ & 0.66 & 3.815 \\
c44-w1 & 0.29 & 32.553 & -8.562 & 11.58 & 32.575 & -8.560 & 78.4 & $0.25^{+0.12}_{-0.11}$ & $971.1^{+342.1}_{-382.5}$ & 0.74 & $0.26^{+0.09}_{-0.08}$ & $10.9^{+5.61}_{-6.85}$ & $6.85^{+9.30}_{-4.78}$ & 0.71 & 4.367 \\
c45-w1 & 0.30 & 32.961 & -9.994 & 4.98 & 32.971 & -9.978 & 68.7 & $0.35^{+0.15}_{-0.16}$ & $\sim 0$ & 3.51 & $0.34^{+0.19}_{-0.17}$ & $\sim 0$ & $\sim 0$ & 3.38 & 4.109 \\
c46-w1 & 0.62 & 32.847 & -9.875 & 3.04 & 32.883 & -9.877 & 129.3 & $0.67^{+0.17}_{-0.16}$ & $521.8^{+133.9}_{-129.1}$ & 1.67 & $0.67^{+0.16}_{-0.18}$ & $0.66^{+0.12}_{-0.14}$ & $0.62^{+0.11}_{-0.13}$ & 1.56 & 4.238 \\
c47-w1 & 0.28 & 33.607 & -6.461 & 9.74 & 33.596 & -6.454 & 47.5 & $0.24^{+0.11}_{-0.10}$ & $698.8^{+198.2}_{-211.7}$ & 1.41 & $0.25^{+0.10}_{-0.10}$ & $2.35^{+1.12}_{-1.09}$ & $2.04^{+0.97}_{-0.95}$ & 1.28 & 3.562 \\
c48-w1 & 0.65 & 34.017 & -6.137 & 11.48 & 34.001 & -6.095 & 161.7 & $0.64^{+0.29}_{-0.14}$ & $152.8^{+162.4}_{-139.6}$ & 0.65 & $0.64^{+0.19}_{-0.15}$ & $0.07^{+0.11}_{-0.10}$ & $0.06^{+0.10}_{-0.09}$ & 0.63 & 3.930 \\
c49-w1 & 0.30 & 33.780 & -5.981 & 7.22 & 33.779 & -5.978 & 11.9 & $0.25^{+0.13}_{-0.13}$ & $547.5^{+120.8}_{-117.9}$ & 2.59 & $0.27^{+0.14}_{-0.11}$ & $0.78^{+0.18}_{-0.16}$ & $0.69^{+0.16}_{-0.14}$ & 1.83 & 4.728 \\
c50-w1 & 0.19 & 33.941 & -5.930 & 3.46 & 33.956 & -5.976 & 174.8 & $0.37^{+0.16}_{-0.17}$ & $667.4^{+156.3}_{-198.7}$ & 1.28 & $0.37^{+0.12}_{-0.14}$ & $2.49^{+2.22}_{-1.55}$ & $2.20^{+1.96}_{-1.37}$ & 1.18 & 3.579 \\
c51-w1 & 0.35 & 33.293 & -5.625 & 4.20 & 33.295 & -5.613 & 40.8 & $0.25^{+0.07}_{-0.10}$ & $629.1^{+98.4}_{-118.7}$ & 1.96 & $0.25^{+0.11}_{-0.13}$ & $1.86^{+2.20}_{-1.44}$ & $1.61^{+1.90}_{-1.25}$ & 1.70 & 4.205 \\
c52-w1 & 0.69 & 33.528 & -7.145 & 5.13 & 33.509 & -7.148 & 67.8 & $0.31^{+0.29}_{-0.30}$ & $\sim 0$ & 7.76 & $0.33^{+0.21}_{-0.22}$ & $\sim 0$ & $\sim 0$ & 6.81 & 3.739 \\
c53-w1 & 0.28 & 33.402 & -6.953 & 3.15 & 33.384 & -6.944 & 70.6 & $0.23^{+0.12}_{-0.11}$ & $694.7^{+133.5}_{-129.8}$ & 0.86 & $0.22^{+0.11}_{-0.11}$ & $1.65^{+0.29}_{-0.21}$ & $1.44^{+0.25}_{-0.18}$ & 0.61 & 4.166 \\
c54-w1 & 0.31 & 33.728 & -6.539 & 4.65 & 33.748 & -6.495 & 171.4 & $0.38^{+0.14}_{-0.15}$ & $193.1^{+109.6}_{-101.3}$ & 1.77 & $0.37^{+0.16}_{-0.14}$ & $0.19^{+0.10}_{-0.11}$ & $0.17^{+0.09}_{-0.10}$ & 1.73 & 3.590 \\
c55-w1 & 0.32 & 33.113 & -9.926 & 9.79 & 33.130 & -9.950 & 106.0 & $0.83^{+2.07}_{-0.30}$ & $606.1^{+163.2}_{-231.8}$ & 1.63 & $0.84^{+0.14}_{-0.13}$ & $0.97^{+0.33}_{-0.29}$ & $0.92^{+0.31}_{-0.28}$ & 1.58 & 3.809 \\
c56-w1 & 0.34 & 33.720 & -9.936 & 3.35 & 33.737 & -9.908 & 118.3 & $0.50^{+0.21}_{-0.19}$ & $597.4^{+147.1}_{-138.2}$ & 0.93 & $0.49^{+0.19}_{-0.19}$ & $1.09^{+0.19}_{-0.20}$ & $0.99^{+0.17}_{-0.18}$ & 0.94 & 3.872 \\
c57-w1 & 0.39 & 33.427 & -9.769 & 4.34 & 33.426 & -9.770 & 4.689 & $0.42^{+0.15}_{-0.14}$ & $514.7^{+128.7}_{-123.5}$ & 0.16 & $0.43^{+0.13}_{-0.15}$ & $0.74^{+0.16}_{-0.18}$ & $0.66^{+0.14}_{-0.16}$ & 0.16 & 3.570 \\
c58-w1 & 0.51 & 34.244 & -4.554 & 6.43 & 34.233 & -4.585 & 118.0 & $0.56^{+0.25}_{-0.16}$ & $449.3^{+211.9}_{-163.0}$ & 1.34 & $0.55^{+0.21}_{-0.18}$ & $0.45^{+0.20}_{-0.14}$ & $0.42^{+0.18}_{-0.13}$ & 1.21 & 6.448 \\
c59-w1 & 0.75 & 34.071 & -4.157 & 9.06 & 34.030 & -4.131 & 174.9 & $0.82^{+0.22}_{-0.23}$ & $\sim 0$ & 3.87 & $0.85^{+0.21}_{-0.21}$ & $\sim 0$ & $\sim 0$ & 3.70 & 3.564 \\
c60-w1 & 0.62 & 34.400 & -3.946 & 13.75 & 34.436 & -3.978 & 176.1 & $0.65^{+0.19}_{-0.17}$ & $830.3^{+196.8}_{-225.4}$ & 1.32 & $0.64^{+0.16}_{-0.17}$ & $4.63^{+1.15}_{-1.17}$ & $4.24^{+1.05}_{-1.07}$ & 1.23 & 4.363 \\
c61-w1 & 0.37 & 34.327 & -3.772 & 3.13 & 34.311 & -3.749 & 100.0 & $0.27^{+0.12}_{-0.14}$ & $\sim 0$ & 3.56 & $0.29^{+0.13}_{-0.12}$ & $\sim 0$ & $\sim 0$ & 3.21 & 3.791 \\
c62-w1 & 0.36 & 34.657 & -5.570 & 4.17 & 34.684 & -5.572 & 98.4 & $0.57^{+0.13}_{-0.14}$ & $909.7^{+61.5}_{-60.1}$ & 1.55 & $0.55^{+0.15}_{-0.13}$ & $5.31^{+0.32}_{-0.30}$ & $4.82^{+0.29}_{-0.27}$ & 1.47 & 4.194 \\
c63-w1 & 0.79 & 34.511 & -6.743 & 15.30 & 34.531 & -6.488 & 89.8 & $0.70^{+0.31}_{-0.28}$ & $\sim 0$ & 4.33 & $0.71^{+0.26}_{-0.25}$ & $\sim 0$ & $\sim 0$ & 4.09 & 3.931 \\
c64-w1 & 0.33 & 34.655 & -5.674 & 9.59 & 34.653 & -5.626 & 170.7 & $0.42^{+0.14}_{-0.11}$ & $\sim 0$ & 4.18 & $0.41^{+0.13}_{-0.10}$ & $\sim 0$ & $\sim 0$ & 3.87 & 4.186 \\
c65-w1 & 0.35 & 34.092 & -7.369 & 5.17 & 34.070 & -7.380 & 87.7 & $0.27^{+0.12}_{-0.12}$ & $\sim 0$ & 4.28 & $0.26^{+0.10}_{-0.09}$ & $\sim 0$ & $\sim 0$ & 4.09 & 5.230 \\
c66-w1 & 0.33 & 34.483 & -6.755 & 3.35 & 34.449 & -6.742 & 129.5 & $0.31^{+0.10}_{-0.19}$ & $\sim 0$ & 3.26 & $0.31^{+0.11}_{-0.10}$ & $\sim 0$ & $\sim 0$ & 3.13 & 4.692 \\
c67-w1 & 0.22 & 34.350 & -6.687 & 3.84 & 34.390 & -6.717 & 180.0 & $0.37^{+0.26}_{-0.19}$ & $794.7^{+235.7}_{-340.1}$ & 1.72 & $0.39^{+0.06}_{-0.05}$ & $3.63^{+0.90}_{-1.71}$ & $3.20^{+0.79}_{-1.51}$ & 1.51 & 4.515 \\
c68-w1 & 0.33 & 34.219 & -8.229 & 7.14 & 34.219 & -8.223 & 21.6 & $0.38^{+0.16}_{-0.17}$ & $\sim 0$ & 3.09 & $0.37^{+0.15}_{-0.16}$ & $\sim 0$ & $\sim 0$ & 3.06 & 3.967 \\
c69-w1 & 0.71 & 34.948 & -8.120 & 13.80 & 34.965 & -8.091 & 117.5 & $0.75^{+0.21}_{-0.18}$ & $514.0^{+123.4}_{-109.9}$ & 1.27 & $0.74^{+0.19}_{-0.18}$ & $0.63^{+0.15}_{-0.14}$ & $0.59^{+0.14}_{-0.13}$ & 1.22 & 3.737 \\
c70-w1 & 0.70 & 34.837 & -7.585 & 13.38 & 34.829 & -7.570 & 59.7 & $0.76^{+0.07}_{-0.08}$ & $545.4^{+94.2}_{-98.4}$ & 0.62 & $0.76^{+0.06}_{-0.08}$ & $0.90^{+0.13}_{-0.17}$ & $0.85^{+0.12}_{-0.16}$ & 0.58 & 4.497 \\
c71-w1 & 0.28 & 34.415 & -9.099 & 6.19 & 34.419 & -9.104 & 21.4 & $0.43^{+0.05}_{-0.06}$ & $1342.7^{+242.7}_{-249.2}$ & 0.95 & $0.44^{+0.06}_{-0.06}$ & $33.1^{+5.21}_{-5.32}$ & $28.7^{+4.51}_{-4.61}$ & 0.85 & 5.520 \\
c72-w1 & 0.31 & 34.476 & -9.849 & 4.14 & 34.439 & -9.837 & 138.6 & $0.31^{+0.13}_{-0.14}$ & $1079.6^{+251.6}_{-266.8}$ & 1.24 & $0.27^{+0.16}_{-0.14}$ & $13.2^{+2.86}_{-2.73}$ & $11.3^{+2.44}_{-2.33}$ & 1.23 & 3.705 \\
c73-w1 & 0.32 & 34.692 & -9.463 & 3.50 & 34.729 & -9.454 & 132.6 & $0.22^{+0.11}_{-0.15}$ & $\sim 0$ & 3.61 & $0.22^{+0.13}_{-0.15}$ & $\sim 0$ & $\sim 0$ & 3.58 & 3.800 \\
c74-w1 & 0.41 & 34.280 & -9.354 & 15.23 & 34.290 & -9.367 & 60.1 & $0.47^{+0.17}_{-0.18}$ & $\sim 0$ & 3.33 & $0.49^{+0.16}_{-0.17}$ & $\sim 0$ & $\sim 0$ & 3.30 & 3.990 \\
c75-w1 & 0.21 & 34.029 & -10.419 & 3.37 & 34.055 & -10.421 & 92.8 & $0.53^{+0.97}_{-0.17}$ & $506.9^{+140.8}_{-207.8}$ & 1.45 & $0.57^{+0.92}_{-0.18}$ & $0.57^{+0.82}_{-0.45}$ & $0.53^{+0.76}_{-0.42}$ & 1.42 & 4.841 \\
c76-w1 & -99.0 & 35.401 & -4.420 & 3.72 & 35.374 & -4.391 & 140.7 & $0.50^{+0.23}_{-0.25}$ & $\sim 0$ & 6.14 & $0.42^{+0.23}_{-0.21}$ & $\sim 0$ & $\sim 0$ & 5.80 & 4.355 \\
c77-w1 & 0.43 & 35.441 & -3.772 & 22.39 & 35.456 & -3.768 & 57.2 & $0.34^{+0.11}_{-0.12}$ & $766.1^{+182.5}_{-229.9}$ & 0.86 & $0.35^{+0.15}_{-0.15}$ & $2.04^{+2.78}_{-1.54}$ & $1.79^{+2.44}_{-1.35}$ & 0.82 & 4.209 \\
c78-w1 & 0.44 & 35.595 & -4.890 & 3.26 & 35.601 & -4.892 & 23.0 & $0.31^{+0.15}_{-0.13}$ & $\sim 0$ & 4.45 & $0.31^{+0.14}_{-0.15}$ & $\sim 0$ & $\sim 0$ & 4.17 & 3.788 \\
c79-w1 & 0.04 & 35.216 & -4.782 & 3.08 & 35.229 & -4.743 & 150.3 & $0.06^{+0.12}_{-0.11}$ & $573.8^{+113.9}_{-108.7}$ & 1.16 & $0.07^{+0.11}_{-0.10}$ & $2.78^{+0.33}_{-0.31}$ & $2.30^{+0.27}_{-0.26}$ & 1.09 & 4.204 \\
c80-w1 & 0.59 & 35.436 & -6.358 & 6.80 & 35.436 & -6.354 & 14.1 & $0.45^{+0.19}_{-0.17}$ & $\sim 0$ & 6.63 & $0.43^{+0.20}_{-0.19}$ & $\sim 0$ & $\sim 0$ & 5.90 & 4.710 \\
c81-w1 & 0.22 & 35.590 & -5.682 & 3.09 & 35.569 & -5.723 & 164.9 & $0.59^{+0.11}_{-0.12}$ & $1266.7^{+108.9}_{-117.3}$ & 1.35 & $0.67^{+0.13}_{-0.11}$ & $10.9^{+1.70}_{-1.50}$ & $9.97^{+1.56}_{-1.37}$ & 1.29 & 4.870 \\
c82-w1 & 0.37 & 35.650 & -7.270 & 3.34 & 35.680 & -7.281 & 112.4 & $0.32^{+0.13}_{-0.11}$ & $628.4^{+106.1}_{-104.6}$ & 1.22 & $0.33^{+0.12}_{-0.10}$ & $1.80^{+0.22}_{-0.25}$ & $1.59^{+0.19}_{-0.22}$ & 1.21 & 3.923 \\
c83-w1 & 0.32 & 35.830 & -7.997 & 5.21 & 35.826 & -7.992 & 21.4 & $0.08^{+0.07}_{-0.09}$ & $954.6^{120.0}_{132.6}$ & 1.68 & $0.10^{+0.04}_{-0.05}$ & $11.1^{+1.26}_{-1.30}$ & $9.10^{+1.03}_{-1.07}$ & 1.67 & 4.287 \\
c84-w1 & 0.02 & 35.924 & -7.989 & 3.40 & 35.931 & -7.994 & 31.5 & $0.16^{+0.04}_{-0.05}$ & $789.9^{+64.3}_{-69.1}$ & 0.15 & $0.12^{+0.02}_{-0.01}$ & $7.50^{+0.82}_{-0.77}$ & $6.21^{+0.68}_{-0.64}$ & 0.19 & 4.127 \\
c85-w1 & 0.35 & 35.486 & -8.950 & 7.92 & 35.486 & -8.937 & 48.4 & $0.25^{+0.15}_{-0.18}$ & $\sim 0$ & 3.98 & $0.24^{+0.17}_{-0.19}$ & $\sim 0$ & $\sim 0$ & 3.91 & 4.372 \\
c86-w1 & 0.27 & 35.868 & -8.865 & 5.82 & 35.883 & -8.864 & 55.4 & $0.34^{+0.16}_{-0.14}$ & $523.9^{+122.8}_{-118.9}$ & 0.65 & $0.36^{+0.15}_{-0.16}$ & $0.79^{+0.12}_{-0.13}$ & $0.70^{+0.11}_{-0.12}$ & 0.41 & 4.779 \\
c87-w1 & 0.32 & 35.821 & -9.347 & 6.25 & 35.805 & -9.344 & 61.1 & $0.38^{+0.19}_{-0.22}$ & $\sim 0$ & 3.52 & $0.39^{+0.18}_{-0.20}$ & $\sim 0$ & $\sim 0$ & 2.94 & 3.690 \\
c88-w1 & 0.24 & 35.941 & -10.924 & 5.53 & 35.935 & -10.888 & 129.8 & $0.26^{+0.08}_{-0.09}$ & $538.0^{+112.3}_{-107.2}$ & 0.27 & $0.25^{+0.10}_{-0.08}$ & $0.90^{+0.15}_{-0.16}$ & $0.78^{+0.13}_{-0.14}$ & 0.22 & 3.543 \\
c89-w1 & 0.34 & 35.827 & -10.399 & 3.79 & 35.828 & -10.406 & 26.4 & $0.43^{+0.17}_{-0.12}$ & $\sim 0$ & 4.77 & $0.42^{+0.16}_{-0.13}$ & $\sim 0$ & $\sim 0$ & 3.70 & 5.260 \\
c90-w1 & 0.19 & 35.520 & -10.346 & 4.55 & 35.488 & -10.353 & 118.1 & $0.08^{+0.11}_{-0.10}$ & $\sim 0$ & 2.96 & $0.10^{+0.12}_{-0.10}$ & $\sim 0$ & $\sim 0$ & 3.11 & 3.640 \\
c91-w1 & 0.20 & 35.284 & -10.330 & 4.14 & 35.317 & -10.305 & 149.6 & $0.28^{+0.13}_{-0.14}$ & $\sim 0$ & 4.19 & $0.26^{+0.14}_{-0.12}$ & $\sim 0$ & $\sim 0$ & 3.72 & 4.422 \\
c92-w1 & 0.65 & 36.372 & -4.258 & 4.79 & 36.372 & -4.248 & 36.9 & $0.65^{+0.26}_{-0.26}$ & $\sim 0$ & 5.69 & $0.66^{+0.22}_{-0.20}$ & $\sim 0$ & $\sim 0$ & 4.63 & 4.548 \\
c93-w1 & 0.30 & 36.121 & -4.165 & 3.36 & 36.097 & -4.167 & 87.5 & $0.26^{+0.13}_{-0.12}$ & $359.6^{+131.2}_{-121.7}$ & 0.25 & $0.25^{+0.11}_{-0.11}$ & $0.77^{+0.17}_{-0.18}$ & $0.67^{+0.15}_{-0.16}$ & 0.20 & 4.063 \\
c94-w1 & 0.55 & 36.108 & -5.088 & 3.13 & 36.077 & -5.102 & 122.2 & $0.20^{+0.17}_{-0.15}$ & $344.8^{+97.5}_{-89.1}$ & 0.78 & $0.20^{+0.18}_{-0.16}$ & $0.23^{+0.06}_{-0.04}$ & $0.20^{+0.05}_{-0.04}$ & 0.81 & 3.521 \\
c95-w1 & 0.35 & 36.617 & -4.998 & 10.12 & 36.636 & -4.990 & 72.8 & $0.22^{+0.14}_{-0.15}$ & $\sim 0$ & 4.55 & $0.21^{+0.13}_{-0.14}$ & $\sim 0$ & $\sim 0$ & 4.15 & 3.948 \\
c96-w1 & 0.50 & 36.121 & -4.821 & 8.92 & 36.116 & -4.852 & 114.5 & $0.47^{+0.12}_{-0.14}$ & $701.3^{+131.3}_{-129.0}$ & 0.61 & $0.47^{+0.13}_{-0.12}$ & $2.10^{+0.31}_{-0.27}$ & $1.89^{+0.28}_{-0.24}$ & 0.57 & 4.211 \\
c97-w1 & 0.29 & 36.455 & -5.896 & 10.48 & 36.465 & -5.892 & 38.6 & $0.35^{+0.11}_{-0.12}$ & $430.4^{+116.3}_{-120.2}$ & 0.79 & $0.35^{+0.13}_{-0.12}$ & $0.53^{+0.12}_{-0.11}$ & $0.47^{+0.11}_{-0.10}$ & 0.76 & 4.505 \\
c98-w1 & 0.32 & 36.631 & -5.695 & 3.81 & 36.635 & -5.692 & 17.3 & $0.23^{+0.10}_{-0.11}$ & $\sim 0$ & 4.40 & $0.24^{+0.12}_{-0.15}$ & $\sim 0$ & $\sim 0$ & 4.36 & 4.147 \\
c99-w1 & 0.51 & 36.890 & -7.462 & 3.34 & 36.903 & -7.483 & 88.4 & $0.48^{+0.13}_{-0.19}$ & $1055.6^{+263.4}_{-305.1}$ & 1.73 & $0.49^{+0.14}_{-0.18}$ & $6.03^{+1.27}_{-1.32}$ & $5.41^{+1.14}_{-1.18}$ & 1.51 & 3.546 \\
c100-w1 & 0.31 & 35.996 & -8.595 & 18.74 & 36.008 & -8.599 & 45.3 & $0.36^{+0.16}_{-0.14}$ & $905.3^{+182.9}_{-193.5}$ & 0.87 & $0.35^{+0.15}_{-0.16}$ & $4.53^{+0.85}_{-0.91}$ & $3.92^{+0.74}_{-0.79}$ & 0.73 & 4.631 \\
c101-w1 & 0.34 & 36.405 & -9.775 & 5.22 & 36.415 & -9.758 & 72.7 & $0.30^{+0.09}_{-0.07}$ & $670.8^{+103.7}_{-92.8}$ & 1.36 & $0.31^{+0.07}_{-0.06}$ & $1.26^{+0.19}_{-0.22}$ & $1.11^{+0.17}_{-0.19}$ & 1.23 & 3.856 \\
c102-w1 & 0.33 & 36.402 & -11.158 & 6.97 & 36.398 & -11.166 & 30.2 & $0.24^{+0.11}_{-0.19}$ & $\sim 0$ & 2.95 & $0.27^{+0.12}_{-0.11}$ & $\sim 0$ & $\sim 0$ & 3.08 & 4.075 \\
c103-w1 & 0.27 & 36.785 & -11.058 & 4.54 & 36.791 & -11.053 & 27.7 & $0.29^{+0.12}_{-0.15}$ & $301.5^{+152.2}_{-158.7}$ & 0.59 & $0.29^{+0.14}_{-0.13}$ & $0.15^{+0.09}_{-0.11}$ & $0.13^{+0.08}_{-0.10}$ & 0.50 & 3.725 \\
c104-w1 & 0.29 & 36.473 & -10.992 & 3.88 & 36.476 & -10.988 & 15.2 & $0.23^{+0.05}_{-0.05}$ & $1567.2^{+188.5}_{-185.9}$ & 0.96 & $0.24^{+0.06}_{-0.05}$ & $46.6^{+3.32}_{-2.45}$ & $38.9^{+2.77}_{-2.04}$ & 0.90 & 3.838 \\
c105-w1 & 0.29 & 37.199 & -5.589 & 6.37 & 37.200 & -5.618 & 103.7 & $0.32^{+0.07}_{-0.08}$ & $729.8^{+93.6}_{-89.8}$ & 0.19 & $0.33^{+0.08}_{-0.08}$ & $3.43^{+0.31}_{-0.22}$ & $2.97^{+0.27}_{-0.19}$ & 0.18 & 3.713 \\
c106-w1 & 0.32 & 37.662 & -4.991 & 10.71 & 37.670 & -4.988 & 36.7 & $0.79^{+3.20}_{-0.26}$ & $700.2^{+231.9}_{-322.5}$ & 1.53 & $0.75^{+1.23}_{-0.22}$ & $1.77^{+2.77}_{-1.56}$ & $1.66^{+2.60}_{-1.47}$ & 1.60 & 4.449 \\
c107-w1 & 0.34 & 37.720 & -4.855 & 7.63 & 37.724 & -4.863 & 30.6 & $0.27^{+0.08}_{-0.06}$ & $324.6^{+92.8}_{-83.7}$ & 0.69 & $0.26^{+0.08}_{-0.07}$ & $0.26^{+0.10}_{-0.11}$ & $0.23^{+0.09}_{-0.10}$ & 0.62 & 4.784 \\
c108-w1 & 0.30 & 37.358 & -4.816 & 3.08 & 37.359 & -4.835 & 71.2 & $0.33^{+0.13}_{-0.09}$ & $348.6^{+113.3}_{-104.2}$ & 0.56 & $0.33^{+0.12}_{-0.10}$ & $0.24^{+0.07}_{-0.09}$ & $0.21^{+0.06}_{-0.08}$ & 0.53 & 3.655 \\
c109-w1 & -99.0 & 37.000 & -6.397 & 3.17 & 37.022 & -6.429 & 117.5 & $0.46^{+0.16}_{-0.09}$ & $741.4^{+152.6}_{-194.8}$ & 1.72 & $0.43^{+0.16}_{-0.11}$ & $3.17^{+3.32}_{-2.07}$ & $2.82^{+2.95}_{-1.84}$ & 1.25 & 3.946 \\
c110-w1 & 0.49 & 37.812 & -5.572 & 3.10 & 37.784 & -5.587 & 116.9 & $0.52^{+0.17}_{-0.15}$ & $303.5^{+141.2}_{-133.1}$ & 0.89 & $0.52^{+0.15}_{-0.13}$ & $0.16^{+0.10}_{-0.07}$ & $0.15^{+0.09}_{-0.06}$ & 0.83 & 4.335 \\
c111-w1 & 0.28 & 37.767 & -7.269 & 10.60 & 37.776 & -7.272 & 33.9 & $0.37^{+0.12}_{-0.10}$ & $\sim 0$ & 3.17 & $0.36^{+0.11}_{-0.11}$ & $\sim 0$ & $\sim 0$ & 3.18 & 3.708 \\
c112-w1 & 0.36 & 37.541 & -7.536 & 3.04 & 37.540 & -7.513 & 81.0 & $0.42^{+0.16}_{-0.18}$ & $817.2^{+164.2}_{-177.0}$ & 0.87 & $0.43^{+0.17}_{-0.17}$ & $3.76^{+0.72}_{-0.74}$ & $3.34^{+0.64}_{-0.66}$ & 0.81 & 4.646 \\
c113-w1 & 0.19 & 37.354 & -7.494 & 5.18 & 37.340 & -7.486 & 58.8 & $0.39^{+0.09}_{-0.10}$ & $\sim 0$ & 4.83 & $0.38^{+0.10}_{-0.10}$ & $\sim 0$ & $\sim 0$ & 4.41 & 3.725 \\
c114-w1 & 0.37 & 37.088 & -9.226 & 3.35 & 37.096 & -9.242 & 62.4 & $0.35^{+0.11}_{-0.13}$ & $672.0^{+114.3}_{-119.5}$ & 1.21 & $0.35^{+0.12}_{-0.13}$ & $2.71^{+0.56}_{-0.51}$ & $2.39^{+0.49}_{-0.45}$ & 1.10 & 3.616 \\
c115-w1 & 0.27 & 37.299 & -8.901 & 3.48 & 37.294 & -8.892 & 35.3 & $0.31^{+0.12}_{-0.14}$ & $947.2^{+155.7}_{-168.3}$ & 2.47 & $0.32^{+0.13}_{-0.14}$ & $5.55^{+0.72}_{-0.75}$ & $4.77^{+0.62}_{-0.64}$ & 2.44 & 4.734 \\
c116-w1 & 0.28 & 37.355 & -8.841 & 5.10 & 37.347 & -8.835 & 34.1 & $0.26^{+0.16}_{-0.15}$ & $174.9^{+194.5}_{-189.3}$ & 2.81 & $0.25^{+0.16}_{-0.17}$ & $0.08^{+0.13}_{-0.14}$ & $0.07^{+0.12}_{-0.13}$ & 2.67 & 4.513 \\
c117-w1 & 0.64 & 37.828 & -11.147 & 5.49 & 37.802 & -11.140 & 93.2 & $0.77^{+0.27}_{-0.25}$ & $\sim 0$ & 5.64 & $0.74^{+0.24}_{-0.25}$ & $\sim 0$ & $\sim 0$ & 5.46 & 3.684 \\
c118-w1 & 0.27 & 37.922 & -4.883 & 12.07 & 37.936 & -4.880 & 54.2 & $0.29^{+0.13}_{-0.12}$ & $311.2^{+103.4}_{-97.4}$ & 0.56 & $0.30^{+0.11}_{-0.12}$ & $0.14^{+0.05}_{-0.06}$ & $0.12^{+0.04}_{-0.05}$ & 0.53 & 3.555 \\
c119-w1 & 0.49 & 38.525 & -4.728 & 3.14 & 38.483 & -4.737 & 153.6 & $0.46^{+0.14}_{-0.11}$ & $422.1^{+126.1}_{-135.6}$ & 0.48 & $0.46^{+0.13}_{-0.14}$ & $0.24^{+0.11}_{-0.08}$ & $0.22^{+0.10}_{-0.07}$ & 0.43 & 3.567 \\
c120-w1 & 0.31 & 38.048 & -6.491 & 5.70 & 38.022 & -6.501 & 102.0 & $0.12^{+0.11}_{-0.12}$ & $\sim 0$ & 3.77 & $0.16^{+0.12}_{-0.11}$ & $\sim 0$ & $\sim 0$ & 3.71 & 3.850 \\
c121-w1 & 0.33 & 37.925 & -7.790 & 3.65 & 37.938 & -7.828 & 145.0 & $0.37^{+0.12}_{-0.11}$ & $725.9^{+152.9}_{-143.8}$ & 0.16 & $0.37^{+0.14}_{-0.12}$ & $2.12^{+0.39}_{-0.43}$ & $1.86^{+0.34}_{-0.38}$ & 0.16 & 3.853 \\
c122-w1 & 0.43 & 37.998 & -7.624 & 3.44 & 37.986 & -7.658 & 130.0 & $0.33^{+0.15}_{-0.17}$ & $\sim 0$ & 3.83 & $0.32^{+0.16}_{-0.15}$ & $\sim 0$ & $\sim 0$ & 3.74 & 3.725 \\
c123-w1 & 0.37 & 38.688 & -8.801 & 3.62 & 38.737 & -8.797 & 172.1 & $0.41^{+0.13}_{-0.13}$ & $547.5^{+117.8}_{-102.7}$ & 1.19 & $0.40^{+0.12}_{-0.13}$ & $1.81^{+0.29}_{-0.31}$ & $1.60^{+0.26}_{-0.27}$ & 1.13 & 3.563 \\
c124-w1 & 0.23 & 38.669 & -9.835 & 3.70 & 38.630 & -9.843 & 141.1 & $0.26^{+0.15}_{-0.15}$ & $756.0^{+143.0}_{-139.6}$ & 0.57 & $0.26^{+0.15}_{-0.14}$ & $7.14^{+1.32}_{-1.27}$ & $6.09^{+1.13}_{-1.08}$ & 0.56 & 4.467 \\
c125-w1 & 0.63 & 37.902 & -9.562 & 4.82 & 37.907 & -9.595 & 117.5 & $0.59^{+0.29}_{-0.18}$ & $814.3^{+313.2}_{-187.3}$ & 1.53 & $0.57^{+0.33}_{-0.15}$ & $3.00^{+3.14}_{-1.90}$ & $2.74^{+2.86}_{-1.73}$ & 1.47 & 4.915 \\
c126-w1 & 0.26 & 38.639 & -10.472 & 5.31 & 38.647 & -10.458 & 57.3 & $0.44^{+0.10}_{-0.10}$ & $536.5^{+102.1}_{-108.7}$ & 1.62 & $0.40^{+0.11}_{-0.10}$ & $0.96^{+0.15}_{-0.13}$ & $0.86^{+0.13}_{-0.12}$ & 1.16 & 4.764 \\
\hline
\end{longtable}
\end{center}
\end{turnpage}

\newpage
\begin{sidewaystable*}
\begin{center}
\tabletypesize{\tiny}
\begin{longtable}{cccccccccccccccccccc}
\caption{\rm Catalog of the Matching Groups/Clusters between $\nu>3.5$
Convergence Peaks and X-Ray/K2-detected Groups/Clusters in CFHTLS-Wide W1 Fields. K2
redshift denotes the median redshift of bright ($i \le 20$) cluster members. X-ray redshift
denotes the photometric redshifts of X-ray clusters (Adami et al.\ 2011). The parameters $d_{\rm xray}$ and
$d_{\rm k2}$ are the offset between X-ray/optical and weak lensing center, respectively. \label{tab:matchxray}} \\
\hline
 ID & $\alpha_{\rm xray}$ & $\delta_{\rm xray}$ & $\alpha_{\rm K2}$ & $\delta_{\rm K2}$ & $\alpha$ & $\delta$ & $d_{\rm xray}$ & $d_{\rm k2}$ & $\nu$ & $z_{\rm K2}$ & $z_{\rm xray}$ & $\sigma_{v-xray}$ & $z_{\rm SIS}$ & $\sigma_{v}$ & $\chi^2_{\rm SIS}$ & $z_{\rm NFW}$ & $m_{\rm NFW}$ & $m_{200}$ & $\chi^2_{\rm NFW}$ \\
\hline
  &  &  &  &  & &  &  $\rm arcsec$ & $\rm arcsec$ & &  &  &  $\rm km/s$ &  & $\rm km/s$ & & & $10^{14}M_{\odot}/h$ & $10^{14}M_{\odot}/h$ & \\
\hline
J022145.2-034617 & 35.438 & -3.772 & 35.441 & -3.772 & 35.456 & -3.768 & 66.5 & 57.2 & 4.209 & 0.43 & 0.429$\pm$0.001 & 977$\pm$157 & $0.34^{+0.11}_{-0.12}$ & $766.1^{+182.5}_{-229.9}$ & 0.86 & $0.35^{+0.15}_{-0.15}$ & $2.04^{+2.78}_{-1.54}$ & $1.79^{+2.44}_{-1.35}$ & 0.82 \\
J022402.0-050525 & 36.008 & -5.090 & 36.108 & -5.088 & 36.077 & -5.101 & 251.1 & 122.2 & 3.521 & 0.55 & 0.324$\pm$0.001 & 364$\pm$69 & $0.20^{+0.17}_{-0.15}$ & $344.8^{+97.5}_{-89.1}$ & 0.78 & $0.20^{+0.18}_{-0.16}$ & $0.23^{+0.06}_{-0.04}$ & $0.20^{+0.05}_{-0.04}$ & 0.81 \\
J022433.8-041405 & 36.141 & -4.234 & 36.121 & -4.165 & 36.097 & -4.161 & 289.1 & 87.5 & 4.063 & 0.30 & 0.262$\pm$0.001 & 483$\pm$100 & $0.26^{+0.13}_{-0.12}$ & $359.6^{+131.2}_{-121.7}$ & 0.25 & $0.25^{+0.11}_{-0.11}$ & $0.77^{+0.17}_{-0.18}$ & $0.67^{+0.15}_{-0.16}$ & 0.20 \\
J022530.6-041420 & 36.377 & -4.239 & 36.372 & -4.258 & 36.372 & -4.248 & 36.3 & 36.9 & 4.548 & 0.65 & 0.14$\pm$0.002 & 899$\pm$218 & $0.65^{+0.26}_{-0.26}$ & $\sim 0$ & 5.69 & $0.66^{+0.22}_{-0.20}$ & $\sim 0$ & $\sim 0$ & 4.63 \\
J021837.0-054028 & 34.654 & -5.675 & 34.655 & -5.674 & 34.653 & -5.626 & 175.7 & 170.7 & 4.186 & 0.33 & 0.275$\pm$0.0 & - & $0.31^{+0.14}_{-0.11}$ & $\sim 0$ & 4.18 & $0.32^{+0.13}_{-0.10}$ & $\sim 0$ & $\sim 0$ & 3.87 \\
J021842.8-053254 & 34.678 & -5.548 & 34.657 & -5.570 & 34.684 & -5.572 & 91.1 & 98.4 & 4.194 & 0.36 & 0.38$\pm$0.001 & 847$\pm$279 & $0.57^{+0.13}_{-0.14}$ & $909.7^{+61.5}_{-60.1}$ & 1.55 & $0.55^{+0.15}_{-0.13}$ & $5.31^{+0.32}_{-0.30}$ & $4.82^{+0.29}_{-0.27}$ & 1.47 \\
J022632.4-050003 & 36.638 & -5.007 & 36.678 & -4.950 & 36.643 & -4.959 & 62.6 & 72.8 & 3.156 & 0.22 & 0.494$\pm$0.0   & - & $0.23^{+0.19}_{-0.17}$ & $\sim 0$ & 3.80 & $0.25^{+0.16}_{-0.21}$ & $\sim 0$ & $\sim 0$ & 3.81 \\
\hline
\end{longtable}
\end{center}
\end{sidewaystable*}
\normalsize

\end{document}